\documentclass[%
 reprint,
superscriptaddress,
nofootinbib,
 amsmath,amssymb,
 aps,
pra,
floatfix,
]{revtex4-2}
\usepackage[hidelinks]{hyperref}
\usepackage{amsmath,amsfonts,amssymb}
\usepackage{xcolor}
\usepackage{graphicx}
\usepackage{float}
\usepackage{amsthm}
\usepackage{appendix}
\usepackage{blkarray}
\usepackage{multirow}

\usepackage[ruled,vlined]{algorithm2e}

\theoremstyle{definition}
\newtheorem{definition}{Definition}[section]
\newtheorem{theorem}{Theorem}[section]
\newtheorem{lemma}{Lemma}[section]

\newtheorem{corollary}{Corollary}[theorem]

\newtheorem*{theorem*}{Theorem}
\numberwithin{equation}{section}
\newenvironment{hproof}{%
  \proof}{\endproof}

\DeclareMathOperator{\Tr}{Tr}

\begin{document}

\title{Quantum Lego: Building Quantum Error Correction Codes from Tensor Networks}

\author{ChunJun Cao}
 \email{ccj991@gmail.com}
\affiliation{Joint Center for Quantum Information and Computer Science, University of Maryland, College Park, MD, 20742, USA}

\author{Brad Lackey}
\email{Brad.Lackey@Microsoft.com}
\affiliation{Quantum Systems Group, Microsoft, Redmond, WA 98052, USA}

\begin{abstract}
We introduce a flexible and graphically intuitive framework that constructs complex quantum error correction codes from simple codes or states, generalizing code concatenation. More specifically, we represent the complex code constructions as tensor networks built from the tensors of simple codes or states in a modular fashion. Using a set of local moves known as operator pushing, one can derive properties of the more complex codes, such as transversal non-Clifford gates, by tracing the flow of operators in the network. The framework endows a network geometry to any code it builds and is valid for constructing stabilizer codes as well as non-stabilizer codes over qubits and qudits. For a contractible tensor network, the sequence of contractions also constructs a decoding/encoding circuit. To highlight the framework's range of capabilities and to provide a tutorial, we lay out some examples where we glue together simple stabilizer codes to construct non-trivial codes. These examples include the toric code and its variants, a holographic code with transversal non-Clifford operators, a 3d stabilizer code, and other stabilizer codes with interesting properties. Surprisingly, we find that the surface code is equivalent to the 2d Bacon-Shor code after ``dualizing'' its tensor network encoding map.
\end{abstract}

\maketitle
\tableofcontents

\section{Introduction}
Quantum error-correcting codes (QECCs) are critical ingredients for fault-tolerant quantum computation. In light of developing future large scale quantum computers, it is increasingly important to be able to design complex QECCs over a large number of qubit/qudits. Naturally, this points to the need for a simplifying framework that distills essential information from these complex quantum many-body states that are difficult to intuit for designers. Some pioneering examples have been explored in the context of quantum many-body systems and topological quantum computation\cite{toriccode,surfacecode,tqm,Kitaev2006,stringnet,Bombin2006,Bombin2013,HaahCode}. These are often constructed over geometries that are sufficiently regular, although some number of lattice defects have also been discussed \cite{RausHarrington,Fowler2009,BombinDef, Hastings,Nagayama}. 


On the other hand, tensor networks \cite{TNrev1,TNrev2,HoloReview} are tools that efficiently capture the features of complex quantum states. Therefore, they are natural candidates for studying QECCs. Indeed, this connection has been explored by \cite{FP2013} and more recently in the context of quantum gravity inspired by the developments in AdS/CFT \cite{HaPPY, Yang2016,RTN,BEG,DonnellyEdge,ABSC}. More practical aspects of these holographic codes in relation to quantum computing, many-body quantum states, and quantum field theory have also been discussed in  \cite{HoloSteane,HoloSteaneDec,Jahn2017,Jahn2019,Jahncc,HMERA,Jahn2020,Cree2021,HoloReview}. However, we argue that the underlying techniques implicit in these constructions, particularly those of the HaPPY code \cite{HaPPY}, can be extended far beyond the holographic contexts in which they are used. Some of this have been discussed by  \cite{TNC,Farrellypdec} in qubit stabilizer codes with certain restrictions to tensor contractions.


In this paper, we generalize the guiding principles behind the aforementioned works and propose a graphically intuitive and flexible framework for designing quantum error-correcting codes. 
More concretely, the underlying idea is analogous to playing with a lego set where one connects the ``quantum lego blocks'' from smaller codes that have useful and straightforward properties to create larger ``quantum lego structures'', which are complex quantum error-correcting codes\footnote{A paper \cite{Farrelly} with similar idea but somewhat complementary focus also appeared shortly after this work was released on arXiv. }.  This can be understood as a generalization of code concatenation. The properties of the larger codes can then be derived graphically following local operator flows known as operator pushing. 
While the ideas of operator pushing and using tensor network for code building are not new, the literature thus far have been restricted to the case of contracting tensors in directions that are isometric, which is a process that can be dualized to code concatenation. For instance, the operator pushing in \cite{HaPPY} relied on each tensor being an isometry along the radial direction. In our work we lift this restriction and show that operator pushing continues to hold in full generality, which now only relies on the symmetries of the lego blocks and a generalized matching condition that we derive. This significantly expands the previous capabilities of the code-building methods based on tensor networks. For example, by combining the tensors of codes along the directions that are non-isometric, one can easily construct the toric code from the simplest qubit erasure correction codes, which would not have been possible with the isometric restriction in place. In fact, we show that a quantum lego set of simple building blocks is capable of creating any code with our extension --- this universality establishes the framework as a powerful language.

This framework can be applied to study both stabilizer codes and non-stabilizer codes on both qubits and qudits, but when specialized to stabilizer codes, there is also a polynomial time algorithm (Appendix~\ref{app:e}) that determines the corresponding stabilizer generators via operator pushing. Although we do not provide a general decoding algorithm in this work, we emphasize that (exact and efficient) tensor-network-inspired decoding algorithms for certain known subclasses of these codes do exist. In addition, we find that for tensor networks that are efficiently contractible, the sequence of contractions explicitly constructs an encoding/decoding circuit. 

To better understand its capability and to provide explicit tutorials, we construct a few examples that include well-known existing codes as well as new codes by connecting simple building blocks of stabilizer codes over a few qubits. These code-building exercises not only can create new codes geometrically, but also can recast existing codes that did not have obvious geometric interpretations into the graphical language of tensor networks. On a more practical level, because the rules for code building are based on operator matching and are extremely simple, it provides a graphically intuitive language for analyzing and deriving properties of a complex code. Furthermore, their simplicity also accentuates the framework's potential for automated processes of code building, e.g. through machine learning, which are more natural and scalable in the long run.

Because of the flexibility in choosing different tensors and how they can be connected, the quantum lego framework can also customize known QECCs with tailor-made properties. For instance, one can modify the toric code with ease by altering the tensors in its network. This helps create variants, defects, and other customizable boundary conditions that are more irregular and respond to errors asymmetrically. It is particularly useful when one wishes to preserve some overall properties of a code while accommodating certain idiosyncrasies of the quantum hardware or personalized user requirements. 

As a generalization of code concatenation expressible in a graphical form, we hope it will be useful in devising decoding algorithms, deriving bounds for fault-tolerant thresholds, and connecting tensor network properties, e.g. contractibility, with decoding, circuit-building, and beyond. Because it has potential in building non-trivial codes by combining smaller tensors with poor error correction properties, we are also optimistic that this framework can be used to construct interesting (potentially non-additive) codes that support transversal non-Clifford operations from simple tensors on qubits and qudits.

In Sec~\ref{sec:2}, we provide a general overview of the framework where we define the quantum lego blocks in the form of tensors and discuss how they are joined. We show that with the encoding tensors of trivial one qudit codes, 2-qudit repetition codes and the tensors of the $|0\rangle$ state, any quantum code is constructible. When specialized to manipulating stabilizer code legos, we provide two new systematic results (Sec~\ref{subsubsec:stabcodes}) and a convenient method using check matrix manipulations (Appendix~\ref{app:d}).  We also show how decoding circuits/recovery maps may be constructed via tensor contraction in this framework and discuss connections with known error correction schemes.

In Sec~\ref{sec:3}, we construct explicit examples, including three new quantum codes, to highlight how they can be created and modified with relative ease using this framework. We first present a few simple examples for intuition building in Sec~\ref{subsec:3a}. Then we show how the famous toric code can be constructed with a straightforward arrangement by gluing tensors of the $[[4,2,2]]$ codes  (Sec~\ref{subsec:3b}). It is followed by various modifications of the toric/surface code to highlight its potential in customization (Sec~\ref{subsec:3c}). Surprisingly, we show that one can obtain the 2d Bacon-Shor codes from the 2d surface code tensor networks by simply re-interpreting some of the physical legs as logical legs. Therefore, this description unifies the two types of codes in a single  construction and provides a candidate tensor network for the quantum compass model. By extension, it establishes a novel connection between the well-studied $\mathbb{Z}_2$ gauge theory and a Hamiltonian with frustration which can be difficult to analyze.

For the remainder of Section~\ref{sec:3}, we use the quantum lego to construct a few codes that, to the best of our knowledge, have not appeared in literature. In Sec~\ref{subsec:3d} and \ref{subsec:3e}, we consider two codes that are inspired by discussions in quantum gravity. The first is a code with a flat geometry built from perfect tensors and the second is a holographic code that supports a transversal non-Clifford operator.  Both can be applied to magic state distillation. We finish the section with a new 3d subsystem code built from the Steane codes that supports localized stabilizers and fractal-like logical operators (Sec~\ref{subsec:3f}), which may be relevant from the point of view of topological quantum error correction and fractons. Finally in Sec~\ref{sec:4}, we summarize key features of this framework and possible future directions. 




\section{General Formalism}
\label{sec:2}
\subsection{Overview}
The general idea behind the framework is to first convert simple, well-understood building blocks, such as quantum codes or states over small number of qubits, into tensors. Then one connects these tensors into a tensor network (TN) to create more complex QECCs, not unlike building complex and versatile structures using lego blocks.  
Using the graph associated with the TN and known properties of the smaller code building blocks, one can then deduce certain characteristics of the larger code using local moves, also known as operator pushing, on the graph. The framework is not specific to stabilizer codes and is valid for qubits as well as qudits. However, in this work, we provide examples that are stabilizer codes as they have been more thoroughly studied.

More specifically, operator pushing allows one to construct the different representations of logical operators. If the resulting code is also produced by contracting stabilizer codes or states, then we also determine the stabilizer generators of the larger code using a polynomial time algorithm. To a lesser extent, this method provides us graphical clues to construct codes with different network geometries and localized stabilizer generators. Furthermore, if the tensor network is contractible, then the sequence of contractions will also map to an encoding/decoding circuit for the code. On syndrome decoding, one can employ a compatible scheme by \cite{FP2013, TNC} for a subclass of the tensor networks.

As we will soon explain, an isometric tensor can be written as an encoding map for a QECC\footnote{In the error correction language, if a tensor is isometric along certain contractions, it implies that the code it describes can correct any erasure errors at those locations.}. Therefore, a code-building process can be mapped to code concatenation when we only restrict ourselves to contracting tensors on legs that are isometric. Hence, the fully unrestricted network-building exercise we introduce here is a natural graphical generalization of code concatenation. 

For the rest of this work, we will only focus on tensor networks with constant uniform bond dimensions. Namely, the quantum lego blocks we use will be codes over qudits of the same dimensionality. Although the formalism in principle should allow us to build tensors that have varying bond dimensions on different legs, e.g., code word stabilized codes \cite{cwstab}, we will leave them to future work.

\subsection{Tensors as Quantum LEGO Blocks}
Before we can build a tensor network, we first discuss the ``tensors''. At its heart, tensors have been used to describe quantum states and mappings as they are simply the coefficients once a basis in the Hilbert space has been chosen. For a review on the subject, please see \cite{TNrev1,TNrev2}.

For concreteness, let us define an $[[n,k]]$ quantum code\footnote{We also allow the case where $k=0$, encoding a single state, and do not decorate our notation with the bond dimension $d$.} as a mapping $V:\mathcal{H}_d^{\otimes k}\rightarrow \mathcal{H}_d^{\otimes n}$, where $\mathcal{H}_d=\mathbb{C}^d$. For typical encoding maps, $V$ is also an isometry. One can easily convert it to a rank $n+k$ tensor $V_{i_1,\dots,i_{n+k}}$ by decomposing it over a basis

\begin{equation}
    V = \sum_{i_j} V_{i_1\dots i_{n+k}} |i_{k+1},\dots,i_{k+n}\rangle\langle i_{1}\dots,i_{k}|.
    \label{eqn:encodingV}
\end{equation}

\begin{figure}
    \centering
    \includegraphics[width=0.4\textwidth]{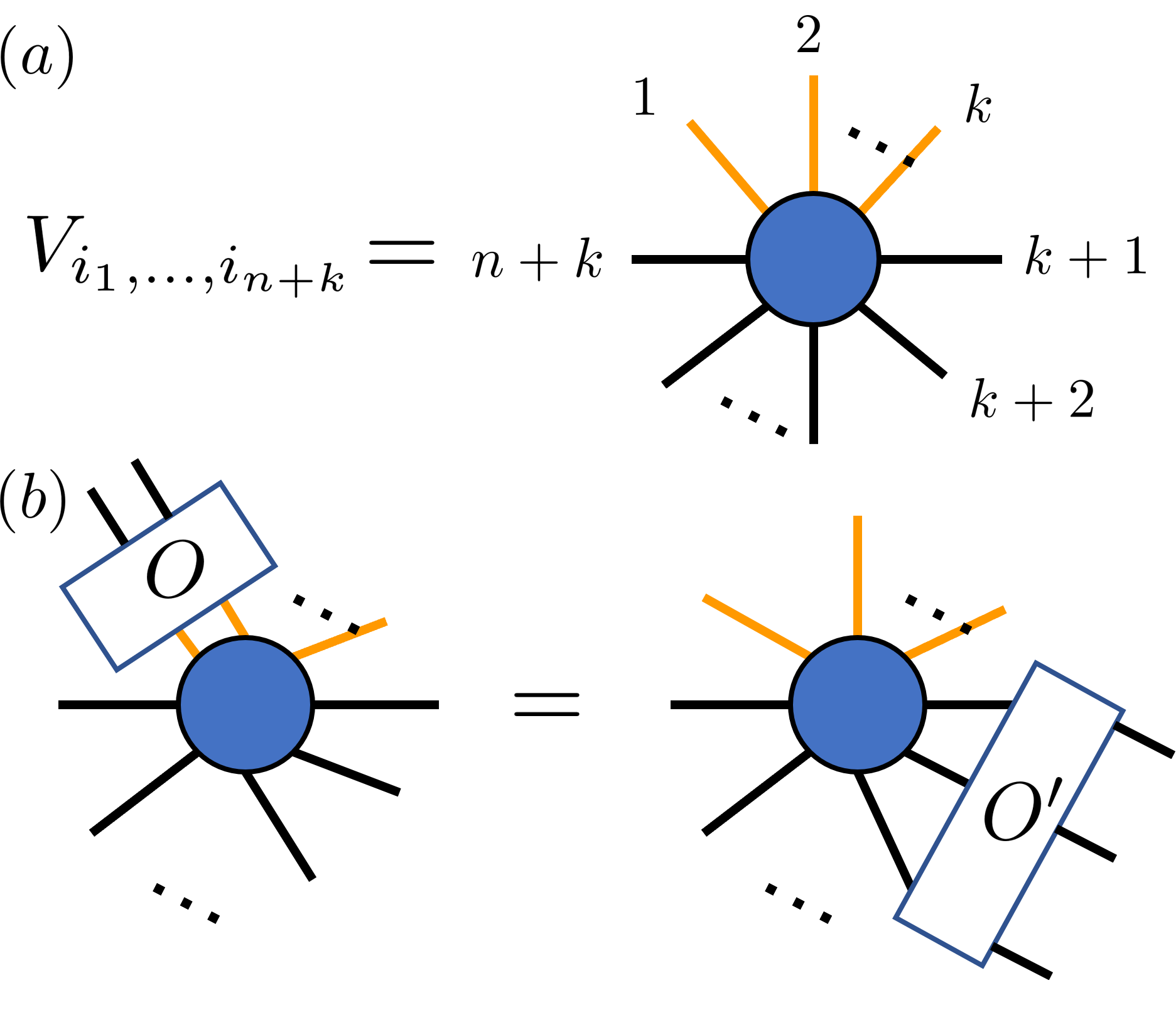}
    \caption{(a) A graphical representation of a tensor $V$ obtained from an encoding map. We have coloured the ``logical'' legs orange while the physical legs black. (b) Operator $O$ can be pushed through a tensor to $O'$ as long as $|V\rangle$ is a $+1$ eigenstate of $O^{-1}\otimes O'$ (Lemma~\ref{lemma:b2}). The same pushing rule also tells us that as a code, $O'$ acting on the physical qubits performs a logical operation $Q(O)$ defined in Lemma~\ref{lemma:b2} on the encoded qubits. If $O=I$ and $O'$ is Pauli, then clearly $O'$ is a stabilizer. }
    \label{fig:tensor_push}
\end{figure}

For each index $i_j$, one can assign a dangling edge. This tensor has a graphical interpretation (Figure~\ref{fig:tensor_push}a), where the dangling legs with bond dimension $d$ denote the total number of qubit/qudits the state or the mapping is over. The tensor is also an isometric tensor if $V$ is an isometric map. More detailed discussions of isometric tensors are found in Appendix~\ref{app:a}.

A careful reader will notice that the tensor $V_{i_1,\dots i_{n+k}}$ by itself does not specify whether it should be a mapping like (\ref{eqn:encodingV}) or a state of the form 
\begin{equation}
    |V\rangle = \sum_{i_j} V_{i_1\dots i_{n+k}} |i_1,\dots,i_{n+k}\rangle
\end{equation}
unless we provide additional information about the basis over which we sums over. 

For example, for an encoding isometry $V$ as in Figure~\ref{fig:tensor_push}a, one can divide the dangling legs of the tensor $V_{i_1\dots}$, using the decomposition we are given, into two types --- the physical legs, which denotes the number of physical qubits(qudits) over which the code is defined, and the logical legs, which denotes the number of logical qubits(qudits) it encodes. 
 One need not insist over this particular interpretation, however. Depending on the user's preference, it can be equally useful to convert the tensor into a state $|V\rangle$ over $n+k$ qubits(qudits) or a mapping from $k'$ to $n+k-k'$ qubits(qudits). For concrete examples, see the perfect tensor \cite{HaPPY} and perfect code \cite{Laflamme:1996iw} where this inter-conversion is done for the $[[5,1,3]]$ code.  

Indeed, these different manifestations of the same tensor can be interchanged through the channel-state duality\cite{choi,Jam}. For operations on the tensors, one can remain agnostic about what each dangling leg represents until the final step where we assign meaning to them. In this work, we will specify how we interpret these tensor legs in the tensor network as we derive the code properties\footnote{We can also choose $\{|i\rangle\}$ to be the computational basis unless otherwise specified although this choice is more or less irrelevant as we will never work with these tensor components directly in this paper.}. We will also make use of this flexibility to create new codes.

A tensor $V_{i_1,\dots,i_{n+k}}$ can be expensive to describe as it has exponentially many components. In this work, we are less interested in the tensor itself, but rather its underlying structures. We describe such structures by finding the unitary stabilizers of the state $|V\rangle$ described by the tensor. They encode the symmetries of the tensor. 
\begin{definition}
For any state $|V\rangle \in \mathcal{H}$, a unitary $U$ that satisfies $U|V\rangle=|V\rangle$ is called a unitary stabilizer of $|V\rangle$. Additionally, if they also satisfy $\mathcal{H}=\mathcal{H}_d^{\otimes N}$ for some prime $d$ and $U=\bigotimes_i^N U_i$, then $U$ is also a unitary product stabilizer (UPS).
\end{definition}
For example, when $U$ is also an element of the Pauli group, then it is nothing but a stabilizer of $|V\rangle$ in the usual sense. 

The set of unitary stabilizers of a tensor/code is of particular interest to us because they can be easily converted to logical operators of a code defined by the same tensor $V$. More importantly, they tell us whether operators acting on certain legs of the tensor can be ``pushed'' (Figure~\ref{fig:tensor_push}b) to some other operators supported on (a subset of) the complementary legs \footnote{More elaborate examples how it is used in holographic codes is discussed in\cite{HaPPY,ABSC}.}.
 See Lemma \ref{lemma:b2} and Appendix~\ref{app:b} for details. 
These pushing rules will allow us to navigate the larger tensor network and will help us greatly when creating codes with transversal operators.


It can be difficult to identify these unitary stabilizers for a general tensor $V$ due to computational costs. However, the problem becomes far more tractable if we take $V$ from relatively small or well-known quantum error correction codes, as their unitary stabilizers are either known or are relatively easy to find even by brute force. For stabilizer codes, it is sufficient to simply work with its stabilizers and logical Pauli operators, which are clearly unitary product stabilizers. Sometimes it is also beneficial to keep track of its non-Clifford operators in addition to Pauli operators, as we will see in later sections. Similarly, if there exist transversal single qubit/qudit logical operators in any non-stabilizer code constructions, then one can also convert them to UPS's of the state $|V\rangle$ generated by its encoding tensor. 
With non-stabilizer codes in mind, we also list a few categories of such tensors that will be pertinent to our current discussion.

The first are isometries. These are tensors where certain subsystems of the dual state $|V\rangle$ is maximally mixed. For instance, this is true for any subset of the orange legs for $V$ in Figure~\ref{fig:tensor_push} if it is an encoding isometry. See Appendix~\ref{app:a} for further definitions and properties. For these tensors, it was shown in \cite{HaPPY} that any unitary operator $O$ incident on legs that are maximally mixed can be pushed to an equivalent unitary operator acting on the other edges. This thus ensures a class of unitary stabilizers through Lemma~\ref{lemma:b2}.

More generally, we may not care about being able to push all operators $O$, but only particular operators, e.g. $YX$ and $TT$. In such cases, one has more flexibility in code constructions.  They admit unitary stabilizers that are tensor products of particular incoming operators and outgoing ones. Some examples are found in Sec~\ref{subsec:3b}, \ref{subsec:3f} and the double trace code in Sec \ref{subsec:3a}, but these unitary stabilizers need not be Paulis.

Finally, it is often desirable to build codes where certain logical operators are transversal. In this framework, transversality of (tensor product of) single qubit (qudit) logical operators\footnote{By multi-qubit gates, we refer to gates such as CNOT or Toffoli where they are non-trivial superpositions of tensor product of single qubit gates. } can often be read off in a straightforward manner provided we use tensors with suitable \textit{unitary product stabilizers} (UPS). These unitary stabilizers are simple tensor product of operators acting on each tensor leg. 
Tensors of such properties can be especially useful when constructing codes that have transversal non-Clifford gates because one can easily read off transversality of, e.g. T-gates, of the larger code via operator pushing once we understand the properties of the individual quantum lego blocks. See examples in Sec~\ref{subsec:3a} and \ref{subsec:3e}.

Tensors created from stabilizer states are examples of these latter two cases, although many also belong to the first category if they serve as isometric encoding maps with $k>0$ or correct erasure errors. Examples of more general tensors beyond tensors of stabilizer states have been discussed in the context of TNs with global symmetries \cite{Singh2010}. Such tensors are also in the intersection of the second and the third category without additional constraints.




\subsection{Conjoining the quantum lego blocks from tracing}
With suitable tensors in hand, we can now combine the blocks together. Graphically one can connect them by gluing some of the dangling legs. Algebraically, this ``tracing'' corresponds to summing over the indices of the legs that are glued. Equivalently, to create a connected edge, one can rewrite the tensors as states and project two qubits(qudits) to be glued to a maximally entangled state 
\begin{equation}
|\Phi^+\rangle = \sum_{i=0}^{d-1}|ii\rangle
\end{equation}
and rescale. Physically, it can be implemented via Bell measurements and post-selections. Here $d$ is again the bond dimension. See Appendix~\ref{app:b} for details. Such tracing operation eliminates some pairs of the physical/logical degrees of freedom in the tensor components and creates a new state or mapping which can define a quantum code. 

Recall that unitary stabilizers contain crucial information about a tensor and thus the code it defines, such as the logical operators and stabilizers, we wish to understand how they transform under the gluing operation. Let us now derive them for the larger tensor network from the unitary stabilizers of the individual tensors following a sequence of local moves called operator pushing. 
For the sake of simplicity and clarity, we will focus on a subset of unitary stabilizers called unitary product stabilizers as they are good for generating transversal operators. They will also be sufficient for understanding all the examples we discuss in Sec~\ref{sec:3}. For interested readers, we give a more detailed account on pushing general unitary stabilizers in Appendix~\ref{app:b}. 

\begin{figure}
    \centering
    \includegraphics[width=0.47\textwidth]{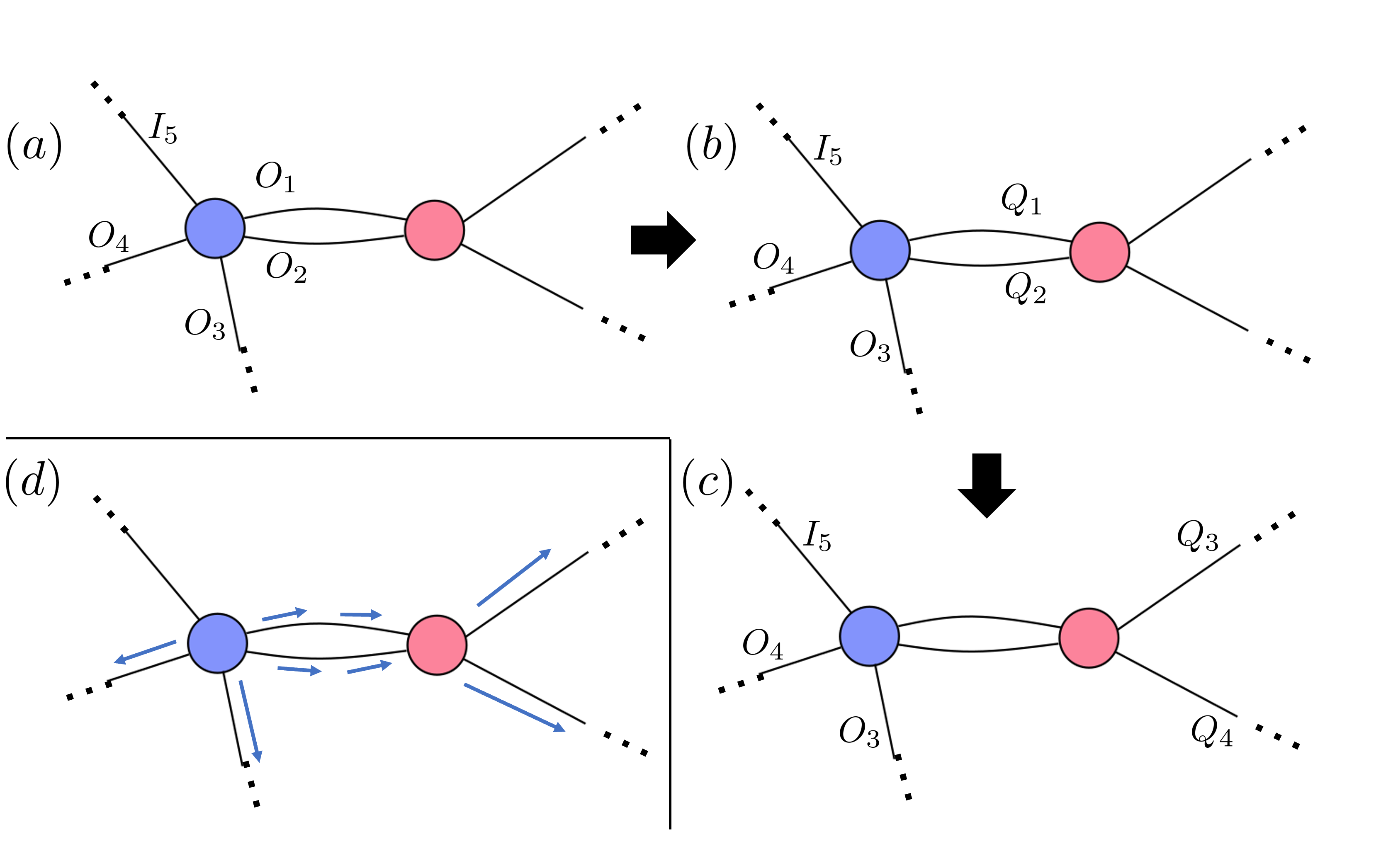}
    \caption{Consider two tensors adjacent to each other in a larger tensor network. (a) To find the UPS of a larger network, one inserts a UPS for a local tensor (blue), and (b) pushes it to the matching operators $Q_1,Q_2$ over connected edges. Then $Q_1,Q_2$ are pushed to the remaining legs of the adjacent tensor (pink) using the UPS of the pink tensor (c). (d) The whole process may be denoted as an operator flow, where the sequence of matching and pushing non-identity local operators are denoted by arrows along the edges.}
    \label{fig:TN_pushing}
\end{figure}

To begin, we insert a UPS of an individual tensor in the network (Figure~\ref{fig:TN_pushing}a). If such an operator only has non-trivial support over the dangling edges, then the resulting operator acting on the dangling edges is a unitary stabilizer of the larger tensor network. If such an operator has non-trivial support over a connected edge $e$ in the form of $O_e\ne I$, we look for a matching operator $Q_e$ such that 

\begin{equation}
    \langle\Phi^+|O_e\otimes Q_e=\langle\Phi^+|.
    \label{eqn:matching}
\end{equation}
We call $(O_e, Q_e)$ a matching set. If multiple edges connect the same two adjacent tensors, we find $\{Q_e\}$ for all such edges $e$. In other words, $O_e$ is pushed to $Q_e$ over each connected edge $e$ (Figure~\ref{fig:TN_pushing}b)\footnote{For Pauli operator over qubits, $Q_e$ and $O_e$ are identical up to a phase, therefore the operator stays the same over a connected edge. However, one should be more careful when pushing a general operator.}.

Then we push $\bigotimes_e Q_e$ through to the remaining legs of the adjacent tensor and off of the connected edges $e$ using unitary product stabilizers (UPS) of the adjacent tensors (Figure~\ref{fig:TN_pushing}c). The algorithm succeeds in finding a UPS of the larger tensor network if one can repeatedly perform this operation until the support of the inserted UPS of a tensor can be consistently ``pushed'' to only the dangling legs.  The process terminates in failure if a matching set can not be found in any of the connected edges or $\bigotimes_e Q_e$ cannot be cleaned off of connected edges via operator pushing. One can then repeat this exercise for other UPS insertions on the same tensor as well as on other tensors. A local example is shown in Figure~\ref{fig:TN_pushing}abc. We often drop the operator insertions to avoid clutter, but denote the sequence of such operator pushings and matchings through a flow diagram (Figure~\ref{fig:TN_pushing}d). An alternative description of the same process is to insert a UPS of each local tensor. If the operators inserted form matching sets on all connected edges according to (\ref{eqn:matching}), then we keep the resulting operator acting on the dangling edges as a UPS of the larger tensor network. We then repeat this process until we have exhausted all combinations of local UPS insertions. 

This procedure acquires for us the unitary (product) stabilizers of this larger tensor network. If we now designate the dangling legs of the tensor network to be physical and logical legs, we can re-interpret the network as an encoding map of a QECC. Using (\ref{cor:us_log}), we can then convert these unitary stabilizers to logical operators (and stabilizers) of the quantum error correction code. 

A particularly instructive subclass of codes with such UPS pushing are stabilizer codes, which we will discuss at length in the examples. The UPS we track are the stabilizers and logical operators, which fully characterize a given code. Occasionally we will also track non-Pauli operator in the interest of producing transversal non-Clifford gates (Sec~\ref{subsec:3e}). Because the number of stabilizer generators scale linearly with system size, operator matching/pushing can be tracked and performed more efficiently. As stabilizer codes can be specified by their check matrices, we also provide an equivalent description for the conjoining/tensor gluing over check matrices in Appendix~\ref{app:d}. In addition to our graphical guideline for deriving its stabilizers, we also provide an equivalent polynomial time algorithm based on check matrix operations that identifies the stabilizer generators and logical operators of the larger code (Appendix~\ref{app:e}). The latter description may be easier to implement in a machine. 

There is also incentive in creating TNs where consistent pushing can be easily identified beyond the pushing and matching of UPS's. (See Appendix~\ref{subapp:optpush}). For instance, there are toy models of holography where ``non-stabilizerness'' is crucial while transversality is less important \cite{White2020}. Therefore, in those cases, it is reasonable to consider general networks built from non-stabilizer isometries without many UPS's such that operator pushing can result in high weight non-transversal logical operators that are superpositions of Pauli operators. For general purpose tensor network computations, we also may not care that a particular operator has to be a tensor product of operators acting on each physical qudit. Instead, operators pushed through a network need not be fully transversal as long as the overall weight allows for efficient network contractions and computations of certain expectation values. For example, they can be the tensor product of low weight operators.

\subsubsection{Stabilizer Codes}
\label{subsubsec:stabcodes}
Stabilizer codes are a special but popular type of quantum code. Here we prove a few general statements on tracing that are also practically useful. Let us call the corresponding tensor of a stabilizer state a \textit{stabilizer tensor}. Since any stabilizer encoding map can be converted to a stabilizer state using the channel-state duality, it suffices to examine the tracing of stabilizer states or tensors. 


\begin{theorem}
\label{lm:stabtrace}
Tracing stabilizer tensor(s) produce another stabilizer tensor.
\end{theorem}
The proof is obvious from the check matrix operations~(Appendix \ref{app:d}), where the tracing produce check full rank matrices that can be used to describe the corresponding stabilizer state, and hence its associated tensor. This is also expected because tracing is a Clifford operation that does not increase stabilizer rank. As a tensor can be used as an encoding map with a potentially non-trivial kernel, it also describes a quantum code. 
Therefore, by iterating the above procedure, we are assured a method for producing stabilizer codes from quantum lego blocks that are themselves (smaller) stabilizer codes.

Calderbank-Shor-Steane (CSS) codes are a special class of stabilizer codes where the stabilizer generators are Pauli strings that are tensor products of $X$ and $I$ (or $Z$ and $I$) only. These CSS codes can be particularly desirable because of the transveral properties of certain Clifford operations. Let the tensor from a (self-dual) CSS code, including the case of $k=0$, be a \textit{(self-dual) CSS tensor}.
\begin{theorem}
\label{lm:csstrace}
Tracing (self-dual) CSS tensor(s) produce another (self-dual) CSS tensor. 
\end{theorem}
Because tracing correspond to operator matching where the $X$ and $Z$ type generators have to be matched to their own types, we can perform the matching on the $X$ and $Z$ sections of the check matrix separately. The resulting code is also a CSS code. If the initial codes are self-dual, then so will be the resulting code. Therefore, one can easily construct  CSS codes by gluing together  CSS tensors. Two such examples are shown in Sections~\ref{sec:3} with the toric code and the 3d code.  A proof of the above theorems can be found in Appendix~\ref{subapp:dproof}.

\subsection{Atomic Legos and expressivity}
A key question to ask is how general of a code can quantum legos construct. For example, can one find a simple set ``atomic codes'' or elementary lego blocks that can produce all QECCs using this framework? 

\begin{theorem}
\label{thm:atomic}
Any qudit quantum error correction code can be built from the following atomic lego blocks
\begin{itemize}
    \item Encoding tensor of a 2-qudit repetition code (rank 3)
    \item Encoding tensor of any 1-qudit code (rank 2)
    \item Tensor of $|0\rangle$ (rank 1)
\end{itemize}
\end{theorem}
A detailed proof is given in Appendix~\ref{app:atomiclego}, but the intuition behind the above statement is quite simple: a quantum gate can be seen as a tensor contraction. For example, tracing a particular tensor leg with the encoding tensor of a 1-qudit code corresponds to applying a single qudit unitary to that leg or qudit. Single-qudit measurements can be captured by a restricted form of contracting the rank-1 $|0\rangle$ and rank-2 tensors. By contracting the rank-3 encoding tensors with a rank-2 encoding tensor in a particular pattern, one can also re-express it as implementing a two-qudit gate. Therefore, a subset of the atomic lego contraction patterns map to a universal gate set. By combining them sequentially, one can then construct a quantum circuit that can be used to build up any state, and by extension, encoding map.

It immediately follows that if we restrict ourselves to the atomic legos that correspond to Clifford operations, e.g., the encoding tensor of repetition codes, rank 2 tensors from single qudit Clifford unitaries and rank 1 tensors from $|0\rangle$, then one can construct any stabilizer state, and by channel-state duality, the encoding tensor of any stabilizer code.

When tensors are restricted to the particular contractions patterns to form gates, then we see that the framework is at least as expressive as the circuit model. However, these tensor contractions are far more versatile in general, as they can be grouped differently than the ones given above. More general contraction patterns also need not be convertible to unitary time evolutions. 

\subsection{Decoding and Error Correction}
For codes whose tensor networks can be efficiently evaluated through sequential contraction of isometries, we can also construct its explicit local unitary decoding/encoding circuit $U$ based on this contraction sequence. Explicit examples of such contractible networks or codes include the MERA \cite{Vidal2008}, branching MERA \cite{branchingmera,bmeracode}, holographic codes \cite{HaPPY,Yang2016,HoloSteane,ABSC}, tree-tensor network or concatenated codes \cite{TNrev1,TNrev2,FP2013}. However, it is also straightforward to construct other different networks that have not been discussed widely using Algorithm~\ref{alg:TNbuild} in Appendix~\ref{app:c}. 

\begin{figure*}
    \centering
    \includegraphics[width=0.9\textwidth]{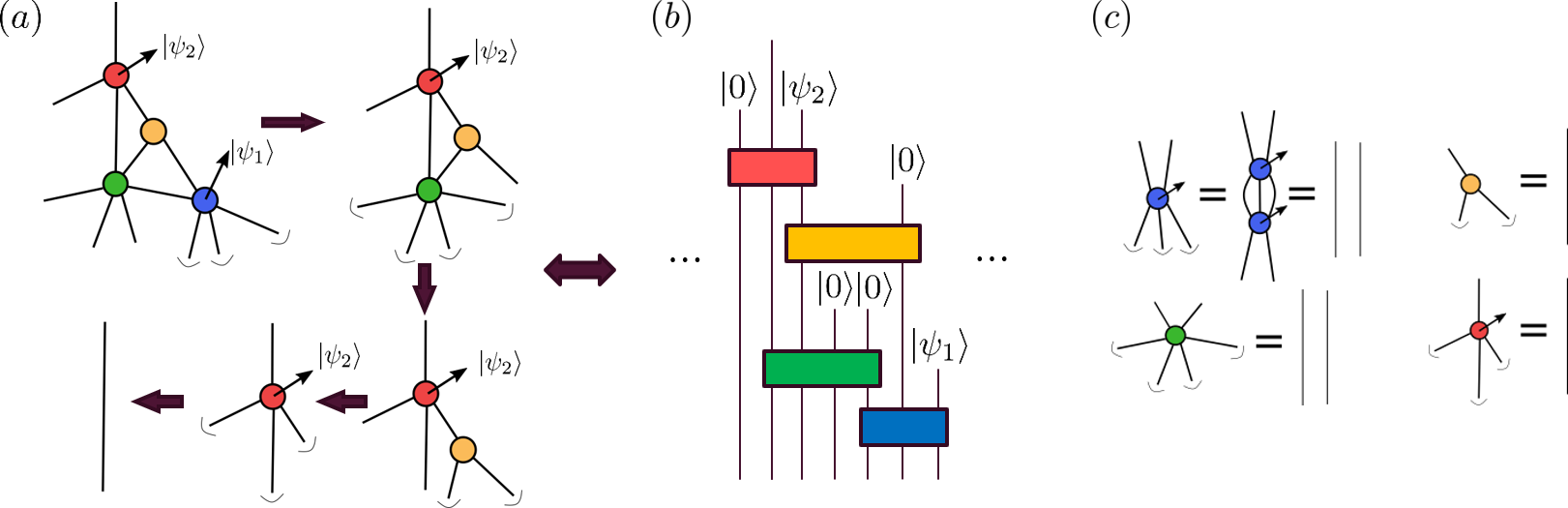}
    \caption{(a) The contraction sequence for a portion of the tensor network corresponds to (b) a unitary decoding circuit where each unitary gate can be independently derived from the tensor/code of the same colour. The arrows denote the encoded logical degrees of freedom\footnote{This is a notation similar to say, the HaPPY pentagon code, where the arrow corresponds to the bulk degree of freedom. From a tensor network perspective, the specific tensor that describes the isometry, e.g. the blue tensor, will depend on the encoded state. $|\psi_{1,2}\rangle$ are dummy states we use for notational clarity. The actual encoded state need not be a product state.} (c) The relevant tensors are isometries in that the contraction of certain edges reduce them to identity operators. A shorthand notation is used to avoid clutter. }
    \label{fig:cont_circuit}
\end{figure*}

To build the decoding circuit, one first lay down a wire for each dangling physical leg in the tensor network. The contraction of a certain tensor acting on a set of dangling physical legs maps to acting a unitary on the same set of wires in the circuit picture. Thus, each step in the contraction sequence maps to a time step in the quantum circuit (Figure~\ref{fig:cont_circuit}).
The corresponding unitary gates are guaranteed to exist and can be explicitly determined from each contracted tensor.  One can also terminate the contraction sequence and obtain a partial circuit if we only want to extract a subset of the data qubits. For explicit examples see Sec 5.4 of \cite{ABSC}. Detailed definition of the unitary gates and their existence is found in Appendix~\ref{app:a}. Information regarding the contraction sequence and circuit building is found in Appendix~\ref{app:c}. 

When the tensor networks described above also have bond dimension 2, we have a code on qubits whose encoding unitary $U$ is known. On the front of error correction from syndrome measurements, one can easily deploy known decoding algorithms for such codes by Ferris and Poulin \cite{FP2013}. In this case, the encoding unitary compatible with \cite{FP2013} can be constructed as in (\ref{eqn:fpcompU}). When contracted with tensors representing errors and syndromes, it produces the relevant conditional probability $Q(E|s)$ of error $E$ given syndrome $s$. When the tensors in the network have bounded and small degrees such that they are computationally tractable, then $Q(E|s)$ may also be efficiently computable. 

If the tensors further correspond to stabilizer codes, then one may also use the maximum likelihood decoder introduced in \cite{TNC,Farrellypdec}. In particular, all bond dimension 2 tensor networks generated by Algorithm~\ref{alg:TNbuild} using stabilizer code tensors will 
contain the tensor network codes (TNC) in \cite{TNC} as a subclass\footnote{However, note that  the algorithm does not cover all QECCs produced by the quantum lego method, especially those that are not contractions of isometries. PEPS, for example, are often difficult to contract. Interestingly, certain systems with topological order, such as the toric code, admit alternative descriptions in the form of exact MERA tensor networks, which are efficiently contractible. }. 
Therefore, we can apply the efficient decoder in \cite{TNC} for such kind of quantum lego codes.  Note that the tensor $T(L)$ in \cite{TNC} is different from our tensor --- the former is a bond dimension 4 tensor that records the logical operator Pauli strings and equivalent representations while the latter is a bond dimension 2 encoding map. Nevertheless, it is straightforward to connect the two formalisms using either encoding unitaries as shown in Appendix A of \cite{TNC} or more directly by enumerating the different representations of logical operators. For instance, one can reproduce the tensors $T(L)$ by enumerating all Pauli string representations of $L$ through row additions in the check matrices (See Appendices~\ref{app:d} and \ref{subapp:c3}). 

For some codes, although the exact contractions of individual isometries may not possible, it may still be viable to contract groups of these tensors into isometries or contract them approximately. In such cases, the above algorithm still applies modulo the differences in grouping the tensors. 
For other tensor networks that do not fall into the above subclasses and are not obviously contractible in the same way, such as the codes in Sec~\ref{subsec:3b} and \ref{subsec:3f}, decoding may still be devised on a case-by-case basis \cite{AguadoVidal}. Here we do not provide a comprehensive guideline for decoding in all such tensor networks attainable using the quantum lego, however, these are interesting future directions to be explored.






\section{Examples}
\label{sec:3}
Now with the quantum lego set in hand, we provide a few tutorials for building ``quantum lego structures'' using the tensor network technique. In addition to providing some hands-on exercises, we point out how such codes, some of which are highly non-trivial, can be constructed and modified with ease in the quantum lego framework. This suggests that the graphical framework can be a more convenient and intuitive way of creating and understanding QECCs.

We first start with a couple of simple examples for intuition building, then we build the toric code and the surface code to show that the technique is capable of constructing codes with non-trivial properties. Then we construct a few variants of surface codes, including the 2d Bacon-Shor code, using the high level of flexibility and customizability of the quantum lego. In particular, they can be useful in modifying known codes in a highly asymmetric manner in which defects and local properties can be tailored to the needs of the user. Finally, we construct and comment on a few potentially new codes which may be of interest in quantum error correction and possibly quantum gravity. We do not prove detailed code properties of these codes, but rather give their constructions in our formalism and outline a few interesting features that are relevant to our discussion. Further analysis of their properties, variants, and viability as good quantum codes are left as future work.

\subsection{Simple Codes}
\label{subsec:3a}
\begin{figure}
    \centering
    \includegraphics[width=0.48\textwidth]{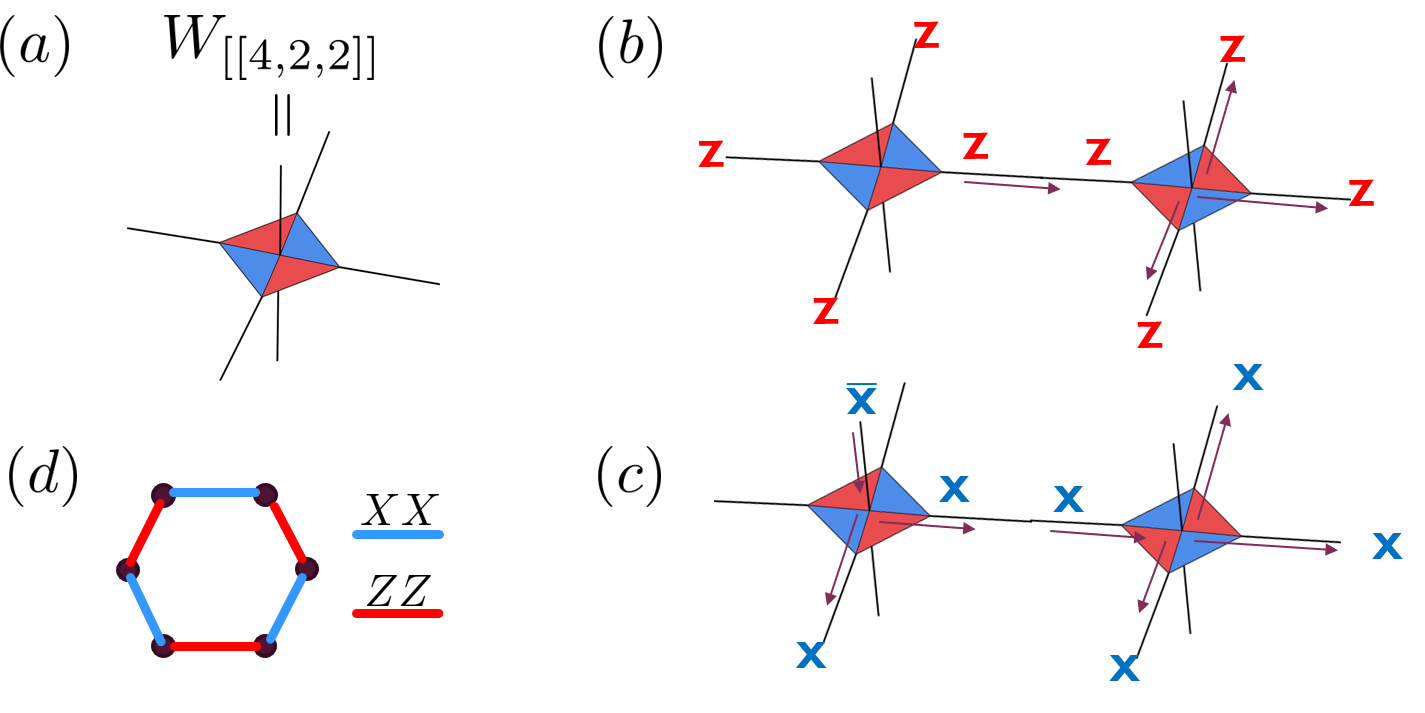}
    \caption{(a) Encoding map of the $[[4,2,2]]$ code as a rank 6 tensor. We naturally represent it as a 6 legged tensor where the in-plane legs correspond to the 4 physical qubits and the other 2 legs correspond to logical degrees of freedom. The colours are used to denote its orientation. (b) Generating a stabilizer operator $ZZZZZZ$ from pushing. (c) Physical representation of logical $\bar{X}_1\bar{I}_2\bar{I}_3\bar{I}_4$ by pushing $\bar{X}$ of the left tensor across the connected edge to the right tensor using a stabilizer (logical identity) of the right tensor. (d) The resulting $[[6,2,2]]$ gauge code where the generators of the gauge group are $XX,ZZ$ operators that act on the endpoints of the coloured edges.}
    \label{fig:422suite}
\end{figure}

\textit{Single-Trace 4-Qubit Code:}
 The encoding isometry $W_{[[4,2,2]]}:\mathcal{H}_2^{\otimes 2}\rightarrow \mathcal{H}_2^{\otimes 4}$ of a simple $[[4,2,2]]$ stabilizer code can be written as a rank 6 tensor with bond dimension 2 (Figure~\ref{fig:422suite}a). Its stabilizer group is $\langle XXXX,ZZZZ\rangle$. Logical $X$ and $Z$ operators are weight two products of $X$ and $Z$ respectively. This allows us to determine a characterizing set of unitary product stabilizers of the tensor. Some of them are shown in a graphical language in Figure~\ref{fig:422stabilizer}.  The particular colour-coding of the tensor is also useful in denoting its orientation --- see Appendix~\ref{subapp:422toric} for details.

Using these as  building blocks, we perform a single trace between two such tensors (Figure~\ref{fig:422suite}bc). One can treat the new tensor network as an encoding map that maps 4 logical degrees of freedom (the 4 legs pointing up or down) to 6 physical degrees of freedom (the 6 in-plane legs). Because $k>0$ and each $[[4,2,2]]$ code corrects one located error on any qubit, the resulting 6-qubit code has a distance no larger than 2. 
 
To characterize the new code, we find its UPS's from operator pushing. Here the relevant ones are the stabilizers (logical identity) and logical operators. Some practical guidelines for gluing stabilizer codes can be found in Appendix~\ref{subapp:optpush_practice}. 
To determine the stabilizers of the larger code, we insert the stabilizers of each tensor such that the operators acting on the contracted edge satisfy~(\ref{eqn:matching}). In this case, it translates into having two Paulis of the same type acting on the one connected edge. Equivalently, one can see it as inserting a stabilizer on the left tensor (without loss of generality) and ``pushing'' the Pauli operator on the connected edge through the right tensor by multiplying one of its stabilizers (Figure~\ref{fig:422suite}b). This produces an operator that has Pauli $Z$s acting on all the dangling edges. As a result, we  can conclude that $ZZZZZZ$ is a stabilizer. A similar exercise can be repeated for the all $X$ and all $Y$ stabilizers.

To produce logical operators of this code, one can repeat the above exercise by inserting some combinations of UPS's that correspond to logical and stabilizer operators (Figure~\ref{fig:422suite}c). This produces a $[[6,4,2]]$ stabilizer code where the 4 tensor legs pointing downward or upward correspond to the logical degrees of freedom. If one takes the downward pointing legs to be gauge qubits, then it is a $[[6,2,2]]$ gauge code where the generators of the gauge group are given by Figure~\ref{fig:422suite}d. This is nothing but a 6 qubit implementation of the generalized Bacon-Shor code \cite{BravyiGBSC,Zhang6qubit}.

One can also grow the chain of tensors in Figure~\ref{fig:422suite}c by gluing another $[[4,2,2]]$ tensor, say, onto the leg extending to the right. Such results in an $[[8,6,2]]$ code. In general, for a chain generated by gluing $m-1$ $[[4,2,2]]$ tensors, we arrive at a $[[2m,2m-2,2]]$ code\cite{Faist}. This particular way of generating codes by gluing legs where at least one of them corrects an erasure error is discussed in \cite{TNC}. In this case, the logical legs represent independent degrees of freedom.

{This series of tracing can also be understood as code concatenation. For instance, we can treat the right tensor in Figure~\ref{fig:422suite}bc as an (encoding) isometry which maps one qubit to five others.}

\begin{figure*}
    \centering
    \includegraphics[width=0.67\textwidth]{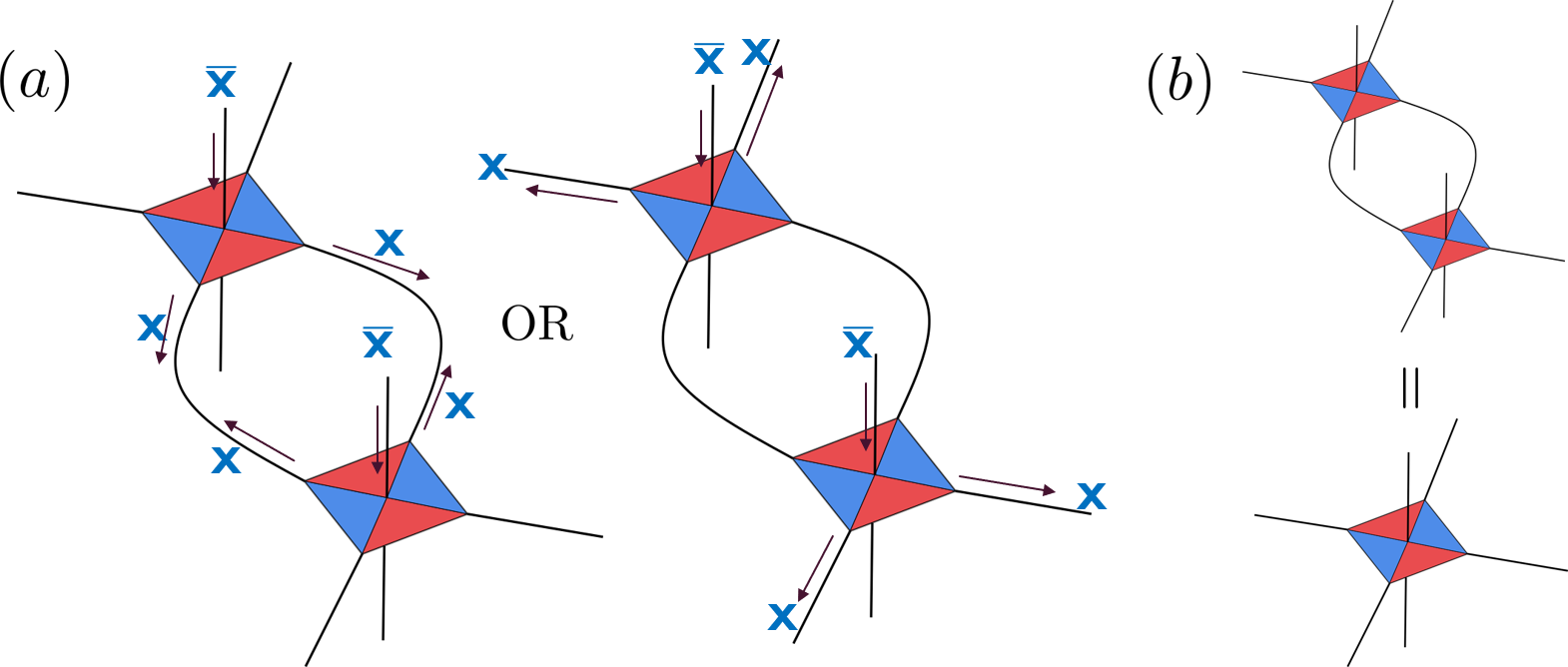}
    \caption{(a) Pushing $\bar{X}\bar{X}$ operators leads to a stabilizer because the two traced legs are not correctable errors. (b) The double-trace tensor network actually reduces to the original $[[4,2,2]]$ code where the two upward (or downward) pointing logical legs are identified because of the stabilizer constraint. }
    \label{fig:doubletrace}
\end{figure*}

\textit{Double-Trace 4-Qubit Code: } 
{However, the tensor tracing is not always equivalent to code concatenation. This is especially apparent when we pass physical qubits through a tensor that does not behave as an isometry. It occurs when we are tracing edges whose erasures are not correctable. } For instance, we can trace two edges in the above example (Figure~\ref{fig:doubletrace}b). Repeating the same operator-pushing exercise, we would find that the stabilizers are generated by $XXXX$ and $ZZZZ$. However, now the dangling edges in the tensor network that denote logical degrees of freedom are no longer independent, i.e., the encoding map has a non-trivial kernel. This can be seen as the $\bar{P}\bar{P}$ operator, where $P$ being a Pauli operator, is represented as a stabilizer on the physical qubits.  This serves as a constraint on the apparent logical degrees of freedom. Therefore, after subtracting the number of constraints from the apparent logical degrees of freedom, we are left with two total encoded qubit degree of freedom in the tensor network. {Because the non-trivial constraint here posed by the loop introduces a new stabilizer this procedure, it is not the same as concatenation.}
Indeed, one can verify either graphically, or algebraically through the check matrices\footnote{See Appendix D.2.1 of \cite{ABSC}.}, that this tensor network reduces to the original 6-legged encoding map (Figure~\ref{fig:doubletrace}b) after we mod out the kernel.

\textit{Steane Code from 4-Qubit Codes: } Thus far, we have kept two particular legs as logical from the tensor of the $[[4,2,2]]$ code. However, as we have pointed out in Sec~\ref{sec:2}, their designation as logical degrees of freedom in the tensor network is completely artificial. This freedom of choice can be seen as a consequence of the Choi-Jamiolkowski isomorphism (See Appendix~\ref{subapp:cj_dual}). 

\begin{figure*}
    \centering
    \includegraphics[width=0.8\textwidth]{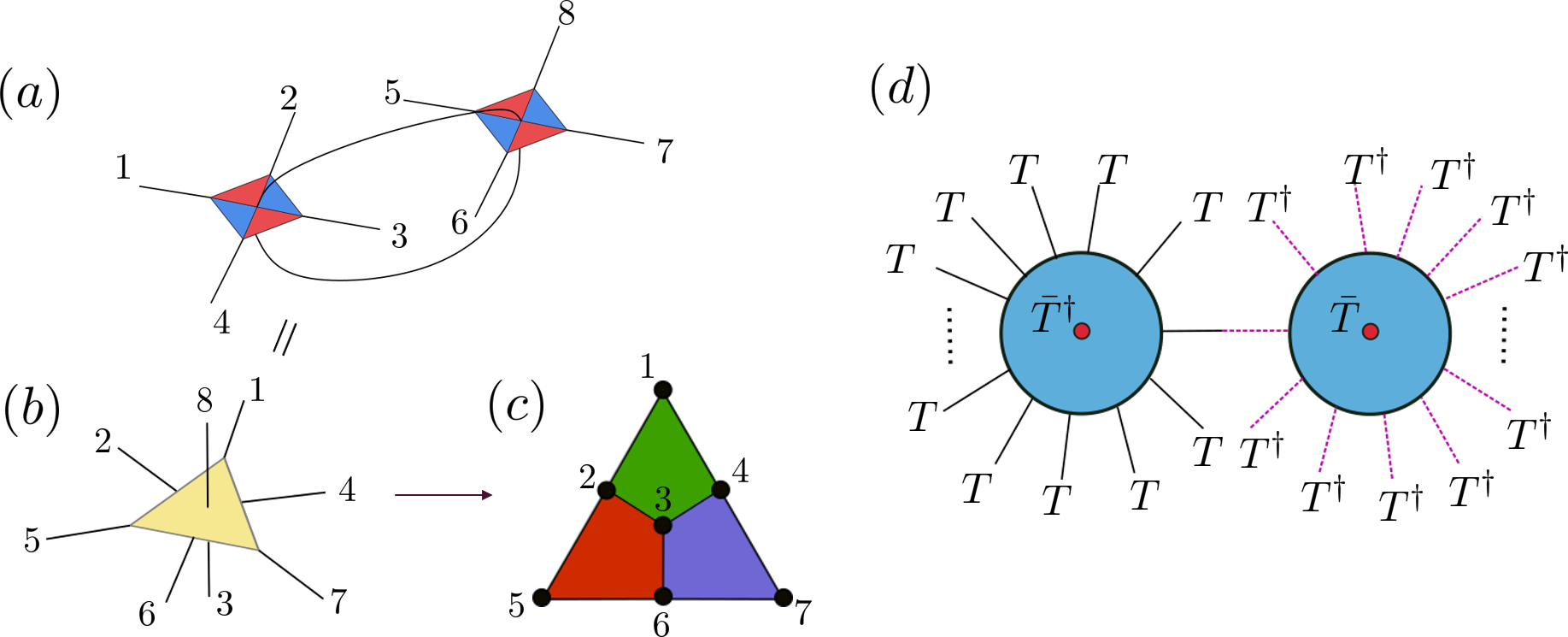}
    \caption{(a) Tracing together the ``logical'' degrees of freedom in two $[[4,2,2]]$ codes produces (b) an 8-legged tensor. Without loss of generality, we choose leg number 8 to be a logical degree of freedom. Then it becomes an encoding isometry of the Steane code \cite{SteaneCode} (c) where stabilizer generators are $XXXX,ZZZZ$ acting on the 4 vertices of each of the coloured faces. Logical X and Z operators are all X and all Z operators acting on all 7 vertices. (d) Because the 15-qubit Reed-Muller code has a transversal T gate, we can easily deduce from operator pushing that the code by tracing two such tensors would also have a transversal non-Clifford gate. Here the red dots represent dangling logical legs and $\bar{T}^{\dagger}\bar{T}$ can be pushed to the boundary dangling legs where $T$ acts on the black solid lines and $T^{\dagger}$ acts on the purple dashed lines.}
    \label{fig:steane_RM}
\end{figure*}

In this example, staying completely agnostic about the meaning of each dangling leg before tracing also allows us to construct codes with higher distances using exactly the same tensor component. 
Let us illustrate a new construction of the Steane code using these tensors from the 4 qubit codes. Here we simply trace over the dangling edges that correspond to the ``logical'' degrees of freedoms in the $[[4,2,2]]$ code  (Figure~\ref{fig:steane_RM}a). The resulting tensor network acts on 8 dangling legs (Figure~\ref{fig:steane_RM}b). Using operator pushing, we arrive at an 8-qubit stabilizer state or a 7-qubit stabilizer code if we choose any one leg to be the logical degree of freedom. One can easily check that this is nothing but the $[[7,1,3]]$ Steane code (Figure~\ref{fig:steane_RM}c), where the stabilizer generators are $XXXX$ and $ZZZZ$ that act on each vertex of the colored faces\footnote{Detailed derivation is left as an exercise for the reader.}.

\textit{Traced Reed-Muller Code with Transversal Non-Clifford Gate: } While it is also straightforward to analyze the resulting stabilizer codes from tracing using check matrices (Appendix~\ref{app:d}), the graphical analysis of certain operators in the code has distinct advantages. One particular aspect is designing codes that support certain types of transversal operators. Because tracing together tensors that are stabilized by the tensor product of unitary operators also lead to a new tensor of a similar property (Corollary~\ref{cor:transversal}), this allows us to produce, by simple observation, more complex codes with transversal non-Clifford gates by picking and gluing the right kinds of tensors. For a somewhat trivial example, we can trace together two identical $[[15,1,3]]$ codes \cite{1996Knill,Anderson2014} with transversal $\bar{T}=T^{\otimes 5}$. For a non-Clifford gate, we can push $\bar{T}$ from one logical leg and $\bar{T}^{\dagger}$ on the other. This produces a transversal $\bar{T}\bar{T}^{\dagger}=T^{\otimes 14}\otimes (T^{\dagger})^{\otimes 14}$ (Figure~\ref{fig:steane_RM}d) such that $T$ and $T^{-1}$ act on the connected edge, as required by operator pushing (\ref{eqn:matching}). It is also easy to build a code with the same physical operator but implements $\bar{T}\bar{T}$ by attaching an extra $X$ gate on the logical leg on one of the tensors\footnote{In a way, this is similar to modifying the tensors in the XZZX code except there we attach Hadamards. See Sec~\ref{subsec:3c} and Appendix \ref{subsubapp:xzzx}.}. In the same way, it is easy to build other subsystem codes with transversal $T$ operators. For example, we give a slightly more complex construction of a holographic code that has a transversal $T$ gate in Sec~\ref{subsec:3e}.

\subsection{Toric Code}
\label{subsec:3b}
\begin{figure*}
    \centering
    \includegraphics[width=\textwidth]{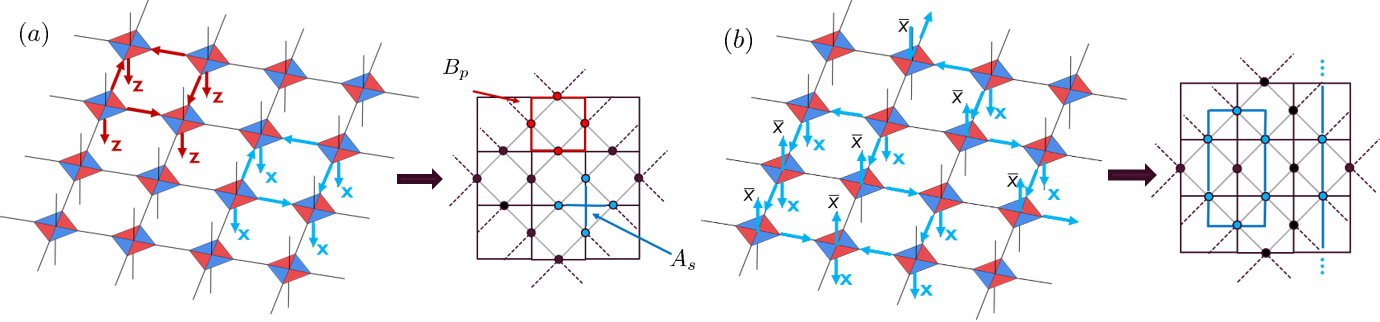}
    \caption{(a) One can recover the star $A_s$ and plaquette $B_p$ operators by pushing the stabilizers of local tensor modules. (b) Similarly, one can push through a combination of stabilizers and logical operators to obtain closed contractible loops (stabilizers/logical identity) and uncontractible loops (logical operators). The lattice of the tensor network is rotated by 45 degrees relative to the conventional toric code lattice. When the tensor network is overlaid with the toric code lattice, we view the network from the top down such that the logical legs point out of the page and the physical legs point into the page. A vertex on the toric code lattice is coloured red (or blue) if a $Z$ (or $X$) acts on the physical qubit.}
    \label{fig:toric_code_network}
\end{figure*}

Using  the $[[4,2,2]]$ tensor, one can also construct more geometric-looking codes with highly non-trivial properties. One such example is the toric code \cite{toriccode}. By connecting the tensors in a square lattice (Figure~\ref{fig:toric_code_network}), we create a PEPS-like tensor network where a physical qubit lies on each vertex\footnote{The form of the tensor network is similar to that of \cite{PEPStoric} but has additional ``logical legs'' that are also dangling, which are necessary for us to indicate that it is an encoding map. }. 

 The tensors that make up of the PEPS\footnote{It can be easily converted into a toric code ground state by contracting, say, $|0\rangle$s with the logical inputs. This is also similar to \cite{Brell2011} if the projection onto Bell pairs during tensor gluing is enforced by a two-body Hamiltonian and treating the four qubit code as a subsystem code.} have two orientations, which are related to each other by a $\pi/2$ rotation. To make it periodic, one can then identify the sides of the network such that the unrotated lattice (right diagram in Figure~\ref{fig:toric_code_network}a) is periodic in the horizontal and vertical directions. We have chosen  the dangling leg pointing up as logical and the one pointing down as physical. Note that because we are tracing over legs that are not correctable erasure errors, the apparent logical degrees of freedom in the resulting tensor network are not independent, not unlike the double trace example in Sec~\ref{subsec:3a}.  As such, pushing some logical operators of local tensors result in stabilizers, which serve as constraints, or a kernel in our logical space. After subtracting the dimension of these constraints, we can reduce the $2L^2$ number of apparent logical degrees of freedom into two. Similar to the double trace example above (Figure~\ref{fig:doubletrace}), we see that the toric code construction again contains a large number of loops and non-trivial constraints from contracting tensors that are not acting as isometric encoding maps. This is a more sophisticated example where the tensor network construction is going beyond code concatenation.  We further elaborate its construction and degrees of freedom counting in Appendix~\ref{subapp:422toric}. 


To find the stabilizers of the larger tensor network, we insert the stabilizers of a 6 legged tensor treating it as a $[[5,1,2]]$ code\footnote{These are stabilizers of a $[[5,1,2]]$ code because we take 5 of the legs to be physical and the remaining one pointing up as logical.} and pushing it around the network using stabilizers of the neighbouring tensors. The resulting stabilizer generators can be organized into the product of all X (or all Z) acting on four corners of a square (Figure~\ref{fig:toric_code_network}a). However, these are nothing but the star and plaquette operators in the toric code.  
A similar exercise applies for the logical operator, where a logical operator of a six-legged tensor is injected through a logical dangling leg of the tensor network, and pushed using logical or stabilizer operators (Figure~\ref{fig:toric_code_network}b).

While this is clearly a tensor network construction of the toric code ground space, it is also interesting that just by connecting these somewhat trivial codes with low distances, and by pushing unitary stabilizers of individual tensors, we can graphically reconstruct the relevant operators of the toric code in a simple and intuitive manner. This example also highlights the difference between the quantum lego method and the more generic tensor network approaches to many-body quantum states --- working with operator pushing is easier than dealing with the tensor components directly when tackling specific problems like code-building.

\subsection{Surface code and variants}
\label{subsec:3c}
\begin{figure}
    \centering
    \includegraphics[width=0.48\textwidth]{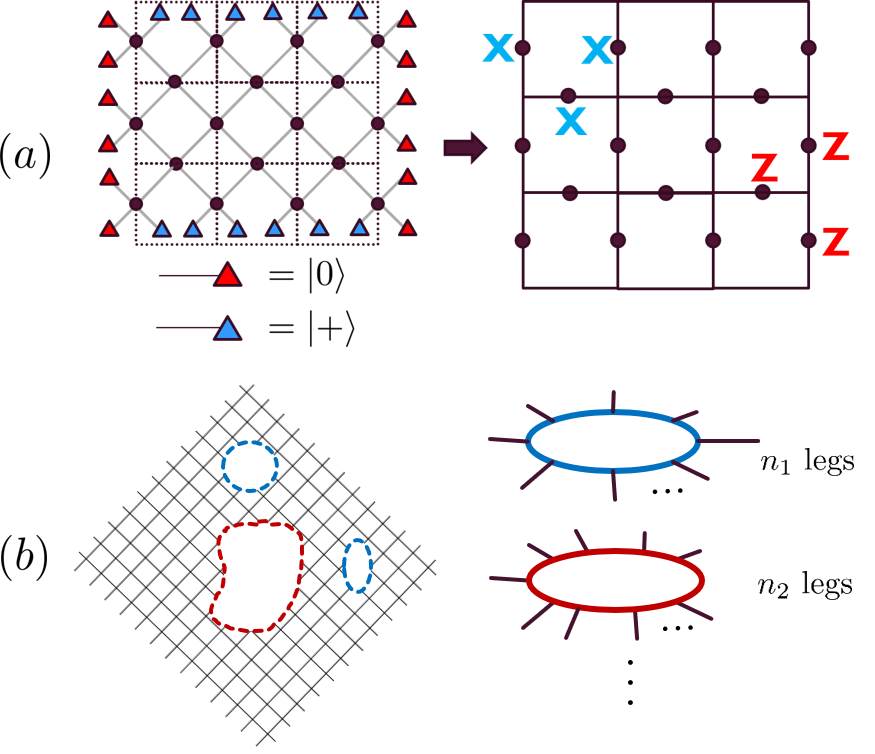}
    \caption{(a) By contracting the dangling legs of a patch of the tensor network with the X and Z eigenstates, one can confirm using operator pushing that we recreate the 2d surface code with the correct number of qubits and the different boundary conditions. The grey solid lines represent the contracted in-plane edges. For each site/tensor marked by a black dot, there are also two dangling legs pointing into and out of the page. These legs are not shown to avoid clutter. (b) One can string together different repetition codes in a tensor network (as denoted by the different colours) and connect some of them to the dangling legs of the 2d tensor network with holes. Each coloured boundary represents contraction with a  boundary tensor of the same colour on the right.}
    \label{fig:surface_code}
\end{figure}

One can also construct different variants by taking a patch of the above tensor network and impose different ``boundary conditions'' on the tensors. As there is a wide variety of tensors that can be used to create distinct boundary conditions, we here only list a few examples. For instance, the bare boundary condition where we leave all boundary legs open creates a subsystem code that also has weight-3 stabilizer generators along the boundary of the 2d surface. This is not the typical surface code, as we have tripled the number of boundary degrees of freedom. Another straightforward example is the construction of the conventional surface code with the rough and smooth boundaries by contracting with the ``stopper'' tensors, which are simply $X$ or $Z$ eigenstates (Figure~\ref{fig:surface_code}a). 
The stopper tensors are essentially disconnected nodes. For a boundary tensor that has more connectivity, one can also connect the boundary dangling legs to tensors obtained from different repetitions codes (Figure~\ref{fig:surface_code}b). This creates subsystem codes that have potentially different properties and defects. See Appendix~\ref{subapp:defects} for more detailed constructions. 

\begin{figure}
    \centering
    \includegraphics[width=0.47\textwidth]{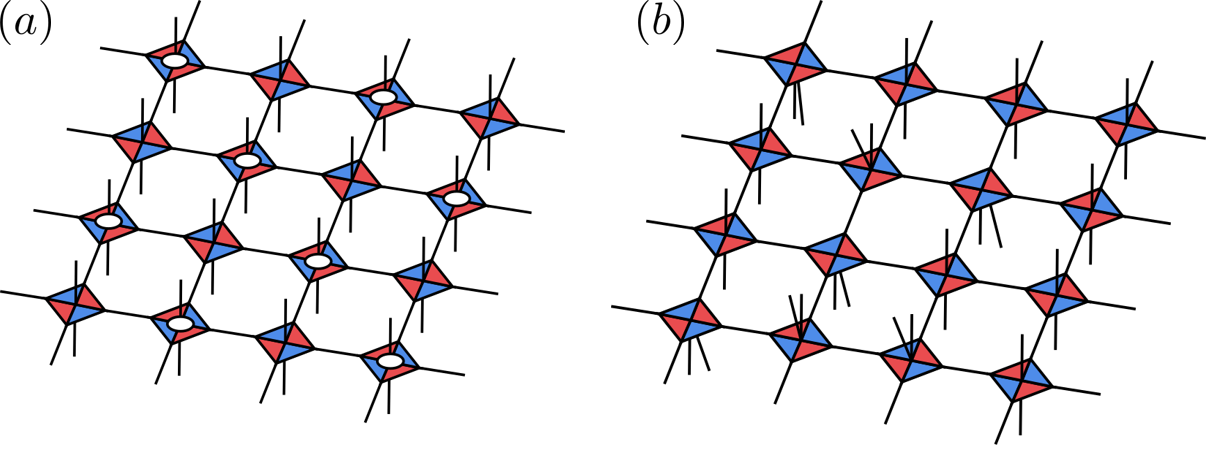}
    \caption{Local patches of the full tensor network that describes a surface code variant: (a) Tensors with a white circuit differ from the original tensor by a Hadamard acting on the downward pointing leg, this changes an all $Z$ or all $X$ stabilizer to the form of $XZZX$ as needed. All other interpretations of the network remains unchanged. (b) We have chosen, non-uniformly, to have some legs (upward pointing) to be logical and some legs (downward pointing) to be physical.}
    \label{fig:bulk_mod}
\end{figure}

It is also straightforward to modify the bulk properties by using slightly different tensors in the interior of the network as long as one preserves certain operator pushing properties. Two such examples have been constructed in detail in Appendix~\ref{app:f}. For instance, by replacing every other tensor with a modified counterpart that differs by a local Hadamard, the resulting network is nothing but an XZZX surface code \cite{xzzx_wen,xzzx_qm,xzzx_threshold} which has been shown to have a better (symmetric) error threshold compared to the conventional surface code (Figure~\ref{fig:bulk_mod}). We assume the proper stopper tensors are contracted on the boundary to produce the correct boundary stabilizers (e.g. Figure~\ref{fig:xzzx_app}). 
A more elaborate modification (Figure~\ref{fig:twistcode}) introduces a defect by adding another 4-qubit code with stabilizer group $\langle XXIX, ZIZZ, YYYI\rangle$. This produces the so-called triangle code \cite{surfacecode_twist}. 

Additionally, because what we call physical versus logical is artificial, one can also build codes by converting some of the physical legs to logical ones and vice versa. One extreme example is where all the bulk degrees of freedoms are logical, which reduces the network into a 1d code (see Figure~\ref{fig:1dcode} in Appendix~\ref{subapp:1dcode}). When all bulk dangling legs are physical, we are left with a stabilizer state on $2L^2$ qubits. It is also interesting to explore the code properties of a tensor that is somewhere in between, where the choice of logical vs physical legs can be somewhat non-uniform (Figure~\ref{fig:bulk_mod}b). By assigning the legs to inputs (or outputs) differently, one also creates new codes distinct from the toric code using the same tensor network.

\begin{figure}
    \centering
    \includegraphics[width=0.48\textwidth]{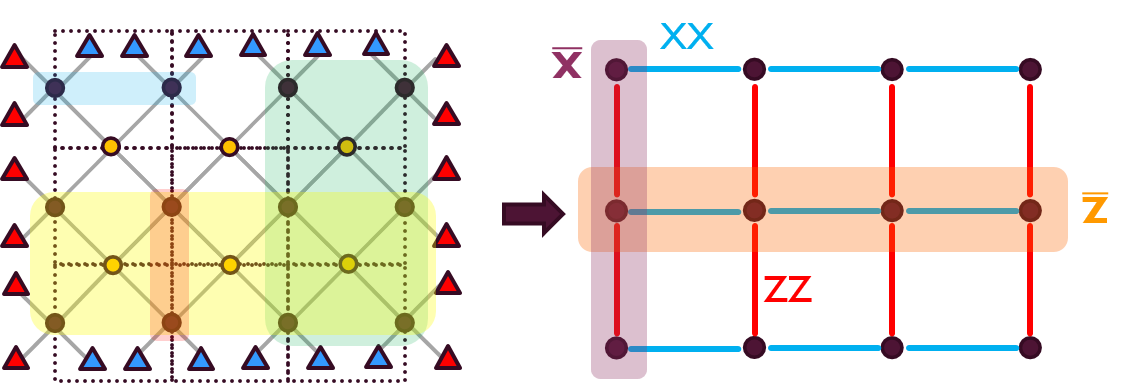}
    \caption{A $3\times 4$ Bacon-Shor code with distance 3 from the surface code tensor network. Tensors every other row are coloured yellow, meaning that both on-site dangling legs are logical (pointing out of the page) while the black sites have one dangling leg pointing into the page (physical) while the other out of the page (logical). The red and blue triangles are again stopper tensors we used for the surface code, which are Z and X $+1$ eigenstates respectively. The weight 2 shaded operators are gauge generators (blue: X type, red: Z type) while some stabilizers are shaded in yellow (Z type) and green (X type). They correspond to acting Pauli X or Zs on the physical qubits (black). Blue and red edges in the right diagram mark the gauge generators. A representative of the logical $X$ ($Z$) operator has $X$ ($Z$) acting on the purple (orange) shaded qubits.}
    \label{fig:2dbsc_planar}
\end{figure}

For an explicit example, by promoting the dangling physical legs to logical legs in every other row in a surface code tensor network, one obtains the 2d Bacon-Shor codes \cite{Shor95,Bacon2006}. A $3\times 4$ example is shown in Figure~\ref{fig:2dbsc_planar}. The gauge generators are weight 2 $XX$ or $ZZ$ acting on the qubits that reside on the end points of the horizontal (blue) and vertical bonds (red) respectively. A more detailed description and derivation is given in Appendix~\ref{subsec:2dbsc}. 
Importantly, this surprising duality through our tensor network connects the low energy subspace of two distinct systems --- a topological quantum field theory in the form of the surface code and the quantum compass model whose ground space is frustrated. Therefore, it can provide exact analytical insights for the low energy behaviour of such frustrated Hamiltonians through well-known results in the surface code.

\subsection{Perfect Code in Flat Geometry}
\label{subsec:3d}
Here, and the next section, we introduce two new codes that to the best of our knowledge have not been discussed in previous literature. Other than their connections to quantum gravity, both also have utility in quantum information. For example, both can be used for magic state distillation: the first supports a transversal $SH$ gate while the second supports a transversal $T$ gate.

The perfect tensor, or the associated $[[5,1,3]]$ code, has been used to create holographic codes \cite{HaPPY}. However, one can also construct codes that have a flat geometry in a way similar to the toric code except that we are using the $[[5,1,3]]$ instead of a $[[5,1,2]]$ code as base tensors. Such codes may be relevant for the bulk entanglement gravity approaches \cite{Cao2017,BEG} where one considers flat emergent geometries and gravity directly from the bulk, as opposed to the usual holographic perspective. 

\begin{figure}
    \centering
    \includegraphics[width=0.3\textwidth]{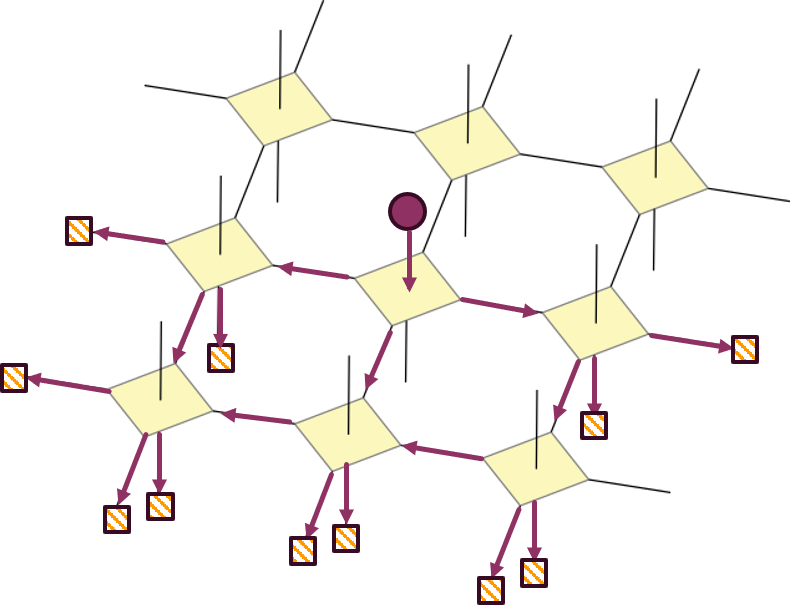}
    \caption{A tensor network with flat geometry created by connecting rank-6 perfect tensors (yellow). This creates a code where each logical degree of freedom is localized to the ``bulk'' legs pointing up. The physical representation and support of a logical operator can be found via pushing just like the holographic code.}
    \label{fig:perf_flat}
\end{figure}

Its tensor network construction is shown in Figure~\ref{fig:perf_flat}. One can build it using Algorithm \ref{alg:TNbuild} from Appendix~\ref{subapp:tnbuild}, therefore the tensor network is also efficiently contractible and admits a straightforward encoding/decoding circuit. Despite having the same network architecture as that of the toric code, its logical legs in this code are fully independent of each other.
We can also identify the support of a logical operator through operator pushing. This codes does not have the same kind of localized stabilizer operators or straightforward string-like logical operators because of the differences in its base rank-6 tensor. The support of one such logical operator is shown in Figure~\ref{fig:perf_flat}.

This is an example that is somewhat in between the holographic code and the toric/surface code in that it introduces extra physical degrees of freedom in the bulk like surface code, while the encoded degrees of freedom are still localized to the bulk regions. It is clear that the same network architecture can lead to codes with different properties because the tensors are different.

\subsection{Holographic code with transversal non-Clifford gate}
\label{subsec:3e}
\begin{figure}
    \centering
    \includegraphics[width=0.3\textwidth]{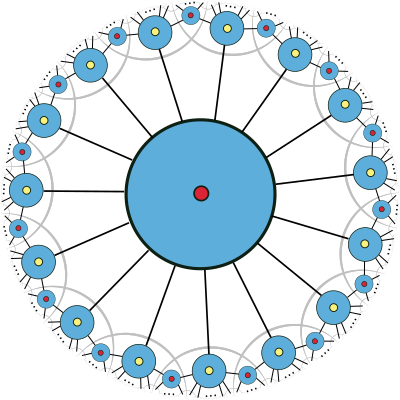}
    \caption{A holographic Reed-Muller code with transversal non-Clifford gate. The ellipsis represents the remaining boundary-facing legs that we drop to avoid clutter. Dangling logical legs in the bulk are denoted as red or yellow dots}
    \label{fig:holoRM}
\end{figure}

Recently, \cite{Cree2021} pointed out that holographic stabilizer codes with good complementary error correction properties can not support transversal non-Clifford gates. While these complementary properties are desirable in holography, they may not be strictly necessary for quantum error correction.  Continuing our earlier example using the $[[15,1,3]]$ Reed-Muller code, consider a $\{15,4\}$ tessellation of the Poincar\'e disk. Because the base tensor is a non-degenerate distance 3 code\footnote{It can be checked that the stabilizer weight is at least 4 for any non-identity element, therefore the code is non-degenerate \cite{csstranst}.}, it is a 2-isometry in that any two legs of the tensor is maximally mixed. Therefore, to produce a holographic code, one can connect these tensors in the manner shown in Figure~\ref{fig:holoRM} using the methods prescribed by \cite{HMERA}. The logical degrees of freedom in each tile are independent of each other as their logical operators can be pushed to the boundary\footnote{Alternatively, one can apply thereom 1 of \cite{TNC} because in building the network we only need to contract pairs of edges where at least one is a correctable erasure error.}. This is not unlike the original HaPPY code. However, we see that it is not a very good erasure correction code such that a small number of erasures lead to a disconnected bulk entanglement wedge.

Using features of the $[[15,1,3]]$ code outlined in Sec~\ref{subsec:3a}, we can verify that by pushing $\bar{T}^{\dagger}$ (or $\bar{T}$) through the logical legs that are marked red (respectively, yellow), we produce a physical operator that is a tensor product of $T$ and $T^{\dagger}$ on the boundary such that operators on the internal edges match properly\footnote{Similarly, by contracting extra phase gates on the bulk/logical legs marked red, the same physical operator implements the all $\bar{T}$ logical operation for a slightly modified version of the code.}. Therefore there is at least one transversal logical non-Clifford operator consisting of the tensor products of $\bar{T}$ and $\bar{T}^{\dagger}$. Again we use overlines to indicate logical operation.
Although the transversal non-Clifford gate acts on all encoded qubits simultaneously, the code may be useful for as a subsystem code that uses code switching \cite{codeswitch} or for magic state distillation \cite{Bravyi_2005,BravyiHaah}, especially in manybody quantum states \cite{EntMSD}.

\subsection{3d Code from the Steane code}
\label{subsec:3f}
\begin{figure}
    \centering
    \includegraphics[width=0.45\textwidth]{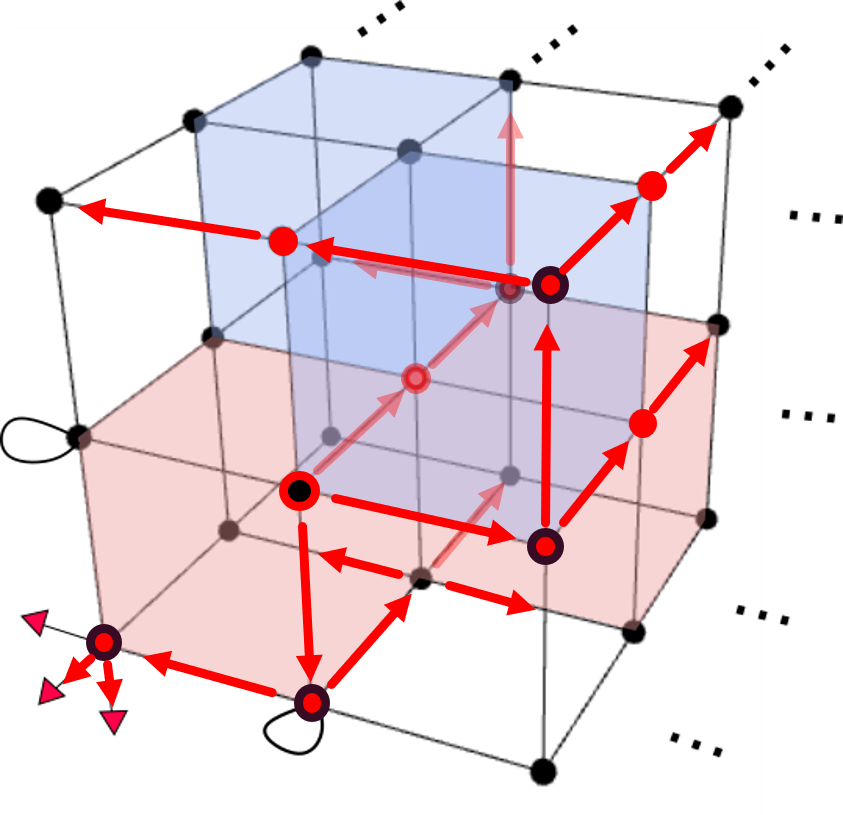}
    \caption{A tree-like operator resulting from pushing logical operators in the tensor network. Red denotes the pushing of Pauli $Z$s. We neglect some of the dangling legs for tensors on the boundary of the cube because that depends on how we choose the boundary condition. For example, the 3 dangling legs bottom left corner are contracted with Z-type stopper tensors and two on the nearby edges are contracted among themselves. Red arrow marks the path of operator flow.}
    \label{fig:3d_log}
\end{figure}

Thus far we have focused on codes that are built on 2 dimensional geometries such as flat or the 2d hyperbolic space. One can also construct 3d codes with localized stabilizer generators using the Steane code by orienting the tensors appropriately. This creates a cubic lattice tensor network where a physical qubit is located on each vertex (Figure~\ref{fig:3d_log}). Each Steane code tensor is represented as a tensor with 6 contracted legs that connect the vertex to its neighbours. The remaining 7th physical leg and the logical leg are located on the vertices. They are not shown in the diagrams to avoid clutter. Instead, we use the following notation: a vertex with coloured boundary circle denotes a non-identity logical operator being pushed into the logical leg at that vertex, while a black boundary indicates logical identity is pushed. Similarly, vertices with coloured interior indicate non-trivial Paulis acting on the dangling physical legs on-site while a black interior indicates the identity operator acting on the physical qubits located at the vertices. 

By pushing stabilizers of each Steane code tensor, one can show that there are localized stabilizers consisting of weight 8 all $X$ or all $Z$ operators that act on all corners of the coloured cubes (coloured red or blue in Figure ~\ref{fig:3d_log}) in the lattice. The existence and form of other stabilizer generators will heavily depend on the boundary condition, which is not unlike our earlier construction of the toric/surface code. 

Potential logical operators can be generated by pushing a combination of logical and stabilizer elements of the Steane code tensors, where some can take on localized geometric shapes like boxes or quasilocal fractal structures like trees. A couple of examples are shown in Figure~\ref{fig:3d_log} and Figure~\ref{fig:3d_latt_box}. 

This 3d subsystem code is another non-trivial example for generalization beyond concatenation, wherein lifting the isometry contraction allows us to construct codes with spatially local stabilizers with the abundance of loops in the network. As far as we know, this code has not been constructed in literature. Further details of the construction and operator pushing can be found in Appendix~\ref{subapp:3dcode}. 

Unlike the toric code example, we do not claim that the stabilizer group is generated solely by stabilizers localized to the coloured boxes. It is also unclear that how many logical degrees of freedom the code encodes or what forms non-trivial logical operators should take. We will leave its analysis with different boundary conditions to future work.

\section{Discussion}
\label{sec:4}

Inspired by the work of \cite{HaPPY}, we proposed a graphical framework applicable to any QECC, which constructs complex codes from simple ones using tensor networks.
While holographic codes were among its first concrete implementations, we here have found that the fully extended framework which we call quantum lego is far more capable and can generate a range of different codes with different properties. In particular, it connects tensor network geometry with error-correcting codes and can deduce properties of the larger code from those of the simpler components using local moves like operator pushing. Compared to past literature, we have generalized the idea in \cite{HaPPY} to all types of codes over qubits and qudits. More importantly, we lifted a highly restrictive condition that only joins tensors in directions that are isometric (or error correcting). Our work significantly extended the range of constructible codes that can be studied with operator pushing and past techniques. In doing so, we also provide a genuine generalization of code concatenation and show that any quantum code can be constructed in such a manner with simple atomic lego blocks. When applied to stabilizer codes, we proved systematic statements on how the code transforms under tensor gluing and how to efficiently capture it using a new type of operation over their check matrices. Additionally, we have constructed new QECCs and brought about novel insight for existing codes by applying the quantum lego language. We find that the likes of toric code can be easily attained and modified using this framework. We also identified a surprising duality which maps the 2d Bacon-Shor code to the surface code, thus connecting the low energy subspace of a well-known gapped local Hamiltonian to that of a frustrated system.
We note a few distinct advantages of this approach and then comment on some potential future directions.
\subsection{Summary of Features}

\begin{figure}
    \centering
    \includegraphics[width=0.45\textwidth]{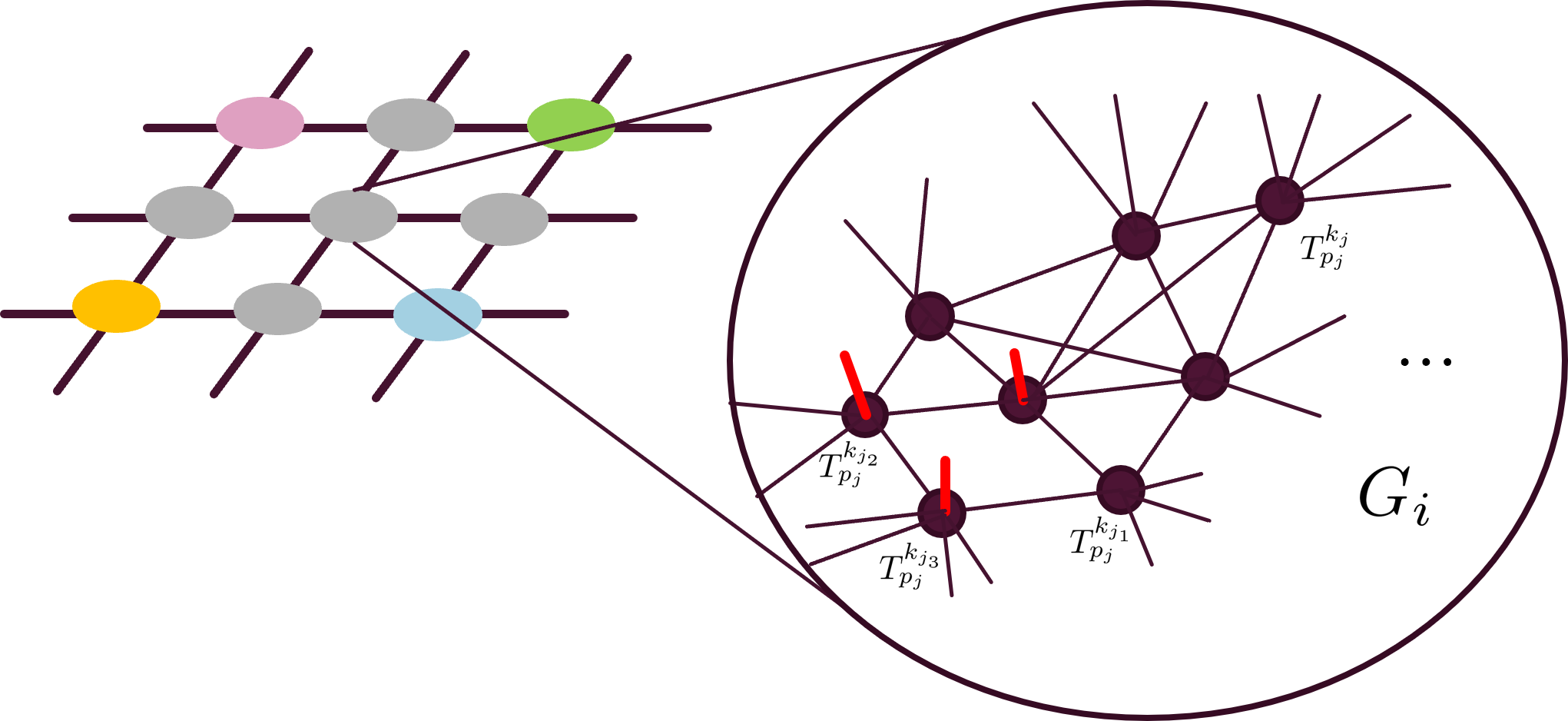}
    \caption{A network architecture where the subgraph $G_i$ may have all-to-all connectivity below a certain scale even though the connection is sparse and local on a sufficiently coarse-grained level. Coloured edges represent logical legs. }
    \label{fig:cluster}
\end{figure}

\textit{Geometry:}
All codes created in this manner have a natural association with network geometries. While we are able to create codes whose geometry is ``regular'' like those of the square lattice and of the hyperbolic plane, we are also able to accommodate more flexible geometries that are more natural in different hardwares. For instance, in addition to spatially local constructions on lattices, one may imagine a network of ion traps that correspond to graphs of clusters where there is all-to-all connectivity below a certain scale but spatially local at a more coarse grained level (Figure~\ref{fig:cluster}).
Alternatively, one could construct hybrid architectures by gluing together networks of different geometries. These instances provide a fertile testing ground for this framework. By reproducing known codes using this approach it also enables us to interpret  existing codes geometrically. For cases where the code properties may be difficult to analyze otherwise, a graph geometry may facilitate such an analysis.

\textit{Flexibility and Customizability:} 
Distinct from highly idealized constructs, we may also have varied demands for codes that serve different purposes. In particular, they may have irregularities or defects in their geometries and forms of logical operations. These somewhat specialized differences can originate in both design and engineering of the quantum hardware. For instance, in machines that rely on spatial locality, lattice defects and dead qubits may also require that a flexible and customizable code be tailor-made for these specific purposes. As we see in some of the earlier examples, the tensor network approach allows local modifications of tensors which easily creates variations of the base surface code. In other cases, one may require a code that corrects error asymmetrically, e.g. correcting bit flip errors better than phase errors in different parts of the machine. As shown in Sec~\ref{subsec:3c} and Appendix~\ref{subapp:defects}, tensor network approach may allow us to modify a base code with the bulk of the desirable properties and create different variants of the same code by taking out and putting in customized tensors at certain locations. 

\textit{Accessibility:} Importantly, this framework works very minimally with the specific expression of the tensors and requires little background on tensor networks. At its core, this method of code-building is but a gluing and operator pushing/matching exercise once the unitary stabilizers of each tensor have been identified. In particular, the idea of operator pushing also distinguishes this framework from generic tensor network approaches because we only need to track a discrete set of elements by following the flow of (Pauli) operators instead of dealing with the multi-component tensors numerically. Just like how one can build lego houses without being able to manufacture the lego blocks, the framework provides an intuitive and accessible approach that allows users to generate interesting codes without heavy involvement or expert knowledge in the area of tensor network or quantum error correction. Furthermore, the simplicity in the stabilizer codes created using this method may also allow us to automate the graphical code creation process using techniques like machine learning.

\textit{Tensor Network, Codes, and Physics:} This effort further strengthens the connection between tensor networks and quantum error-correcting codes. As such, tensor network can also serve as a bridge between QECCs and areas of high energy physics (c.f. simulations of quantum field theories), condensed matter physics and quantum gravity. In light of past productive connections made between tensor network and these areas, further exploration can be beneficial for both quantum information/computing and our understanding of physics in the upcoming quantum information age. For instance, finding the tensor network constructions of well-known gauge codes using the quantum lego method may help us better understand the low energy behaviour of Hamiltonians with frustration in the ground space. It is also curious whether the idea of operator pushing can help facilitate certain numerical analyses  when computing correlations in the low energy states of quantum manybody systems.

\subsection{Future Directions}
\textit{Framework Improvement:}
While we have summarized some critical properties and constructions of the framework, many gaps of knowledge remain to be explored. In particular, we need to understand the full range of capabilities and limitations of this framework, as well as its connections with other approaches for building QECCs. For instance, it is not clear how it can be related to other graph-based quantum codes like quantum LDPC codes \cite{Hastings2020}. We would also like to understand if tensor networks provide a graphical way that simplifies the determination of code distances or related notions. Interesting problems remain in decoding for tensor networks that may not be efficiently or exactly contractible. Relatedly, we would like to explore more efficient ways to perform active error correction from syndrome measurements using tensor networks. Seeing how spatial locality may be important in QECC, we also like to develop a more comprehensive guideline for creating codes that have mostly spatially local stabilizer generators or other geometric features instead of relying mostly on human intuitions. 

\textit{Tensor Contraction, Decoding and Fault-tolerant Thresholds:}
A crucial part of quantum error correction lies in the syndrome decoding algorithms. Although the quantum lego is capable of creating codes in the form of tensor networks, it remains unclear whether, or how, one can devise useful decoding algorithms from it. However, because building a tensor network via tensor contraction can be considered as a generalization of code concatenation for which decoding and the derivation of threshold is better understood, it is interesting to understand whether useful techniques for concatenated codes can be extended to our quantum lego codes. If so, this may help us determine decoding algorithms and bounds for fault-tolerant thresholds of the larger tensor network since the properties of the individual tensors (smaller codes) are well-known.

We have also addressed decoding circuit and how it is related to the contraction of tensors in certain networks. This connection offers us clues as to how the decoding circuit transforms as one grows the tensor network (or performs generalized code concatenation). However, it has not been made clear when the encoding map has a non-trivial kernel, i.e. when the logical legs are not independent. As many non-trivial codes, e.g. toric code, carry this property, it is also crucial to extend the network-circuit connection to such encoding maps.

\textit{Stabilizer Code Generation and Characterization:}
We have provided several stabilizer QECCs created using tensor networks in this work. A few examples, e.g. the 3d code and some surface code variants, are new and their code properties have not yet been fully characterized. Therefore, we would like to determine whether these codes or variations can have interesting properties, such as having distance scaling favourably with system size or desirable transversal operators. In particular, it will be fruitful to further explore the connections between the 3d code and the Haah code \cite{HaahCode} and fractons.  We would also like to generate other non-trivial  stabilizer codes (or subsystem codes) that have interesting network geometries and mostly local stabilizer generators as way to circumvent some no-go theorems \cite{BravyiTerhal}. These might be 1d or 2d codes that have a few global/non-local stabilizer generators. We would also like to better understand the connection between the surface code and the Bacon-Shor code. Since one can be gradually deformed to the other by ``bending'' some of the dangling legs, it is curious what type of codes one creates if we only partially deform the surface code. Beyond the regular lattice geometries like the examples in Sec~\ref{subsec:3b}, one can also explore other more general network geometries such as the one in Figure~\ref{fig:cluster}. 

We also hope to reproduce other known families of codes of interest, such as color codes, (generalized) Bacon-Shor codes, and LDPC codes. This helps us glean further insights from their tensor networks, like distilling the underlying principles that have led to their desirable properties and generating variants or customized versions of such codes.

\textit{Non-additive Code and Transversal Gates:}
In Sec~\ref{sec:3}, we have focused on stabilizer codes generated by the tensor network framework. However, the framework itself is also valid for generating non-additive codes as the same tracing operation applies generically to any tensor and the operator matching process holds for any tensor that admits a unitary (product) stabilizer. For instance, one way to create larger non-stabilizer codes is to convert known non-additive codes with transversal operators into tensors and then trace them together.

We can also ask whether the tensor network language can help us construct non-stabilizer quantum lego blocks in the first place. For example, one can find small tensors that have desirable unitary product stabilizers and then convert them into encoding isometries that describe non-additive codes. One approach is to find states/tensors that are $+1$ eigenstates of operators of the form 
\begin{equation}
    \{\bar{O}\otimes O_1\otimes O_2\otimes \dots \::\: \bar{O}\in \mathcal{L}\}
\end{equation}
where we will take the first site of the tensor to be the intended data qubit when the tensor is converted to an encoding isometry, and  
$\mathcal{L}$ is the set of logical operators that we wish to be transversal. This is somewhat similar to finding tensors that have global symmetries \cite{PEPSMPSsym,Singh2010,Hayden17} if all of $O_i, \bar{O}$ are identical, which is a more restrictive condition than the one we impose here.  Although for larger codes the search can be exponentially expensive, it should be possible for tensors of fewer dangling legs. 
Larger codes can then be generated in the usual way by gluing together these smaller components. Using operator pushing properties derived in Appendix~\ref{app:b}, we can also ensure the larger code to have certain transversal operators. 

On a differet note, while it is clear how one may generate transversal single qubit (qudit) operators using operator pushing after gluing the tensors, it is not clear how transversality may or may not be preserved after tracing for multi-qubit gates such as CNOT or CCZ. Further work in this area is needed.

\textit{Asymmetric Error Correction and Customizability}
Sometimes certain errors are more likely than others. Following the hints from examples in Figure~\ref{fig:surface_code}b we would like to understand how effective this framework can be in systematically generating codes with asymmetric error correction properties. More generally, we would like to understand how the flexibility in choosing different tensors in a network can change the code properties locally and globally. Towards building a framework that hopefully allows more user customizability in code building, we wish to explore the extent to which the tensor network approach can create useful tailor-made codes. Some would include modifying well-known codes with good properties and connecting them with different tensors. Some explicit directions are to understand the surface/toric code variants which we have discussed in Sec~\ref{subsec:3c}. 

\textit{Approximate QECC, Random Tensor Network and Classical Codes}
It is also possible to use the self-same technique to construct approximate QECCs --- in this case, one would replace some of the tensors with their ``skewed'' counterparts as shown in \cite{ABSC}. Explicit models of approximate QECCs (AQECC) in the form of tensor networks may also provide further insight in areas like quantum gravity. 

On a related note, random tensor networks of flexible geometry have been discussed in \cite{RTN}. While such codes may be more cumbersome to implement on a quantum computer in the short term, they are powerful theoretical tools that probe the general properties of (A)QECCs. It would be beneficial to further pursue this direction and understand how random tensor networks (RTN) may help with code building. Furthermore, construction of decoding unitaries of RTN codes have not been fully understood. Therefore, it is also interesting to establish connections between RTN and random unitary circuit models of quantum codes \cite{Gullans} including systems that with random measurements \cite{miptqec,miptqec2,miptqec3}. 

While this framework is intended for building quantum codes, one can ask whether it is also useful for classical error-correcting codes. Because its graphical generalization of code concatenation, it may also be fruitful to determine the necessary modifications for generating classical codes using tensor networks.

\textit{Automated Code Generation and Machine Learning}
Thus far, we have applied the quantum lego framework to generate new codes by hand. With reasonable amount of effort, one can trace together these quantum codes and track their properties. A lot of this was easily doable because of the abundant symmetries in the tensors and the network geometry. However, as we move to larger codes with fewer symmetries, it will become increasingly demanding to track the different operator pushings by hand. However, because of the simple pushing rules involved, and the limited number of ways a small tensor can be oriented, such tasks of code generation can be delegated to a machine. Furthermore, it should be computationally feasible and exciting to combine this code building process with machine learning techniques to create codes that have certain desirable features.

\section*{Acknowledgements}
We thank Victor Albert, Michael Beverland, Glen Evenbly, Michael Gullans, Brian Swingle, and Eugene Tang for the helpful discussions and comments. C.C. acknowledges the support by the U.S. Department of Defense and NIST through the Hartree Postdoctoral Fellowship at QuICS, by the Simons Foundation as part of the It From Qubit Collaboration, and by the DOE Office of Science, Office of High Energy Physics, through the grant DE-SC0019380. 

\bibliographystyle{unsrt} 
\bibliography{ref} 

\begin{thebibliography}{10}

\bibitem{toriccode}
A.Y. Kitaev.
\newblock {\em Proceedings of the 3rd International Conference of Quantum
  Communication and Measurement, Ed. O. Hirota, A. S. Holevo, and C. M. Caves}.
\newblock Plenum, New York, 1997.

\bibitem{surfacecode}
S.~B. {Bravyi} and A.~Yu. {Kitaev}.
\newblock {Quantum codes on a lattice with boundary}.
\newblock {\em arXiv e-prints}, pages quant--ph/9811052, November 1998.

\bibitem{tqm}
Eric {Dennis}, Alexei {Kitaev}, Andrew {Landahl}, and John {Preskill}.
\newblock {Topological quantum memory}.
\newblock {\em Journal of Mathematical Physics}, 43(9):4452--4505, September
  2002.

\bibitem{Kitaev2006}
Alexei {Kitaev}.
\newblock {Anyons in an exactly solved model and beyond}.
\newblock {\em Annals of Physics}, 321(1):2--111, January 2006.

\bibitem{stringnet}
Michael~A. {Levin} and Xiao-Gang {Wen}.
\newblock {String-net condensation:{\quad}A physical mechanism for topological
  phases}.
\newblock {\em \prb}, 71(4):045110, January 2005.

\bibitem{Bombin2006}
H.~{Bombin} and M.~A. {Martin-Delgado}.
\newblock {Topological Quantum Distillation}.
\newblock {\em \prl}, 97(18):180501, November 2006.

\bibitem{Bombin2013}
H.~{Bombin}.
\newblock {An Introduction to Topological Quantum Codes}.
\newblock {\em arXiv e-prints}, page arXiv:1311.0277, November 2013.

\bibitem{HaahCode}
Jeongwan {Haah}.
\newblock {Local stabilizer codes in three dimensions without string logical
  operators}.
\newblock {\em \pra}, 83(4):042330, April 2011.

\bibitem{RausHarrington}
Robert {Raussendorf} and Jim {Harrington}.
\newblock {Fault-Tolerant Quantum Computation with High Threshold in Two
  Dimensions}.
\newblock {\em \prl}, 98(19):190504, May 2007.

\bibitem{Fowler2009}
Austin~G. {Fowler}, Ashley~M. {Stephens}, and Peter {Groszkowski}.
\newblock {High-threshold universal quantum computation on the surface code}.
\newblock {\em \pra}, 80(5):052312, November 2009.

\bibitem{BombinDef}
H.~{Bombin}.
\newblock {Topological Order with a Twist: Ising Anyons from an Abelian Model}.
\newblock {\em \prl}, 105(3):030403, July 2010.

\bibitem{Hastings}
M.~B. {Hastings} and A.~{Geller}.
\newblock {Reduced Space-Time and Time Costs Using Dislocation Codes and
  Arbitrary Ancillas}.
\newblock {\em arXiv e-prints}, page arXiv:1408.3379, August 2014.

\bibitem{Nagayama}
Shota {Nagayama}, Austin~G. {Fowler}, Dominic {Horsman}, Simon~J. {Devitt}, and
  Rodney {Van Meter}.
\newblock {Surface code error correction on a defective lattice}.
\newblock {\em New Journal of Physics}, 19(2):023050, February 2017.

\bibitem{TNrev1}
Rom{\'a}n {Or{\'u}s}.
\newblock {A practical introduction to tensor networks: Matrix product states
  and projected entangled pair states}.
\newblock {\em Annals of Physics}, 349:117--158, October 2014.

\bibitem{TNrev2}
Rom{\'a}n {Or{\'u}s}.
\newblock {Tensor networks for complex quantum systems}.
\newblock {\em Nature Reviews Physics}, 1(9):538--550, August 2019.

\bibitem{HoloReview}
Alexander {Jahn} and Jens {Eisert}.
\newblock {Holographic tensor network models and quantum error correction: A
  topical review}.
\newblock {\em arXiv e-prints}, page arXiv:2102.02619, February 2021.

\bibitem{FP2013}
Andrew~J. {Ferris} and David {Poulin}.
\newblock {Tensor Networks and Quantum Error Correction}.
\newblock {\em \prl}, 113(3):030501, July 2014.

\bibitem{HaPPY}
Fernando {Pastawski}, Beni {Yoshida}, Daniel {Harlow}, and John {Preskill}.
\newblock {Holographic quantum error-correcting codes: toy models for the
  bulk/boundary correspondence}.
\newblock {\em Journal of High Energy Physics}, 2015:149, June 2015.

\bibitem{Yang2016}
Zhao {Yang}, Patrick {Hayden}, and Xiao-Liang {Qi}.
\newblock {Bidirectional holographic codes and sub-AdS locality}.
\newblock {\em Journal of High Energy Physics}, 2016:175, January 2016.

\bibitem{RTN}
Patrick {Hayden}, Sepehr {Nezami}, Xiao-Liang {Qi}, Nathaniel {Thomas}, Michael
  {Walter}, and Zhao {Yang}.
\newblock {Holographic duality from random tensor networks}.
\newblock {\em Journal of High Energy Physics}, 2016(11):9, November 2016.

\bibitem{BEG}
ChunJun {Cao} and Sean~M. {Carroll}.
\newblock {Bulk entanglement gravity without a boundary: Towards finding
  Einstein's equation in Hilbert space}.
\newblock {\em \prd}, 97(8):086003, April 2018.

\bibitem{DonnellyEdge}
William {Donnelly}, Donald {Marolf}, Ben {Michel}, and Jason {Wien}.
\newblock {Living on the edge: a toy model for holographic reconstruction of
  algebras with centers}.
\newblock {\em Journal of High Energy Physics}, 2017(4):93, April 2017.

\bibitem{ABSC}
ChunJun {Cao} and Brad {Lackey}.
\newblock {Approximate Bacon-Shor code and holography}.
\newblock {\em Journal of High Energy Physics}, 2021(5):127, May 2021.

\bibitem{HoloSteane}
Robert~J. {Harris}, Nathan~A. {McMahon}, Gavin~K. {Brennen}, and Thomas~M.
  {Stace}.
\newblock {Calderbank-Steane-Shor Holographic Quantum Error Correcting Codes}.
\newblock {\em arXiv e-prints}, page arXiv:1806.06472, June 2018.

\bibitem{HoloSteaneDec}
Robert~J. {Harris}, Elliot {Coupe}, Nathan~A. {McMahon}, Gavin~K. {Brennen},
  and Thomas~M. {Stace}.
\newblock {Decoding Holographic Codes with an Integer Optimisation Decoder}.
\newblock {\em arXiv e-prints}, page arXiv:2008.10206, August 2020.

\bibitem{Jahn2017}
Alexander {Jahn}, Marek {Gluza}, Fernando {Pastawski}, and Jens {Eisert}.
\newblock {Holography and criticality in matchgate tensor networks}.
\newblock {\em arXiv e-prints}, page arXiv:1711.03109, November 2017.

\bibitem{Jahn2019}
A.~{Jahn}, M.~{Gluza}, F.~{Pastawski}, and J.~{Eisert}.
\newblock {Majorana dimers and holographic quantum error-correcting codes}.
\newblock {\em Physical Review Research}, 1(3):033079, November 2019.

\bibitem{Jahncc}
Alexander {Jahn}, Zolt{\'a}n {Zimbor{\'a}s}, and Jens {Eisert}.
\newblock {Central charges of aperiodic holographic tensor-network models}.
\newblock {\em \pra}, 102(4):042407, October 2020.

\bibitem{HMERA}
ChunJun {Cao}, Jason {Pollack}, and Yixu {Wang}.
\newblock {Hyper-Invariant MERA: Approximate Holographic Error Correction Codes
  with Power-Law Correlations}.
\newblock {\em arXiv e-prints}, page arXiv:2103.08631, March 2021.

\bibitem{Jahn2020}
Alexander {Jahn}, Zolt{\'a}n {Zimbor{\'a}s}, and Jens {Eisert}.
\newblock {Tensor network models of AdS/qCFT}.
\newblock {\em arXiv e-prints}, page arXiv:2004.04173, April 2020.

\bibitem{Cree2021}
Sam {Cree}, Kfir {Dolev}, Vladimir {Calvera}, and Dominic~J. {Williamson}.
\newblock {Fault-tolerant logical gates in holographic stabilizer codes are
  severely restricted}.
\newblock {\em arXiv e-prints}, page arXiv:2103.13404, March 2021.

\bibitem{TNC}
Terry {Farrelly}, Robert~J. {Harris}, Nathan~A. {McMahon}, and Thomas~M.
  {Stace}.
\newblock {Tensor-network codes}.
\newblock {\em arXiv e-prints}, page arXiv:2009.10329, September 2020.

\bibitem{Farrellypdec}
Terry {Farrelly}, Robert~J. {Harris}, Nathan~A. {McMahon}, and Thomas~M.
  {Stace}.
\newblock {Parallel decoding of multiple logical qubits in tensor-network
  codes}.
\newblock {\em arXiv e-prints}, page arXiv:2012.07317, December 2020.

\bibitem{Farrelly}
Terry {Farrelly}, David~K. {Tuckett}, and Thomas~M. {Stace}.
\newblock {Local tensor-network codes}.
\newblock {\em arXiv e-prints}, page arXiv:2109.11996, September 2021.

\bibitem{cwstab}
Andrew {Cross}, Graeme {Smith}, John~A. {Smolin}, and Bei {Zeng}.
\newblock {Codeword Stabilized Quantum Codes}.
\newblock {\em arXiv e-prints}, page arXiv:0708.1021, August 2007.

\bibitem{Laflamme:1996iw}
Raymond Laflamme, Cesar Miquel, Juan~Pablo Paz, and Wojciech~Hubert Zurek.
\newblock {Perfect quantum error correction code}.
\newblock 2 1996.

\bibitem{choi}
Man-Duen Choi.
\newblock Completely positive linear maps on complex matrices.
\newblock {\em Linear Algebra and its Applications}, 10(3):285--290, 1975.

\bibitem{Jam}
A.~Jamiołkowski.
\newblock Linear transformations which preserve trace and positive
  semidefiniteness of operators.
\newblock {\em Reports on Mathematical Physics}, 3(4):275--278, 1972.

\bibitem{Singh2010}
Sukhwinder {Singh}, Robert N.~C. {Pfeifer}, and Guifr{\'e} {Vidal}.
\newblock {Tensor network decompositions in the presence of a global symmetry}.
\newblock {\em \pra}, 82(5):050301, November 2010.

\bibitem{White2020}
Christopher~David {White}, ChunJun {Cao}, and Brian {Swingle}.
\newblock {Conformal field theories are magical}.
\newblock {\em \prb}, 103(7):075145, February 2021.

\bibitem{Vidal2008}
G.~{Vidal}.
\newblock {Class of Quantum Many-Body States That Can Be Efficiently
  Simulated}.
\newblock {\em prl}, 101(11):110501, September 2008.

\bibitem{branchingmera}
G.~{Evenbly} and G.~{Vidal}.
\newblock {Class of Highly Entangled Many-Body States that can be Efficiently
  Simulated}.
\newblock {\em \prl}, 112(24):240502, June 2014.

\bibitem{bmeracode}
Andrew~J. {Ferris} and David {Poulin}.
\newblock {Branching MERA codes: a natural extension of polar codes}.
\newblock {\em arXiv e-prints}, page arXiv:1312.4575, December 2013.

\bibitem{AguadoVidal}
Miguel {Aguado} and Guifr{\'e} {Vidal}.
\newblock {Entanglement Renormalization and Topological Order}.
\newblock {\em \prl}, 100(7):070404, February 2008.

\bibitem{BravyiGBSC}
Sergey {Bravyi}.
\newblock {Subsystem codes with spatially local generators}.
\newblock {\em \pra}, 83(1):012320, January 2011.

\bibitem{Zhang6qubit}
Zhang {Jiang} and Eleanor~G. {Rieffel}.
\newblock {Non-commuting two-local Hamiltonians for quantum error suppression}.
\newblock {\em arXiv e-prints}, page arXiv:1511.01997, November 2015.

\bibitem{Faist}
Philippe {Faist}, Sepehr {Nezami}, Victor~V. {Albert}, Grant {Salton}, Fernando
  {Pastawski}, Patrick {Hayden}, and John {Preskill}.
\newblock {Continuous Symmetries and Approximate Quantum Error Correction}.
\newblock {\em Physical Review X}, 10(4):041018, October 2020.

\bibitem{SteaneCode}
Andrew {Steane}.
\newblock {Multiple-Particle Interference and Quantum Error Correction}.
\newblock {\em Proceedings of the Royal Society of London Series A},
  452(1954):2551--2577, November 1996.

\bibitem{1996Knill}
E.~{Knill}, R.~{Laflamme}, and W.~{Zurek}.
\newblock {Threshold Accuracy for Quantum Computation}.
\newblock {\em arXiv e-prints}, pages quant--ph/9610011, October 1996.

\bibitem{Anderson2014}
Jonas~T. {Anderson}, Guillaume {Duclos-Cianci}, and David {Poulin}.
\newblock {Fault-Tolerant Conversion between the Steane and Reed-Muller Quantum
  Codes}.
\newblock {\em \prl}, 113(8):080501, August 2014.

\bibitem{PEPStoric}
F.~{Verstraete}, M.~M. {Wolf}, D.~{Perez-Garcia}, and J.~I. {Cirac}.
\newblock {Criticality, the Area Law, and the Computational Power of Projected
  Entangled Pair States}.
\newblock {\em \prl}, 96(22):220601, June 2006.

\bibitem{Brell2011}
Courtney~G. {Brell}, Steven~T. {Flammia}, Stephen~D. {Bartlett}, and Andrew~C.
  {Doherty}.
\newblock {Toric codes and quantum doubles from two-body Hamiltonians}.
\newblock {\em New Journal of Physics}, 13(5):053039, May 2011.

\bibitem{xzzx_wen}
Xiao-Gang {Wen}.
\newblock {Quantum Orders in an Exact Soluble Model}.
\newblock {\em \prl}, 90(1):016803, January 2003.

\bibitem{xzzx_qm}
Alastair {Kay}.
\newblock {Capabilities of a Perturbed Toric Code as a Quantum Memory}.
\newblock {\em \prl}, 107(27):270502, December 2011.

\bibitem{xzzx_threshold}
J.~Pablo {Bonilla Ataides}, David~K. {Tuckett}, Stephen~D. {Bartlett},
  Steven~T. {Flammia}, and Benjamin~J. {Brown}.
\newblock {The XZZX surface code}.
\newblock {\em Nature Communications}, 12:2172, January 2021.

\bibitem{surfacecode_twist}
Theodore~J. {Yoder} and Isaac~H. {Kim}.
\newblock {The surface code with a twist}.
\newblock {\em arXiv e-prints}, page arXiv:1612.04795, December 2016.

\bibitem{Shor95}
Peter~W. Shor.
\newblock Scheme for reducing decoherence in quantum computer memory.
\newblock {\em Phys. Rev. A}, 52:R2493--R2496, Oct 1995.

\bibitem{Bacon2006}
Dave {Bacon}.
\newblock {Operator quantum error-correcting subsystems for self-correcting
  quantum memories}.
\newblock {\em pra}, 73(1):012340, January 2006.

\bibitem{Cao2017}
ChunJun {Cao}, Sean~M. {Carroll}, and Spyridon {Michalakis}.
\newblock {Space from Hilbert space: Recovering geometry from bulk
  entanglement}.
\newblock {\em \prd}, 95(2):024031, January 2017.

\bibitem{csstranst}
Narayanan {Rengaswamy}, Robert {Calderbank}, Michael {Newman}, and Henry~D.
  {Pfister}.
\newblock {On Optimality of CSS Codes for Transversal $T$}.
\newblock {\em arXiv e-prints}, page arXiv:1910.09333, October 2019.

\bibitem{codeswitch}
Adam {Paetznick} and Ben~W. {Reichardt}.
\newblock {Universal Fault-Tolerant Quantum Computation with Only Transversal
  Gates and Error Correction}.
\newblock {\em \prl}, 111(9):090505, August 2013.

\bibitem{Bravyi_2005}
Sergey Bravyi and Alexei Kitaev.
\newblock Universal quantum computation with ideal clifford gates and noisy
  ancillas.
\newblock {\em Physical Review A}, 71(2), Feb 2005.

\bibitem{BravyiHaah}
Sergey {Bravyi} and Jeongwan {Haah}.
\newblock {Magic-state distillation with low overhead}.
\newblock {\em \pra}, 86(5):052329, November 2012.

\bibitem{EntMSD}
Ning {Bao}, ChunJun {Cao}, and Vincent~Paul {Su}.
\newblock {Magic State Distillation from Entangled States}.
\newblock {\em arXiv e-prints}, page arXiv:2106.12591, June 2021.

\bibitem{Hastings2020}
Matthew~B. {Hastings}, Jeongwan {Haah}, and Ryan {O'Donnell}.
\newblock {Fiber Bundle Codes: Breaking the $N^{1/2} \operatorname{polylog}(N)$
  Barrier for Quantum LDPC Codes}.
\newblock {\em arXiv e-prints}, page arXiv:2009.03921, September 2020.

\bibitem{BravyiTerhal}
Sergey {Bravyi} and Barbara {Terhal}.
\newblock {A no-go theorem for a two-dimensional self-correcting quantum memory
  based on stabilizer codes}.
\newblock {\em New Journal of Physics}, 11(4):043029, April 2009.

\bibitem{PEPSMPSsym}
Norbert {Schuch}, Ignacio {Cirac}, and David {P{\'e}rez-Garc{\'\i}a}.
\newblock {PEPS as ground states: Degeneracy and topology}.
\newblock {\em Annals of Physics}, 325(10):2153--2192, October 2010.

\bibitem{Hayden17}
Patrick {Hayden}, Sepehr {Nezami}, Sandu {Popescu}, and Grant {Salton}.
\newblock {Error Correction of Quantum Reference Frame Information}.
\newblock {\em arXiv e-prints}, page arXiv:1709.04471, September 2017.

\bibitem{Gullans}
Michael~J. {Gullans}, Stefan {Krastanov}, David~A. {Huse}, Liang {Jiang}, and
  Steven~T. {Flammia}.
\newblock {Quantum coding with low-depth random circuits}.
\newblock {\em arXiv e-prints}, page arXiv:2010.09775, October 2020.

\bibitem{miptqec}
Soonwon {Choi}, Yimu {Bao}, Xiao-Liang {Qi}, and Ehud {Altman}.
\newblock {Quantum Error Correction in Scrambling Dynamics and
  Measurement-Induced Phase Transition}.
\newblock {\em \prl}, 125(3):030505, July 2020.

\bibitem{miptqec2}
Michael~J. {Gullans} and David~A. {Huse}.
\newblock {Dynamical Purification Phase Transition Induced by Quantum
  Measurements}.
\newblock {\em Physical Review X}, 10(4):041020, October 2020.

\bibitem{miptqec3}
Yaodong {Li} and Matthew P.~A. {Fisher}.
\newblock {Statistical mechanics of quantum error correcting codes}.
\newblock {\em \prb}, 103(10):104306, March 2021.

\bibitem{Harlow:2016vwg}
Daniel Harlow.
\newblock {The Ryu\textendash{}Takayanagi Formula from Quantum Error
  Correction}.
\newblock {\em Commun. Math. Phys.}, 354(3):865--912, 2017.

\bibitem{quditstab}
Vlad {Gheorghiu}.
\newblock {Standard form of qudit stabilizer groups}.
\newblock {\em Physics Letters A}, 378(5-6):505--509, January 2014.

\bibitem{Kitaev2003}
A.~Yu. {Kitaev}.
\newblock {Fault-tolerant quantum computation by anyons}.
\newblock {\em Annals of Physics}, 303(1):2--30, January 2003.

\bibitem{compasscode}
Muyuan {Li}, Daniel {Miller}, Michael {Newman}, Yukai {Wu}, and Kenneth~R.
  {Brown}.
\newblock {2D Compass Codes}.
\newblock {\em Physical Review X}, 9(2):021041, April 2019.

\end{thebibliography}
\appendix

\section{Properties of Isometry Tensors}
\label{app:a}
\begin{definition}
A tensor of degree (or rank) $N$ is an $\ell$-isometry if contracting $N-\ell$ legs with its conjugate transpose reduces to the identity $I^{\otimes \ell}$. 
\end{definition}

Note that this contraction is performed over specific $N-\ell$ input legs\footnote{See Figure 1 of \cite{HMERA} for an explicit tensor contraction diagram.}.  By definition, for a tensor of degree $N$, we have $\ell\leq N/2$. We call an isometry permutation invariant if such properties hold for the contraction of any $N-\ell$ legs. For example, the perfect tensor is a permutation invariant 3-isometry and the $[[4,2,2]]$ tensor over 4 legs, after fixing the logical states, is a permutation invariant 1-isometry. However, many isometries are not permutation invariant. One such example is a generic isometry used in MERA, where it contracts to identity only in a particular direction. A unitary is a special case of isometry where $N=2\ell$. 

\begin{lemma}
Consider the state $|\psi_W\rangle$ dual to an $\ell$ isometry $W:\mathcal{H}_B\rightarrow \mathcal{H}_A$ with bond dimension $\chi$ such that $\log_{\chi}\dim\mathcal{H}_B=\ell$
\begin{equation}
    |\psi_W\rangle_{AC} =\frac{1}{\sqrt{\chi^{\ell}}} W_{AB}\sum_{i=1}^{\chi^{\ell}} |ii\rangle_{BC},
\end{equation}
the $\ell$-site subsystem $C$ is maximally mixed. 
\label{lemma21}
\end{lemma}
\begin{hproof}
\begin{align*}
\rho_{A}&=\Tr_C[|\psi_W\rangle\langle\psi_W|]\\
&= \frac{1}{\sqrt{\chi^{\ell}}}\sum_{i,j,k=1}^{\chi^{\ell}} W_{AB}|i\rangle\langle k|i\rangle\langle j|k\rangle \langle j|W_{AB}^{\dagger}\\
&=\frac{1}{\sqrt{\chi^{\ell}}}\sum_{k=1}^{\chi^{\ell}} W|k\rangle\langle k|W^{\dagger}\\
&=\frac{1}{\sqrt{\chi^{\ell}}}\sum_{k=1}^{\chi^{\ell}}|\phi_k\rangle\langle\phi_k|
\end{align*}
where $|\phi_k\rangle = W|k\rangle$. Because $W$ is an isometry and $\{|k\rangle\}$ is an orthonormal basis, $\{|\phi_k\rangle\}$ is also an orthonormal basis. Given that the Schimidt coefficients are equal and have rank $\chi^\ell$, the subsystem $C$ must be maximally mixed.
\end{hproof}



\begin{lemma}
An $\ell$ isometry can encode $k\leq \ell$ logical qubits and corrects  $\ell-k\geq 0$ number of erasures.
\label{lemma22}
\end{lemma}
\begin{hproof}
For the $\ell$-isometry, we first distinguish the $\ell$ legs such that contraction of their complement yields identity. Then we choose $k\leq \ell$ legs and designate them as logical degrees of freedom. In the dual state $|\psi_W\rangle_{AC}$ defined in Lemma~\ref{lemma21}, let us label the $k$ logical legs as $R_k$ and the remaining $\ell-k$ legs as $R_{\ell-k}$ such that $R_k\cup R_{\ell-k}= C$. Now we compute the mutual information between $R_{k}$ and $A$, where $R_k$ now serves as a reference system isomorphic to the code subspace. As long as the mutual information $I(R_k:A)$ is $2|R|=2k$, we have proved the above statement because all encoded information is contained in $A$ and erasure of $R_{\ell-k}$ is in principle correctable. It is clear that $S(R_k)=S(R_k^c)=k$ by Lemma \ref{lemma21}. Similarly, $S(A)=\ell, S(R_{\ell-k})=S(R_k A)=\ell -k$, because the state is pure. Then 
\begin{equation}
    I(A:R_k)= S(A)+S(R_k)-S(AR_k)=\ell +k-(\ell -k)=2k,
\end{equation}
while
\begin{equation}
    I(R_{\ell -k}:R_l) = S(R_{\ell-k})+S(R_k)-S(A) = \ell-k +k -\ell=0.
\end{equation}
Indeed $A$ has full access to the encoded information while $R_{\ell-k}$ has none. Therefore erasure of $R_{\ell-k}$ is correctable.
\end{hproof}

\begin{lemma}
An $\ell$-isometric tensor as an encoding isometry that encodes $k\leq \ell$ qubits into $n$ physical qubits admits an $n-(\ell-k)$ local decoding unitary that decodes the $k$ data bits of logical information. If one fixes the encoded state and treats the resulting rank $n$ tensor as an $\ell-k$ isometry, then the same unitary transforms the isometry into $(\ell-k)$-identity, the fixed encoded state, and a tensor product of ancilla bits.
\label{lemma:decode}
\end{lemma}
\begin{hproof}
Suppose $W$ is an $\ell$-isometry with total degree $N=n+k$. Let
\begin{equation}
    |\tilde{\psi}\rangle_{AC} = W_{AR}(|\Phi^+\rangle^{\otimes(\ell -k)})_{R_{\ell-k}C} |\psi\rangle_{R_k} = \tilde{W}_{(AC)R_k} |\psi\rangle_{R_k},
\end{equation}
where $R_k\cup R_{\ell-k} = R$ and $R_{\ell-k}\cong C$.
\begin{equation}
    |\Phi^+\rangle = \frac{1}{\sqrt{\chi}}\sum_{i=1}^{\chi}|ii\rangle.
\end{equation}
is a maximally entangled (Bell) state between two isomorphic Hilbert spaces $R_{\ell-k}$ and $C$. $|\psi\rangle$ denotes the logical information we encode. By doing so, we can re-express $W$ as an encoding isometry $\tilde{W}$ that encodes $k$ data qubits (qudits) in $R_k$ to $n$ physical qubits (qudits) in $A\cup C$. 
The state $|\tilde{\psi}\rangle$ is an encoded state and $A$ is the $n-(\ell-k)$ qubit/qudit subsystem whose complement is maximally mixed. By Lemma \ref{lemma22}, it contains the entire encoded information. Thus there exists a unitary decoder $U_A$ that acts only on $A$ and recovers the $k$ data bits of encoded information on $A_1\subset A$ \cite{Harlow:2016vwg}. Thus $U_A$ is at most $n-(\ell-k)$-local. 
\begin{equation}
    U_A|\tilde{\psi}\rangle = |\psi\rangle _{A_1}|\chi\rangle_{(A-A_1)C}.
\end{equation}
Having extracted the information we wanted, we now focus on $|\chi\rangle$ where we know that $C$ of $\ell-k$ qubits/qudits is maximally entangled with $A-A_1$  of $n-\ell\geq \ell-k$ qubits/qudits by Lemma~\ref{lemma22}. Therefore, there must exist Schmidt decomposition
\begin{equation}
    |\chi\rangle = \frac{1}{\sqrt{D}}\sum_{i=1}^{D} |\phi_i\rangle_{C} |\mu_i\rangle_{A-A_1}
\end{equation}
where $D=2^{\ell-k}$ and  $\{|\phi_i\rangle\}$, $\{|\mu_i\rangle_{A-R_k}\}$ are orthonormal. Furthermore, there must be a unitary $U'_{A-A_1}$ on $A-A_1$ such that

\begin{align}
    &U'_{A-A_1} |\chi\rangle_{(A-A_1)C}\nonumber \\
    &= (|\Phi^+\rangle^{\otimes \ell-k})_{C'C} (|0\rangle^{\otimes (n+k-2\ell)})_{A-A_1-C'},
\end{align}
where $C'\subset A-A_1$ and $\dim C'=\dim C$.
Therefore, there exists an $n-(\ell-k)$-local decoding unitary $U_D=U'_{A-A_1}U_A$ with support on $A$  such that

\begin{equation}
    U_D |\tilde\psi\rangle = |\psi\rangle |\Phi^+\rangle^{\otimes(\ell -k)}|0\rangle^{\otimes (n+k-2\ell)}.
\end{equation}

When $|\tilde{\psi}\rangle$ is converted to an isometric mapping between $A$ and $C$, one distills the channel dual to the tensor product of Bell pairs $|\Phi^+\rangle^{\ell-k}$ instead. Indeed, they are the identity operator $I^{\otimes(\ell-k)}$. 
\end{hproof}

\section{Operator Matching and Unitary Stabilizers}
\label{app:b}
\subsection{Unitary Stabilizer and Operator Pushing}
As we have discussed in Sec~\ref{sec:3}, the process of finding unitary product stabilizers can be construed as a form of operator flow in the tensor network known as operator pushing. 



Suppose one has a linear map $V: \mathcal{H}_{A}\rightarrow \mathcal{H}_B$ from a tensor, through channel-state duality, we can turn it into a state by acting it on a maximally entangled state $|\Phi^+\rangle \in \mathcal{H}_A\otimes \mathcal{H}_{A^*}$ such that 

\begin{equation}
    |\psi_V\rangle_{BA^*} = V|\Phi^+\rangle,
\end{equation}
where
\begin{equation}
    |\Phi^+\rangle =\frac{1}{\sqrt{|A|}} \sum_{i=0}^{|A|} |ii\rangle_{AA^*}.
\end{equation}

Here $\{|i\rangle\}$ is a complete orthonormal basis for $\mathcal{H}_A\cong\mathcal{H}_{A^*}$; $|A|$ is the dimension of the Hilbert space $\mathcal{H}_A$.

\begin{lemma}
A linear map $V$ admits operator pushing $VO = O'V$ if and only if the dual state $|\psi_V\rangle$ is a $+1$ eigenstate of $O'\otimes Q$ where $Q$ satisfies $O_A\otimes Q_{A^*}|\Phi^+\rangle_{AA^*}=|\Phi^+\rangle_{AA^*}$.
\label{lemma:b2}
\end{lemma}
\begin{hproof}
In the forward direction, assume operator $O,O'$ can be pushed across $V$, then

\begin{align}
    |\psi_V\rangle &=V|\Phi^+\rangle = V(O_A\otimes Q_{A^*})|\Phi^+\rangle\\
    &= O'_B\otimes Q_{A^*}V|\Phi^+\rangle =O'\otimes Q|\psi_V\rangle.
\end{align}

Conversely, assume $O'\otimes Q$ ``stabilizes'' $|\psi_V\rangle$
\begin{align}
    O'_BQ_{A^*}|\psi_V\rangle &= (O' V)_{BA} Q_{A^*}|\Phi^+\rangle \\
    &= (O' V)_{BA} O_A^{-1}(O_AQ_{A^*}|\Phi^+\rangle) \\
    &= O'_BVO_A^{-1}|\Phi^+\rangle
\end{align}
This implies that $O'_BVO^{-1}_A=V$, or after multiplying both sides by $O_A$, $O'_BV=VO_A$.
\end{hproof}

Therefore, understanding operator pushing in a linear map is the same as understanding the unitary stabilizers of the dual state. The set of such unitary stabilizers is similar to having symmetries to the tensor/state. See \cite{Singh2010} for example. This, of course, applies to our current discussion in the form of encoding isometries. In particular, for tensors with special symmetries, it may be possible to construct transversal logical gates that are non-Clifford using such techniques. 
However, one must take care in selecting the correct operator $Q$ when we dualize between a mapping and a state. In particular, $Q$ and $O$ are equal (up to sign) when they are Pauli operators over qubits. However, even for Pauli operators in higher dimensions, one needs to take care in setting $X\rightarrow X$ and $Z\rightarrow Z^{-1}$ etc when reading off $Q$ from $O$.

\begin{corollary}
Suppose a tensor $V$ admits a unitary stabilizer $O\otimes O'$ such that $O, O'$ only act on the degrees of freedom that we have chosen to be the logical legs and physical legs respectively, then $O'$ is a representation of a logical operator $\bar{Q}$ for the code defined by $V$ when we interpret it as an encoding map. Here $Q$ satisfies $Q\otimes O|\Phi^+\rangle_{AA*}=|\Phi^+\rangle_{AA*}$. 
\label{cor:us_log}
\end{corollary}
\begin{hproof}
From the above lemma we know that $VQ_{B}= O'_A V$. Thus logical operator $Q$ acting on the encoded information in $\mathcal{H}_{B}$ is realized by $O'_A$ with support on the physical Hilbert space $\mathcal{H}_A$.
\end{hproof}

If $Q=I$ and $O'$ is a Pauli operator, then the it produces a stabilizer.
Note that the uniqueness of $Q$ will depend on whether the encoding map $V$ has a non-trivial kernel. For example, the $ZZZZ$ operator in double trace tensor network of two $[[4,2,2]]$ codes corresponds to both $\bar{Z}\bar{Z}$ and the logical identity. However, in a single $[[4,2,2]]$ code, $ZZZZ$ can only correspond to the logical identity\footnote{It seems likely that $Q$ is unique when V is an encoding isometry, i.e. its kernel is trivial.}.

We have provided rules on how to keep track of the form of operators when they are being pushed. However, sometimes we only wish to track the support of these operators during pushing, and not their specific forms. For such purposes, a flow diagram is often used where arrows are drawn over edges on which the operators are supported and the direction of the arrows indicate the direction of operator pushing. Sometimes when the operators are simple enough, we also use different colours to denote the type of operators being pushed through (Figures~\ref{fig:toric_code_network},\ref{fig:3d_log},\ref{fig:twistcode}).

\subsection{Operator Pushing in Connected Tensor Network}
\label{subapp:optpush}
When we start tracing different tensors together, we wish to understand how their respective unitary stabilizers transform under such operations. This allows us to push operators through tensor networks to find new unitary stabilizers of the larger network.

First we make a few comments about tracing.
\begin{lemma}
Let $T_{i_1\dots i_p}, R_{j_1\dots j_q}$ be the tensors of the corresponding states 

\begin{align}
    |T\rangle = \sum_{i_k, k=1,\dots,p} T_{i_1\dots i_p} |i_1,\dots,i_p\rangle\\
    |R\rangle = \sum_{j_k, k=1,\dots q}R_{j_1\dots j_q}|j_1,\dots,j_q\rangle
\end{align}
where $i_k, j_k$ run from $1$ to $\chi$.

Then tracing two edges of the tensors
\begin{equation}
    \sum_{l=1}^{\chi} T_{i_1\dots l\dots i_p} R_{j_1\dots l\dots j_q} 
\end{equation}
corresponds to 
\begin{equation}
    \langle \Phi^+|T\rangle|R\rangle
\end{equation}
where 
\begin{equation}
    |\Phi^+\rangle = \sum_{l=1}^{\chi}|ll\rangle
\end{equation}
is an unnormalized Bell state on the two edges/qudits to be traced. 
\end{lemma}
\begin{hproof}
Without loss of generality, let us trace the first edge of the two tensors
\begin{align}
    \langle \Phi^+|TR\rangle &= \sum_{i_k,j_s}T_{i_1\dots i_p} R_{j_1\dots j_q} \delta_{i_1 l}\delta_{j_1 l}|i_2,\dots,i_p,j_2,\dots,j_q\rangle\\
    &=\sum_{i_k,j_s; k,s\ne 1} \sum_l T_{l\dots i_p} R_{l\dots j_q} |i_2,\dots,i_p,j_2,\dots,j_q\rangle\\
    &=\sum_{i_k,j_s; k,s\ne 1} H_{i_2\dots i_p,j_2\dots j_q} |i_2,\dots,i_p,j_2,\dots,j_q\rangle,
\end{align}
where it is clear that 
\begin{equation}
    H_{i_2\dots i_p,j_2\dots j_q}=\sum_l T_{l\dots i_p} R_{l\dots j_q}
\end{equation}
is created by tracing the two tensors on the correct legs.
\end{hproof}



\begin{theorem}
\label{thm:1}
Consider a unitary stabilizer $S_{A_1}\otimes O_{A_2}\otimes S'_{B_1}\otimes O'_{B_2}$ of a state $|\psi\rangle\in \mathcal{H}_{A_1}\otimes \mathcal{H}_{A_2}\otimes \mathcal{H}_{B_1}\otimes \mathcal{H}_{B_2}$ such that  $\mathcal{H}_{A_2}\cong\mathcal{H}_{B_2}$, and a maximally entangled state 
\begin{equation}
    |\Phi^+\rangle_{A_2B_2} = \sum_{i=1}^{|A_2|}|ii\rangle_{A_2 B_2}.
\end{equation}

If  $\langle \Phi^+|O_{A_2}O'_{B_2}=\langle \Phi^+|$, then the state $\langle \Phi^+|\psi\rangle$, after tracing the two edges corresponding to subsystems $A_2,B_2$, is stabilized by unitary $S_{A_1}\otimes S'_{B_1}$.
\end{theorem}
\begin{hproof}
\begin{align}
    \langle \Phi^+|\psi\rangle &= \langle \Phi^+| S_{A_1}\otimes O_{A_2}\otimes S'_{B_1}\otimes O'_{B_2}|\psi\rangle\\
    &=S_{A_1}\otimes S_{B_1}' (\langle \Phi^+|O_{A_2}\otimes O'_{B_2})|\psi\rangle\\
    &=S_{A_1}\otimes S_{B_1}' \langle \Phi^+|\psi\rangle
\end{align}
Therefore $S_{A_1}\otimes S'_{B_1}$ is a unitary stabilizer of the traced state.
\end{hproof}
When $|\psi\rangle = |\phi\rangle_{A_1 A_2}\otimes |\phi'\rangle_{B_1B_2}$, then this corresponds to tracing two tensors together with a single trace. Alternatively, if $|\psi\rangle$ is entangled, then it corresponds to a self-trace. 
\begin{corollary}
If the unitary stabilizers $S_{A_1}\otimes O_{A_2}, S'_{B_1}\otimes O'_{B_2}$ consist of tensor products of unitary operators acting on each qudit, then so is the resulting unitary stabilizer after tracing $A_2, B_2$.
\label{cor:transversal}
\end{corollary}
We drop the proof as it is obvious. 
As a consequence, it allows us to construct transversal operators through operator pushing/matching. For instance, let there be a logical operator that acts transversally on its qudits and acts with $O_{A_2}$ on a particular leg. One can generate a new code by tracing subsystem $B_2$ of a different tensor in the form of $|\psi'\rangle$ or its dual isometries with $A_2$. As long as $|\psi'\rangle$ is stabilized by $S'_{B_1}\otimes O'_{B_2}$, where $S'$ is the tensor product of local operators on each qudit, and that $\langle \Phi^+|O_{A_2}O_{B_2}=\langle\Phi^+|$, then the resulting logical operator of new code from tracing (or equivalently, operator pushing) remains transversal.

For the special case where $O,O'$ are generalized Pauli operators, the pair of matched operators on $A_2,B_2$ can take the form of $XX, ZZ^{-1}$. When $d=2$ and Pauli operators are over qubits, this is simply matching the same Pauli operators (up to a phase factor) over a trace, hence the form of operator pushing in \cite{HaPPY}. However, this matching extends beyond just Pauli operators. For example, $TT^{-1}$ is also a valid operator matching which is important for determining the existence of transversal non-Clifford operators in such tensor network constructions. More general operator matching rules can be derived by solving for all possible unitaries $O_A,O_B$ such that $O_AO_B|\Phi^+\rangle=|\Phi^+\rangle$.

 Nevertheless, when we push operators through the network, they only remain tensor products when pushing is done by UPS's. More generally, one can also push non-product operators that are unitary stabilizers. The resulting operator from such pushing need not be transversal. While this may not be as desirable in our conventional practices of code building, 
 it is not a problem for general tensor network toy models of holography.
 
 Suppose we are pushing a non-product operator $O$ through $l$ number of connected edges to adjacent tensors. Let us first decompose $O$ in the Pauli basis
 
 \begin{equation}
     \mathcal{O}=\sum_{i_1,\dots,i_l} c_{i_1,\dots,i_l} P_{i_1}\otimes\dots\otimes P_{i_l}.
 \end{equation}

\begin{figure}
    \centering
    \includegraphics[width=0.47\textwidth]{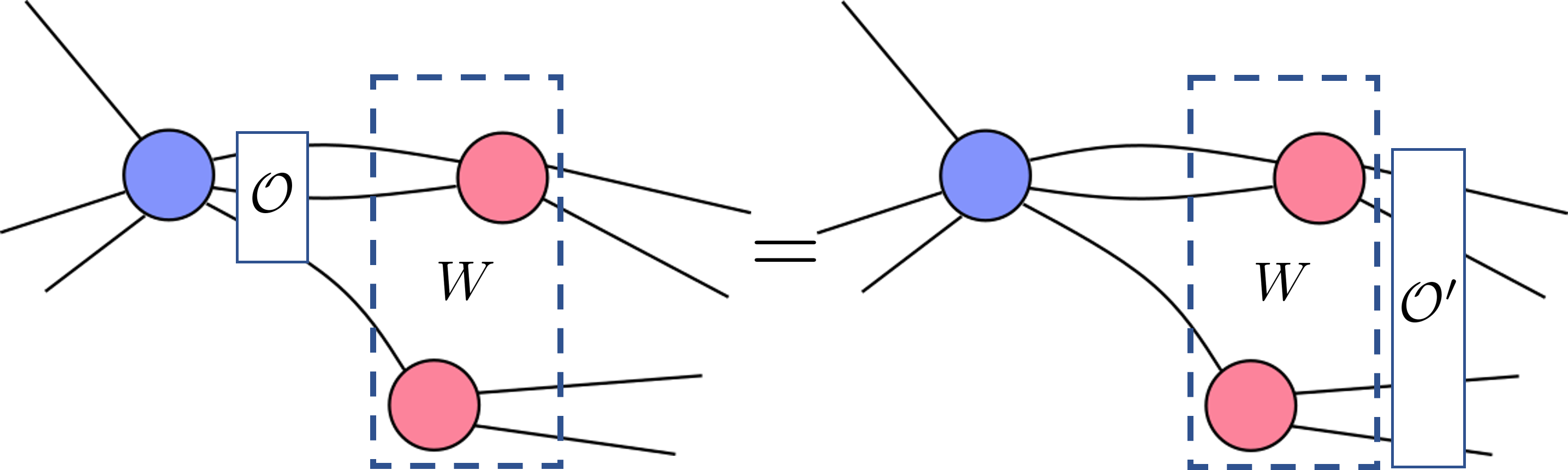}
    \caption{A non-product operator $\mathcal{O}$ can be pushed to the right through the adjacent operators (pink) by treating them collectively as a single tensor $W$ with suitable unitary stabilizers. }
    \label{fig:ent_opt_push}
\end{figure}

If the tensors connected to these edges  can correct these erasure errors, i.e., they collectively act as an isometry such that these $l$ legs are maximally entangled with other degrees of freedoms on the adjacent tensors (e.g. pink tensors in Figure~\ref{fig:ent_opt_push}), then any Pauli operators can be pushed through. Operationally, we can simply apply operator pushing term by term for $\mathcal{O}$ similar to the UPS case and resum them once all terms have been pushed through.
Suppose each term $P_{i_1}\otimes\dots\otimes P_{i_l}$ pushes to an operator $O'_{i_1,\dots, i_l}$ supported on a set of legs on the adjacent tensors which in general need not be a product operator, then the final operator we obtain through pushing $\mathcal{O}$ is 
\begin{equation}
    \mathcal{O}'= \sum_{i_1,\dots,i_l} c_{i_1,\dots,i_l} O'_{i_1,\dots,i_l}.
\end{equation}

See Sec 3.2 of \cite{ABSC} for an explicit example of operator pushing.

If the adjacent tensors do not correct erasure errors on these legs, i.e., not all operators can be pushed through, then we first identify the set of unitary stabilizers these adjacent tensors do have, which act non-trvially as a set of $\{Q^{m}= Q^m_1\otimes\dots\otimes Q^m_l \::\: m\in M\}$ on the $l$ connected edges. We use $M$ to label the set of all such operators acting on the connected edges. Then we find their corresponding matching operators $\{O^{m}= O^m_1\otimes\dots\otimes O^m_l \::\: m\in M\}$ for each $l$ and $m\in M$ such that, 
\begin{align*}
    &(\langle\Phi^+|)^{\otimes l}(O^m\otimes Q^m)\\
    =&(\langle\Phi^+|O^m_1\otimes Q^m_1)\otimes \dots\otimes (\langle\Phi^+|O^m_l\otimes Q^m_l)\\
    =&(\langle\Phi^+|)^{\otimes l}. 
\end{align*}

If $O$ can be expanded in terms of $\{O^m\}$, then we again repeat the same term by term pushing above and add up the pushed operators in the end. 

A more intuitive way to think about this is to treat the collection of adjacent tensors as a mapping $W$. If $W$ admits a suitable unitary stabilizer then one can push $\mathcal{O}$ to the right as $\mathcal{O}'$ (Figure~\ref{fig:ent_opt_push}).

\subsection{Operator Pushing in Practice}
\label{subapp:optpush_practice}
Here we summarize a few practical tips for operator pushing.

 Assume that we start with a number of uncontracted tensors generated by stabilizer states and designate a set of physical and logical legs. From their unitary stabilizers, we can determine the logical operators and stabilizers using Lemma~\ref{lemma:b2} and Corollary~\ref{cor:us_log}. Then we contract these tensors into a network of our choice.

From Theorem~\ref{thm:1}, we know that by inserting a stabilizer of a tensor and push it around with stabilizers of nearby tensors, we will produce a stabilizer of the final tensor network. This is always true and we can rely on such methods to characterize stabilizer groups of such QECCs. 

If, on the other hand, we have pushed through a logical operator of a tensor, one can push it around using both stabilizers and logical operators and other nearby tensors. The nature of the resulting operators will now depend on the tensor network itself. 

If we have contracted the tensors in a way such that these connected edges can be treated as correctable erasure errors in the adjacent tensors, then their logical legs represent independent degrees of freedom. More precisely, the logical legs remain independent degrees of freedom if tensors are contracted in a way described in Theorem 1 of \cite{TNC} or equivalently by Algorithm ~\ref{alg:TNbuild} in  Appendix~\ref{subapp:c2}. In this case, such kind of operators pushing produce logical operators on the larger tensor network. If we further restrict ourselves to bond dimension 2, then we reproduce the tensor network code (TNC) proposed by \cite{TNC}. 

However, if the above is false, then sometimes pushing of logical operators can also produce stabilizer elements. If that happens, then the logical legs in the tensor network are not independent encoded qubits/qudits. Some obvious examples are given in Sec~\ref{subsec:3b} and Sec~\ref{subsec:3f}. In that case, naive counting of the apparent number of logical legs does not correspond to the number of encoded qubits/qudits.

There are also cases in which we are not as interested in producing stabilizer codes. This is particularly important for toy models of holography where non-stabilizerness is crucial. In these scenarios, we are often interested in the support of logical operators but less so about its specific form or (the lack of) transversality. Therefore, a tensor network constructed through contracting general isometries not dual to stabilizer states is also useful. We can understand such codes created by Algorithm \ref{alg:TNbuild} as generalizations of \cite{TNC}, where the tensors of individual stabilizer encoding maps are replaced by non-stabilizer higher dimensional counterparts.
In all of these codes, one can easily derive the support logical operators by following the flow of operator pushing.
Such tensor networks are also efficiently contractible and are thus desirable for numerical computations. Contractibility is nice because we can give a procedure that constructs the explicit decoding circuit for each such network. For a simple illustration see Figure~\ref{fig:cont_circuit}.


\section{General Network and decoding unitary}
\label{app:c}
Some of the tensor networks are efficiently contractible through isometries. In these tensor networks, it is much easier to compute the expectation values of certain operators through tensor contractions. In this section, we describe how tensor contraction steps can be mapped to a decoding unitary circuit of the tensor network quantum code. We also give a construction for the type of networks that are guaranteed to have such explicit encoding/decoding circuits. For all our discussions, we assume the bond dimension to be constant through out the network.  
\subsection{General Methodology}
Here we focus on tensor networks that can be locally efficiently contracted through isometries\cite{HMERA}. Common examples include \cite{Vidal2008,branchingmera,HaPPY,AguadoVidal}. In such tensor networks, when the network is contracted with its conjugate transpose, one can contract the individual tensors sequentially to identity operators. Such properties facilitates the computation of local observables as they only depend on a relatively small part of the entire network. For example, in MERA, computing the expectation value of a local operator would only involve contracting tensors its causal cone.

Suppose we are given one such tensor network $G_0$ and a contraction sequence $S_G=\{\eta_i: G_i\rightarrow G_{i+1}\}$ through the contraction of isometries. One can think of each step of the contraction as a mapping from the initial graph $G_i$ before contraction and the final graph $G_{i+1}$ after. A more concrete expression of this sequence is found in Figure~\ref{fig:cont_circuit}c. When the tensor network acts as an encoding map, we assume the logical dangling legs network are never touched during the contraction sequence. For concreteness, one can fix these logical legs to certain states so that the overall network represents an encoded state. The contraction sequence does not depend on what states they are fixed to.

Explicitly, such networks can be written as

\begin{equation}
    V=\prod_{i=0}^{N} W_i.
\end{equation}
Each $W_i$ is a set of isometries that one can contract in a step of the sequence. In general it may also be written as a tensor product of isometries
\begin{equation}
    W_i=\bigotimes_{\alpha_i}W_i^{\alpha_i},
\end{equation}
where $W_i^{\alpha}$ correspond to the individual isometric tensors that appear in the tensor network. 
To construct the circuit, we begin with a number of wires equal to the total number of physical degrees of freedom in the tensor network. 

Then for each contraction of an isometry into identity, it involves an isometric tensor acting on a particular set of wires that are either dangling (non-logical) legs or contracted legs. Let us treat such an isometry tensor as an encoding map where the dangling wires are physical qubits the user has control and the other contracted wires are erasure errors. Using lemma \ref{lemma:decode}, we can find a decoding unitary that decodes the encoded information by acting only on the uncontracted legs. This unitary converts each ($\ell$) isometry tensor $W_i^{\alpha_i}$ to the tensor product of ($k\leq \ell$) decoded data qubits (qudits), an ancillary all $|0\rangle$ state, and the identity operator that acts on the remaining ($\ell-k$) physical degrees of freedom.
Let $U_{i}^{\alpha_i}$ be the decoding unitary that corresponds to isometry $W_i^{\alpha_i}$. At the $0$-th step
\begin{align}
    \bigotimes_{\alpha_0} U_0^{\alpha_0}W_0W_1\dots W_N\bigotimes_{i=0}^N|\psi_i\rangle =\nonumber\\
    |\psi_0\rangle|\Omega_0\rangle \mathbb{I} (W_1\dots W_N)\bigotimes_{i=1}^N|\psi_i\rangle,
\end{align}
where, respectively, $|\psi_0\rangle, |\Omega_0\rangle$ are the decoded information and the ancillary all 0 state from isometries contractible at step 0 of the sequence. For the sake of clarity in the presentation, we fix the encoded states to be a product state. If the encoded state is entangled, then we recover the a mixed state. Proceeding similarly for $i\geq 1$, we then obtain the decoding circuit

\begin{equation}
    \underbrace{(\bigotimes_{\alpha_N}U_N^{\alpha_N})\dots(\bigotimes_{\alpha_0}U_0^{\alpha_0})}_{\rm unitary~decoding~circuit} \overbrace{V|\Psi\rangle}^{\rm encoded~state} =\bigotimes_{i=0}^N\underbrace{|\psi_i\rangle}_{\rm decoded~info}|\Omega_i\rangle,
\end{equation}
where $\bigotimes_i |\Omega_i\rangle=|0\rangle\otimes |0\rangle \otimes\dots$ are ancillary states.

Graphically we place the corresponding the decoding unitary $U_i^{\alpha_i}$ of a tensor on the same set of wires in the quantum circuit. For the wires that contract to identity in the tensor network, they will continue on to a later time in the circuit. The other wires will terminate on either an ancilla or a decoded state in the circuit. See Figure~\ref{fig:cont_circuit} for an example. Note that for small enough isometry tensors, we assume that these decoding unitaries can be independently derived and are thus known.

Readers may recall that in MERA one can convert the tensor network with isometries into a unitary quantum circuit where the isometries are replaced with a unitary with a fixed ancilla output. A similar process takes place here except one also recovers a non-trivial $|\Psi\rangle$ in addition to the ancillary $|0\rangle$s. 

Another point to note is that the contraction sequence given a tensor network may not be unique. Different sequences thus give rise to different realizations of the decoding process. For encoding, one can run the circuit backwards.

Sometimes, the tensor network may only be approximately contractible, i.e., the tensors contract to identity up to a small error\cite{ABSC}. In such cases, the quantum circuit from exact contraction sequence can still be useful, which encodes/decodes the logical information at the expense of small errors. 

\subsection{Construction of contractible tensor network codes}
\label{subapp:tnbuild}
Here we give a general construction of tensor networks that are contractible in the above description. As such, it always admits an explicit encoding/decoding circuit construction from the contraction sequence above. Algorithm~\ref{alg:TNbuild} is sufficient for producing a network that is locally contractible, and thus resulting in a local decoding quantum circuit. 

\begin{algorithm}
\label{subapp:c2}
\SetAlgoLined
 initialization: start with an $\ell_0$-isometry with no more than $\ell_0$ logical legs;\\
 \For{$0<i<$cut-off}{Begin with the $\ell_{i-1}$-isometry tensor network with $k_{i-1}$ logical legs and $n_{i-1}$ total non-logical external legs\\
 Select another $r_i$-isometry and designate $\kappa_i< r_i$ logical legs and total $m_i$ non-logical external legs\\
 Contract no more than $c_i=r_i-\kappa_i$ legs with the non-logical external legs of the $\ell_{i-1}$-isometry\\
 Produce a new network which is a $\ell_i$-isometry with $k_i=k_{i-1}+\kappa_{i}$ logical legs and $n_i=n_{i-1}+m_i-2c_i$ non-logical external legs
 }

 \caption{Tensor Network Building Algorithm}
 \label{alg:TNbuild}
\end{algorithm}
For isometries that are obtained from stabilizer codes with bond dimension 2, this is simply an iterative procedure that applies Theorem 1 of \cite{TNC} repeatedly provided we pre-assign logical and physical legs to the tensors before contraction and then never contract the logical legs in the individual tensors. Codes generated that way are known as tensor network codes (TNC). If we permit the contraction of logical legs, as in the example shown in Figure~\ref{fig:steane_RM}a, then the procedure also produces non-TNCs. 

\subsection{Compatibility with Other known Decoding Methods on Tensor Networks}
\label{subapp:c3}
Let us refer to the codes built by the current framework as quantum lego codes (QLC). There are several subclasses of QLC whose decoding have been discussed. Therefore, one can reuse some of these methods when we fall into one of these classes. For instance, 
\begin{equation}
    \mathrm{QLC}\supset \mathrm{QbQLC}\supset \mathrm{TNC} \supset \mathrm{HSbC},
\end{equation}
where HsBC stands for holographic stabilizer codes of which HaPPY code is a member.

The approach by Ferris and Poulin (FP) \cite{FP2013} is based on qubit code decoding unitaries which map operators to operators. Let us refer to QLCs over qubits whose encoding/decoding unitaries are known as qubit QLCs (QbQLC). More precisely, let $U$ be an encoding unitary that maps a pre-encoded state on $n$ qubits to an encoded state on $n$ qubits.
\begin{equation}
    |\tilde{\psi}\rangle = U|\psi\rangle_k |0\rangle^{\otimes {n-k}}
\end{equation}
where $|\psi\rangle_k$ is an encoded message over $k$ data qubits. This unitary is known for tensor networks described in the previous section. 

However, it is not exactly the unitary used by \cite{FP2013} to compute the conditional probability distributions $Q(E|s)$ of error $E$ given syndrome $s$, which is a tensor network of bond dimension 4. To do so, we introduce the FP-compatible encoding unitary $\mathcal{U}$ by contracting with tensors of Pauli operators over qubits $\sigma_{\alpha\beta}^i$ where $\alpha,\beta=0,1$ and $i=0,1,2,3$.
\begin{equation}
    \mathcal{U}^{j_1,\dots,j_n}_{i_1,\dots,i_n}= \Tr[U(\sigma^{j_1}\otimes \dots\otimes \sigma^{j_n})U^{\dagger}(\sigma^{i_1}\otimes \dots\otimes \sigma^{i_n})].
    \label{eqn:fpcompU}
\end{equation}
This unitary can then be contracted into a tensor network as described by \cite{FP2013} to compute $Q(E|s)$. 

If the unitary $U$ describes a stabilizer code on qubits, then it is a tensor network code (TNC) for which the maximum likelihood decoder was proposed \cite{TNC}. In that case, a tensor $T(L)_{g_1,\dots,g_n}$, where $g_i=0,1,2,3$, is used in \cite{TNC} to track all the unitary product stabilizers that are also Pauli operators. Then it is straightforward to enumerate all stabilizer equivalent representations of some logical Pauli operator $L$ and construct the tensor based on equation (2) in \cite{TNC}. 

Let us assume that the augmented check matrix of an $[[n,k]]$ stabilizer code (see Appendix~\ref{app:d} below) is known, then we enumerate all stabilizers generated by adding the rows vectors $\vec{v}_i=(v_i)^j = H_{ij}$ on the check matrix 
\begin{equation}
    \mathcal{S}=\{\vec{v}_b=\sum_i \vec{v}_i b^i \::\: \forall b^i\}
\end{equation}
where $b_i$ are binary strings of length $n+k$ and $|\mathcal{S}|=2^{n+k}$. Then for each $\vec{v}_b\in \mathcal{S}$ that corresponds to a particular Pauli $L$ in the augmented section, we identify the corresponding pure stabilizer section, which yields a $2n$-entry row vector. Then for each $i=1,\dots,n$, we look at the 2-tuple given by $i$th and the $i+n$-th entry. From it we generate a number $g_i$ using the mapping below
\begin{align*}
    (0,0) &\leftrightarrow g_i = 0\\
    (1,0) &\leftrightarrow g_i =1\\
    (0,1) &\leftrightarrow g_i= 3\\
    (1,1) &\leftrightarrow g_i = 2.
\end{align*}
These $g_i$ label the $n$ tensor indices of $T(L)_{g_1,g_2,\dots,g_n}$.
Then we set $T(L)_{g_1,g_2,\dots,g_n}=1$. Overall, this sets  $2^{n+k}$ elements of the tensor $T(L)$ to 1 if the corresponding element is in $\mathcal{S}$. We set the remaining elements to 0.
\section{Operator Matching for Stabilizer Codes}
\label{app:d}
The total number of unitary stabilizers of a tensor grows exponentially with system size, however they can be characterized from a few generators. This is especially apparent for $[[n,k]]$ stabilizer codes where the number of the stabilizers grows as $2^{n-k}$ while there are only $n-k$ generators. Therefore, it is informative to track the stabilizer generators under tensor gluing in addition to the graphical description. 
Here we provide an efficient way of describing the operator matching process in the form of check matrix operations so that we only track the stabilizer generators. A cursory version is first discussed in \cite{ABSC} on qubits. Here we also generalize the procedure to prime dimension qudits. 
\subsection{Conjoining Operations}
\label{subapp:d1}

Consider two qudit stabilizer codes with local dimension $D$. Write their check matrices as $H_1, H_2$, which have entries in the cyclic group $\mathbb{Z}_D$. For the sake of clarity, we take $D$ to be a prime\footnote{For generalization to composite dimensions on $n$ qudits, one may have up to $2n$ stabilizer generators as opposed to $n$, but no other modifications are needed \cite{quditstab}.}. The graphical tensor gluing operation can be rephrased as the conjoining operation on check matrices,

\begin{equation}
   H= H_1 \wedge_{J} H_2 \sim H_1 \wedge_{\{e\}} H_2
\end{equation}
where $J$ is a set of columns in the respective check matrices that correspond to the set of qudits over which the tensors are glued. This operation is exactly equivalent to operator matching/pushing, but is more succinct in the language of matrix operations. Equivalently, we denote the conjoining by the corresponding gluing edges $\{e\}$ in the tensor network. Overall, the conjoining operation is a mapping that takes two check matrix inputs and output a single check matrix that corresponds to the stabilizer code of a glued tensor network. The specific construction of this mapping can be translated into a sequence of operations of the check matrices that includes (not necessarily in order) direct sum, row operations, row reductions, and row/column deletions. 

Any conjoining operation can be broken down into a single trace operation, which glues one leg from two disconnected tensors, and a number of possible subsequent self-trace operations, which corresponds to gluing one leg of the tensor with another on the same tensor (or on the connected tensor network). Therefore, it is sufficient to define these elementary operations separately. If one of the matrices is trivial/empty, then the conjoining just correspond to a number of self-traces. These operations are sufficient in creating a check matrix description for any stabilizer tensor network.

Practically, when we apply this to study the gluing of stabilizer tensor networks, it is convenient to treat each tensor as a $k=0$ code such that their respective check matrix descriptions have full rank. In such cases, one can easily verify from the operations below that the resulting check matrices remain full rank after each trace.

\textit{Single Trace Operation:}
To perform a single trace, we identify the columns corresponding to the qudit to be traced over. Without loss of generality, suppose this is the first qudit of each code, which corresponds to the 1st and $n+1$-st columns on both matrices. 

If both codes correct any error on the first qudit, then the check matrices can be written as 
\begin{equation}
H_1=
\left(\begin{array}{cc|cc} 
1 & v_1^t & 0 & u_1^t \\
0 & w_1^t & 1 & r_1^t\\
0 & A_1 & 0 & B_1 
\end{array}\right),
\ 
H_2=
\left(\begin{array}{cc|cc} 
1 & v_2^t & 0 & u_2^t \\
0 & w_2^t & 1 & r_2^t\\
0 & A_2 & 0 & B_2 
\end{array}\right)
\label{eqn:gen_check}
\end{equation}
after suitable row reductions.

In this notation, the left half of each check matrix represents $X$-type operators and the right half $Z$-type operators.
If a particular row in $H_1$ whose entry in the 1st column is nonzero (e.g. first row on both $H_1, H_2$), we match it with the row in $H_2$ that has an non-zero entry in its first column. For the $(n+1)$-st column, we match an operator to its inverse. For instance, in the second rows of $H_1, H_2$, we need to match 1 to $-1$ because from previous sections we know that $Z$ should be matched to $Z^{-1}$. We then remove the $1$st and the $n+1$st entries on these matching rows and then concatenate. This produces a row in the new check matrix $H$. In this example, the matching rows $(1~v_1^t|0~u_1^t), (1~v_2^t|0~u_2^t)\rightarrow (v_1^t~v_2^t|u_1^t~u_2^t)$; $(0~w_1^t|1~r_1^t),(0~w_2^t|1~r_2^t)\rightarrow (w_1^t~~ -w_2^t|r_1^t~~-r^t_2)$. For rows whose entries on the 1st and the $(n+1)$-st columns are zero, then we remove those entries and direct sum the remaining block matrices/rows.

Therefore, tracing the first qubit of these two codes correspond to the following operation
\begin{equation}
    H=
    \left(\begin{array}{cc|cc} 
v_1^t & v_2^t & u_1^t & u_2^t\\
w_1^t & -w_2^t & r_1^t & -r_2^t\\
A_1 & 0 & B_1 & 0\\ 
 0 & A_2 & 0 & B_2 \\
\end{array}\right).
\end{equation}

If, on the other hand, the check matrix does not correct a single erasure error on a qudit to be traced, then we are not guaranteed to find a row with matching entries at the $1$st and $(n+1)$-st columns when conjoining the two matrices. In terms of the check matrix, it means that only one row in the row reduced check matrix would contain non-zero entries in the 1st and the $(n+1)$-st positions.

Suppose only one of the two tensors is erasure correcting on the qudit in the tracing position. Without loss of generality, assume $H_1$ corrects erasure error on the first qudit. Then
\begin{equation}
H_1=
\left(\begin{array}{cc|cc} 
1 & v_1^t & 0 & u_1^t \\
0 & w_1^t & 1 & r_1^t\\
0 & A_1 & 0 & B_1 
\end{array}\right),
\ 
H_2=
\left(\begin{array}{cc|cc} 
i & v_2^t & j & u_2^t \\
0 & A_2 & 0 & B_2 
\end{array}\right),
\label{eqn:gen_check2}
\end{equation}
where $(i,j)\ne (0,0)$\footnote{While choosing the ``generators'', we only require that they generate the generalized Pauli group up to a phase.}.
We then perform row operations on $H_1$ such that 
\begin{equation}
H_1\rightarrow \\
\left(\begin{array}{cc|cc} 
i & i v_1^t-jw_1^t & -j & i u_1^t-jr_1^t \\
0 & w_1^t & 1 & r_1^t\\
0 & A_1 & 0 & B_1 
\end{array}\right)
\label{eqn:gen_check3}
\end{equation}
where $iv^t$ denotes scalar multiplication by $i\in\mathbb{Z}_D$. Then we concatenate the rows that have matching 2-tuple at the $1$st and the $n+1$-st columns e.g. $(i,j)$. We drop the rows with nonzero  but non-matching 2-tuples. If the 2-tuple is zero, then we direct sum the block matrices as before. The single-traced check matrix one produces is 
\begin{equation}
    H=
    \left(\begin{array}{cc|cc} 
i v_1^t-jw_1^t & v_2^t & i u_1^t-jr_1^t & u_2^t\\
A_1 & 0 & B_1 & 0\\ 
 0 & A_2 & 0 & B_2 \\
\end{array}\right).
\end{equation}

If neither of the check matrices correct such located Pauli errors, then the check matrices can be reduced to the form
\begin{equation}
H_1=
\left(\begin{array}{cc|cc} 
i_1 & v_1^t & j_1 & u_1^t \\
0 & A_1 & 0 & B_1 
\end{array}\right),
\ 
H_2=
\left(\begin{array}{cc|cc} 
i_2 & v_2^t & j_2 & u_2^t \\
0 & A_2 & 0 & B_2 
\end{array}\right),
\label{eqn:gen_check4}
\end{equation}
where we again assume $(i_k,j_k)\ne (0,0)$. If $(i_1,j_1)=(i_2,-j_2)\cdot l$ for some $l\in\mathbb{Z}_D$, then we concatenate the rows after removing the entries in the 2-tuples. If not, then we drop both rows in making the new check matrix. More precisely, let $f(i_k,j_k)$ be an indicator function such that $f=1$ if $l(i_1,j_1)=(i_2,-j_2), l\in \mathbb{Z}_D$ and $f=0$ if not.  Since all entries are in $\mathbb{Z}_D$, it takes at most $D$ operations to check if the two 2-tuples are matching. The check matrix now reads

\begin{equation}
    H=
    \left(\begin{array}{cc|cc} 
fv_1^t & (fl)v_2^t & fu_1^t & (fl)u_2^t\\
A_1 & 0 & B_1 & 0\\ 
 0 & A_2 & 0 & B_2 \\
\end{array}\right).
\end{equation}

One then performs a row reduction to produce the final check matrix.

\textit{Self-Trace Operation:}
Self-trace is very similar to single-trace in that they are both just the matrix counterparts of operator matching.
To perform a self-trace, we identify the two qudits/columns to be traced over. Again, without loss of generality, let us take them to be the first two columns in the $X$ and $Z$ sections of the the check matrix. If they satisfy the matching rules, then we delete these columns from the check matrix. If there exist non-matching rows, then those rows are also eliminated. Again, without loss of generality, we trace the first two qudits that correspond to the first two columns. If one traces over other qudits, then we can simply shuffle the columns to the first two positions. Suppose the code detects any single qudit errors on the qudits to be traced, then a check matrix can be arranged in the following form
\begin{equation}
H=
\left(\begin{array}{ccc|ccc} 
1 & 0 & v_1^t & 0 & 0 & u_1^t \\
0 & 1 &  v_2^t & 0 & 0 & u_2^t\\
0 & 0 & v_3^t & 1 & 0 & u_3^t\\
0 & 0 & v_4^t & 0 & 1 & u_4^t\\
0 & 0 & A & 0 & 0 & B
\end{array}\right).
\end{equation}

We perform row operations such that the first two entries of a row are equal in the $X$ section or adds up to $D$ in the $Z$ section. Self-trace can be easily performed by removing the first two columns and any rows whose entries in the first two columns do not match. 

\begin{align}
H&\xrightarrow{\mathrm{row~opt.}}
\left(\begin{array}{ccc|ccc} 
1 & 1 & v_1^t+v_2^t & 0 & 0 & u_1^t+u_2^t \\
0 & 1 & v_2^t & 0 & 0 & u_2^t \\
0 & 0 & v_3^t-v_4^t & 1 & -1 & u_3^t-u_4^t\\
0 & 0 & v_4^t & 0 & 1 & u_4^t\\
0 & 0 & A & 0 & 0 & B
\end{array}\right)\\
&\xrightarrow{\mathrm{self-trace}}
\left(\begin{array}{c|c} 
 v^t_1+v_2^t &  u_1^t+u_2^t \\
 v_3^t-v_4^t & u_3^t-u_4^t\\
 A & B
\end{array}\right).
\end{align}


If the code only corrects one located error on the 2 legs we plan to trace together, then the check matrix takes on the form,
\begin{align}
H&=
\left(\begin{array}{ccc|ccc} 
i & 0 & v_1^t & j & 0 & u_1^t\\
\eta_2m & 1 & v_2^t & \eta_2n & 0 & u_2^t \\
\eta_3m & 0 &  v_3^t & \eta_3n & 1 & u_3^t\\
0 & 0 & A & 0 & 0 & B
\end{array}\right)
\end{align}
 for some $(i,j)\ne (0,0)$, where $\eta_2, \eta_3\in \mathbb{Z}_D$, $(i,j)\ne l(m,n)$ for any $l\in \mathbb{Z}_D$.
If $\eta_2=\eta_3=0$, then

\begin{align}
H&=
\left(\begin{array}{ccc|ccc} 
i & 0 & v_1^t & j & 0 & u_1^t \\
0 & 1 &  v_2^t & 0 & 0 & u_2^t\\
0 & 0 & v_3^t & 0 & 1 & u_3^t\\
0 & 0 & A & 0 & 0 & B
\end{array}\right)\\
&\rightarrow
\left(\begin{array}{ccc|ccc} 
i & i & v_1^t+iv_2^t-jv_3^t & j & -j & u_1^t+iu_2^t-ju_3^t \\
0 & 1 &  v_2^t & 0 & 0 & u_2^t\\
0 & 0 & v_3^t & 0 & 1 & u_3^t\\
0 & 0 & A & 0 & 0 & B
\end{array}\right)\\
&\rightarrow 
\left(\begin{array}{c|c} 
 v_1^t+iv_2^t-jv_3^t  & u_1^t+iu_2^t-ju_3^t \\
 A &  B
\end{array}\right).
\end{align}

As before, it must be possible to reduce all other entries in the 1st and the $(n+1)$-st columns to zero. If there exists another element whose 2-tuple is linearly independent of the $(i,j)$, then together with $(i,j)$ they must generate the full Pauli group to be phase factors. This then reduces to the first case where the traced qudit is a correctable erasure error. If certain rows in these columns can not be made to match, we also eliminate the corresponding rows. 

If $\eta_2\neq \eta_3$, then we can rearrange the check matrix into one of the following form via row operations.
\begin{align}
H&=
\left(\begin{array}{ccc|ccc} 
1 & l' & r_1^t & 0 & 0 & w_1^t\\
0 & k' & r_2^t & 1 & 0 & w_2^t \\
0 & j' & r_3^t & 0 & 1 & w_3^t\\
0 & 0 & A & 0 & 0 & B
\end{array}\right)
\end{align}
or 
\begin{align}
H&=
\left(\begin{array}{ccc|ccc} 
1 & 0 & r_1^t & 0 & l' & w_1^t\\
0 & 1 & r_2^t & 0 & j' & w_2^t \\
0 & 0 & r_3^t & 1 & k' & w_3^t\\
0 & 0 & A & 0 & 0 & B
\end{array}\right)
\end{align}
for some arbitrary $l',j',k'\in \mathbb{Z}_D$ and vectors $w, r$. Because the two above matrices are equivalent up to swapping the second columns in the left and right sections followed by row swaps, we will consider the first matrix without loss of generality. 

\begin{align}
H&\rightarrow
\left(\begin{array}{ccc|ccc} 
1 & l' & r_1^t & 0 & 0 & w_1^t\\
0 & k'-j' & r_2^t-r_3^t & 1 & -1 & w_2^t-w_3^t \\
0 & j' & r_3^t & 0 & 1 & w_3^t\\
0 & 0 & A & 0 & 0 & B
\end{array}\right)
\label{eqn:matchingmatrix}
\end{align}
can produce at least one matching operator by adding the first two rows after multiplying them by constants $a=(j'-k')$ and $b=(l'-1)$ respectively. Namely, the row

\begin{equation}
    (a,al'+b(k'-j'),ar_1^t+b(r_2^t-r_3^t)|b,-b,aw_1^t+b(w^t_2-w_3^t))
    \label{eqn:matching row}
\end{equation}
corresponds to a stabilizer generator. 
For some values of $j',k',l'$, we also obtain other matching rows. However, these must be a linear combination of other rows.

Note that the resulting matrix must be full rank after row elimination. The top 3 rows are linearly independent from the remaining rows. If $r,w$ are all parallel to some vectors in $A,B$, then one must find 3 linearly independent vectors that only have non-zero entries on the 2 columns. This is impossible, so at least one of $r_i,w_i$ is linearly independent of the vectors in $A,B$. The matching solution above have support in all 3 such vectors, so it must not be parallel to row vectors in $A,B$ unless (a) $r_2-r_3=0, w_2-w_3=0$ for any $a,b$ or (b) at least one of $a,b$ is zero.

In case (a), the first and second row of (\ref{eqn:matchingmatrix}) only lead to commuting operators if $l'\ne 0$. This means that a suitable multiple of the first row is matching, hence the final check matrix will also include vector $r_1,w_1$. If it were non-zero, then we are done. If it is zero, then the operator on the first row does not commute with the one on the third, which is a contradiction. Therefore $(r_2,w_2)-(r_3,w_3)\ne 0$ and it must be a linearly independent vector. (\ref{eqn:matching row}) has non-trivial support on all three such row vectors unless at least one of $a,b=0$.

For case (b1), suppose $a= j'-k'=0, b\ne 0$. If one of $(r_2,w_2), (r_3,w_3)$ is linearly independent then we are done. If only $(r_1,w_1)$ is the requisite linearly independent vector, then $r_2,w_2,r_3,w_3=0$. If $l'\ne 0$, then some multiple of the first row produces a matching operator, and we are done. If $l'=0$, then it does not commute with the stabilizer on the second row, which is a contradiction. If $b=0$ (case b2), $l'=1$, the first row is matching. Again, we are done if $(r_1,w_1)$ is linearly independent. If not, then $(r_1,w_1)=0$ and the first row does not commute with the third. If $a=b=0$ (case b3), both the first and second rows are matching operators. The second row already produces a linearly independent vector. 

Finally, if neither qudits are correctable erasures, we again need to check whether the two 2-tuples $(i_1,j_1)$ and $(i_2,j_2)$ match using the function $f$ such that $f=1$ if $l(i_1,j_1)=(i_2,-j_2)$ and $f=0$ if no such $l\in \mathbb{Z}_D$ exists. In the latter case, we simply remove the first two rows, then follow by removing the first two columns of the $X$ and $Z$ sections of the check matrix. More succinctly, summarize the operations as 
\begin{align}
&H=
\left(\begin{array}{ccc|ccc} 
i_1 & 0 & v_1^t & j_1 & 0 & u_1^t \\
0 & i_2 &  v_2^t & 0 & j_2 & u_2^t\\
0 & 0 & A & 0 & 0 & B
\end{array}\right)\\
&\rightarrow
\left(\begin{array}{ccc|ccc} 
li_1 & i_2 & lv_1^t+v_2^t & lj_1 & j_2 & lu_1^t+u_2^t \\
0 & i_2 &  v_2^t & 0 & j_2 & u_2^t\\
0 & 0 & A & 0 & 0 & B
\end{array}\right)\\
&\rightarrow 
\left(\begin{array}{c|c} 
 flv_1^t+fv_2^t  & flu_1^t+fu_2^t \\
 A &  B
\end{array}\right).
\end{align}

To obtain the valid forms of check matrices, one perform row reduction to all $H$s we analyzed above and eliminate any zero rows. This is to eliminate extra row vectors that are linearly dependent of other row vectors. 

\subsection{Channel-State Duality}
\label{subapp:cj_dual}
Specifying the stabilizers is helpful, but often not enough for a multi-qudit logical subspace. To keep track of the logical operators under tensor gluing, we simply augment the above check matrices with their logical operators as well. In the tensor network language, we are simply turning these encoding maps into stabilizer states by demoting the logical legs to physical legs.

We first augment the logical operators. For each logical $\bar{X}_i$ with physical representation $O^i_x$, we write $S^i_{\bar{X}}= O^i_x\otimes I_1\otimes\dots \otimes X_i\otimes I_{i+1}\dots$. Similarly, for logical $\bar{Z}= O_z$, we construct $S_{\bar{Z}}=O_z^i\otimes I_1\otimes\dots\otimes Z^{\dagger}_i\otimes \dots$. Suppose the original stabilizer group of the code is $\mathcal{S}_H$; we then construct a new stabilizer group from each of the augmented logical operators
$S_{\bar{X}}^i,S_{\bar{Z}}^i$ and all augmented stabilizers $S_{h}\otimes I_1\otimes I_2\otimes\dots$, where $S_{h}\in \mathcal{S}_H$ are the $n-k$ generators of the original stabilizer group.
\begin{equation}
    \mathcal{S}_{H_A}= \langle S_h\otimes I\otimes \dots, S_{\bar{X}}^i,S_{\bar{Z}}^i \::\: \forall 1\leq i\leq k \rangle
\end{equation}

 It is straightforward to check that $\mathcal{S}_{H_A}$ has $n+k$ generators and now identifies a stabilizer state over $n+k$ qudits.  Then we have converted the encoding isometry of an $[[n,k]]$ stabilizer code into an $[[n+k,0]]$ stabilizer state in a way similar to \cite{HaPPY} where one converts the perfect code into the perfect tensor. 
 
 More explicitly, let the original check matrix of the code be $H=(H_X|H_Z)$. Then the new check matrix of the $[[n+k,0]]$ code/state is
 \[
 H_A=~
 \begin{blockarray}{c|c|c|c}
\begin{block}{(c|c|c|c)}
\begin{block*}{c|c|c|c}
\begin{block*}{c|c|c|c}
H_X & 0 & H_Z & 0\\
\end{block*}\BAhhline{--|--}
\begin{block*}{c|c|c|c}
P_{LX} & L_X & Q_{LX} & 0\\
P_{LZ} & 0 & Q_{LZ} & L_Z\\
\end{block*}
\end{block*}
\end{block}
\end{blockarray}
\]
where $P_{LX,LZ}, Q_{LX,LZ}$ record the respective $X$ and $Z$ matrix entries of the original logical operators over the $n$ original physical qudits. $L_{X,Z}$ are row vectors with $1$ nonzero entry and $k-1$ zero entries which correspond to the specific logical $X$ or $Z$ operator we have dualized. 

The conjoining operations described in the previous section can then be applied to the augmented check matrix. Note that it is perfectly valid to treat the tensor network as a stabilizer state and perform contractions on any tensor legs as needed. Once the gluing/conjoining operations are finished, one can then pick out the qudits of interest to be logical and physical legs. This is done by selecting the corresponding columns and then reversing the above ``tensorization'' procedures. As such, one can turn relevant stabilizer elements back into logical operators.  

However, if one only performs contractions on the physical legs and do not contract any of the logical edges, then it is also convenient to keep track of the original logical edges in the second and the fourth sections of the matrix separately. They remain the logical degrees of freedom after the conjoining operations, and one can revert them back into logical operators using the reverse procedure described above. In particular, if a corresponding logical operator has non-trivial entries in $L_{X,Z}$ but trivial entries in $P, Q$ after row reducing $H_A$, then it implies that certain apparent logical legs in the tensor networks do not encode independent degrees of freedom. These rows in the check matrix $H_A$ then become constraints that mark which logical legs are inter-dependent. Some explicit examples of this is found in Sec~\ref{sec:3}. 

\subsection{Tracing stabilizer codes: proof}
\label{subapp:dproof}
In this section we provide a quick proof for Theorems \ref{lm:stabtrace} and \ref{lm:csstrace}. The theorem statements are produced here for the readers' convenience.

\begin{theorem*}
Tracing stabilizer tensor(s) produce another stabilizer tensor.
\end{theorem*}

\begin{hproof}
For any stabilizer code, consider its encoding stabilizer tensor. Recall that any tracing/conjoining can be written as a sequence of single-trace and self-trace operations. To prove the above statement, it suffices to show that self-traces of any stabilizer tensor remains a stabilizer tensor. This is because for any single-trace operation, we can consider instead the two disconnected tensors as a single larger tensor whereby single trace is now given by a self-trace. 

Now any stabilizer tensor of rank $n$ can be used to describe a stabilizer state that admits a rank $n$ check matrix $H$. Note that from the procedures in Appendix~\ref{subapp:d1} by taking the case of $k=0$, self-trace on this check matrix always produces another check matrix with rank $n-2$, which corresponds to another stabilizer state. Therefore, self-traces of any stabilizer tensor remains a stabilizer tensor. Using the tensors to represent encoding maps, then traces or self-traces of stabilizer codes remain a stabilizer code.
\end{hproof}

\begin{theorem*}
Tracing (self-dual) CSS tensors produce another (self-dual) CSS tensor. 
\end{theorem*}

\begin{hproof}
Similar to the proof above, we treat the CSS tensor as a state and examine how its check matrix transforms under self-trace.
Recall that for an $n$-qudit CSS code, its check matrix can be written in a form

\begin{equation}
    \begin{pmatrix}
    H_X & \mathbf{0} \\
    \mathbf{0} & H_Z
    \end{pmatrix}
\end{equation}
where $H_X, H_Z,\mathbf{0}$ are block matrices with $n$ columns. If the state, and hence the associated tensor, is self-dual, then $H_X=H_Z$. Now to perform a self-trace, we perform row reduction and the subsequent matching separately on $H_X$ and $H_Z$. The conjoining operations requires that for each row we only need to match the two row entries in $H_X$ (or $H_Z$) separately. As the matching in $H_X$ and $H_Z$ are completely independent, this preserves the block structure in the new check matrix 
\begin{equation}
    \begin{pmatrix}
    H'_X & \mathbf{0} \\
    \mathbf{0} & H'_Z
    \end{pmatrix}.
\end{equation}
Hence the state or the $k=0$ code remains CSS. By definition it maps to a CSS tensor. 

If the tensor is self-dual CSS, then one only need to perform the matching or self-trace on $H_X$ to produce $H_X'$ because identical procedures apply to $H_Z=H_X$, which will yield $H_Z'=H_X'$. Hence the resulting tensor is a self-dual CSS tensor.
\end{hproof}

\section{Algorithm for tracking Stabilizers and logical operators}
\label{app:e}

Here we consider the algorithm of tracking the new stabilizer generators under conjoining or tensor gluing. To start, we assume that each bond/edge of a tensor has the same dimension $D$. For clarity, we will again assume that $D$ is a prime, even though for non-prime $D$ a similar argument holds.

It is often helpful to identify the stabilizer generators of a stabilizer code. In addition, we want to identify, for instance, the minimal weight logical operators and stabilizers, for instance. To facilitate such analysis, it is often convenient to first produce the check matrix of the overall stabilizer code generated from the tensor network. One can then perform whatever tools available to stabilizer codes to the check matrix to further ascertain its properties. 

The algorithm is simple: we assume that we are given the check matrices of each tensor prior to gluing, and then connect the edges one-by-one. This corresponds performing the conjoining operations on the relevant check matrices. We keep conjoining until all intended edges of the tensor network have been connected. 

More precisely, let us start with a number of tensors from which we want to generate a new tensor network with internal (glued) edges $E$.

\begin{algorithm}
\SetAlgoLined
 initialization\;
 \For{each edge $e\in E$}{Identify the corresponding columns in relevant check matrices\\
 Swap columns to the left\\
 Convert the matrices into reduced row echelon forms\\
 Produce a new check matrix $H$ through conjoining\\
 Row reduce $H$ and remove all zero rows
 }

 \caption{Check Matrix Building Algorithm}
\end{algorithm}

\textit{Algorithmic Complexity:}

Let the total number of dangling legs from the initial set of tensor modules be $N$. This includes the physical as well as the logical edges. Here we give an upper bound on its (time) complexity. 

One can encapsulate the union of all uncontracted tensors prior to tracing as a single check matrix of size at most $N\times 2N$. Then each tensor contraction simply corresponds to a self-trace operation on this matrix. This requires moving the columns around, which can be achieved in $O(1)$ time through swapping, and Gaussian elimination to prepare the matrix for self-trace. 
If the traced qudits are correctable erasures, then one only require again $O(1)$ operations to arrange the rows and columns in a desirable form. If, on the other hand, they are not, then one has to check whether the appropriate entries are matching. For a finite field of order $D$, it takes at most $D$ iterations. After $O(1)$ operations to eliminate the proper rows and columns, one again performs Gaussian elimination, which is $O(N^3)$ time\footnote{As we are operating on a finite field, we need not be concerned about the bit complexity for integer values.}. Therefore, each tracing procedure is of order $O(N^3+D)$.

Any subsequent self-tracings will operate on a check matrix that has size $n\leq N$, depending on the number of constraints and trivial rows the tracing produces. Because each dangling leg can be glued only once, and there are total of $N$ dangling legs, one can perform such tracing at most $N/2$ times. Then for any tensor network stabilizer codes built this way, the algorithm that produces the final code is at most $O(N^4+ND)$.

\section{Atomic Lego Blocks}
\label{app:atomiclego}

Here we show that atomic legos of rank-3, rank-2 and rank-1 outlined in Theorem \ref{thm:atomic} are sufficient in producing all quantum states, and by extension, encoding maps. 

\begin{hproof}
We first show that any quantum state over qudits can be constructed using these basic components. Let $|\psi\rangle$ be some state over $n$ qudits constructed from these atomic legos represented as a tensor network with $n$ dangling legs. For example, one can choose $|\psi\rangle = |0\rangle^{\otimes n}$ which is a trivial tensor network with $n$ rank-1 tensors. The encoding tensor of any 1-qudit code is a rank-2 tensor that corresponds to a single-qudit unitary. Therefore, subsequent contractions of any such rank-2 tensor on any leg of $|\psi\rangle$ acts as a single-qudit unitary gate. In the same way, contraction of any dangling leg of $|\psi\rangle$ with the rank-1 $|0\rangle$ tensor acts as a projective measurement in the $Z$-basis with eigenvalue $+1$. More generally, by contracting the $|0\rangle$ tensor with an additional rank-2 tensor, then with $|\psi\rangle$, we can also perform projective measurements in any basis. We notice that if we can construct an additional suitable two qubit gate to act on $|\psi\rangle$, then these tensors form a universal gate set and thus one can in principle produce any target quantum state. 

\begin{figure}
    \centering
    \includegraphics[width=0.25\textwidth]{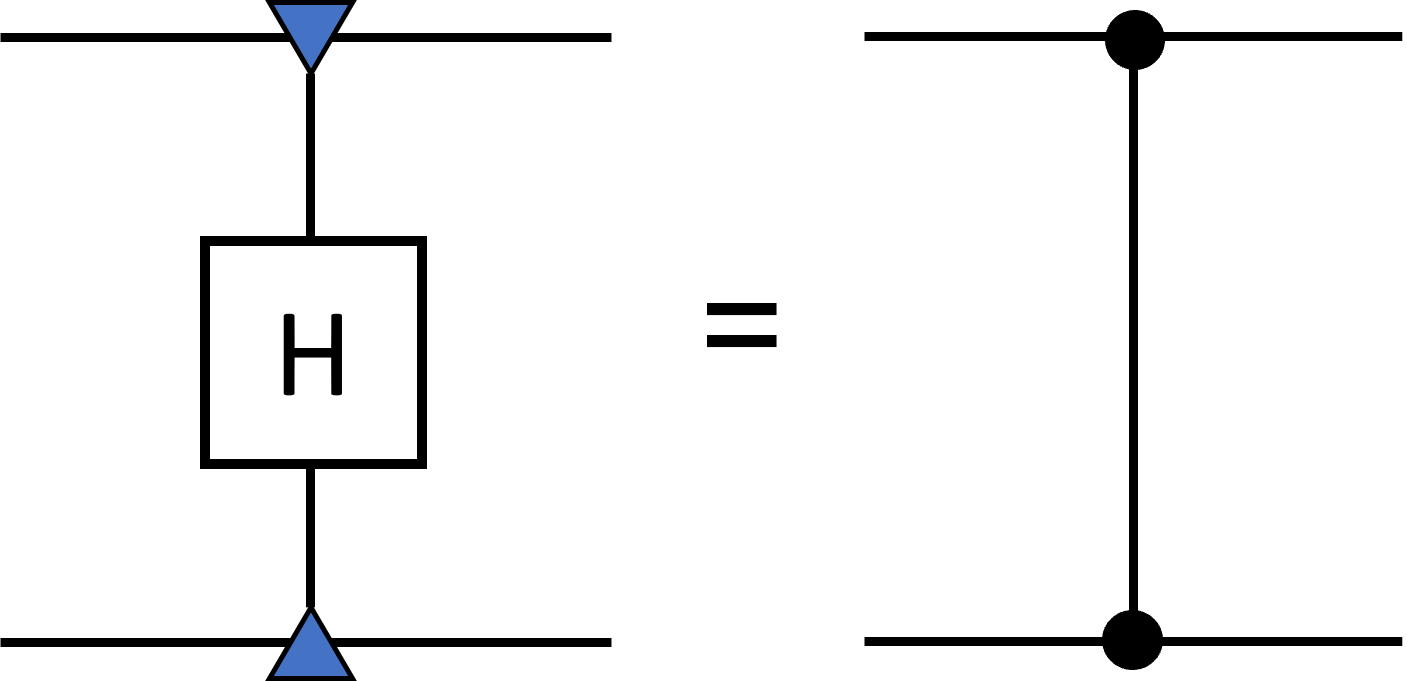}
    \caption{The triangular tensors are the encoding tensors of the repetition code and $H$ is the tensor that corresponds to the Hadamard or Fourier gate. Right figure represents the generalized controlled phase gate, which reduces to the CZ gate on qubits.}
    \label{fig:CZgate}
\end{figure}
For the remaining two-qubit gate, we construct the controlled phase gate, 
\begin{equation}
    \sum_{j=0}^d|j\rangle\langle j|\otimes Z^{j}
\end{equation}
which can be converted to a $\rm{CSUM}$ gate by Hadamard (or Fourier) conjugation on the second qudit \cite{quditstab}. Here $Z=\sum_{j} \omega^j\: |j\rangle\langle j|$ is the generalized Pauli-$Z$ operator on qudits with $\omega=\exp(i2\pi/d)$. The addition of this gate will render our gate set universal.
This gate can be synthesized via the tensor contraction in Figure~\ref{fig:CZgate} using the encoding tensor of the repetition code and a Hadamard. Indeed, let the repetition code encoding map be $\sum_i |ii\rangle\langle i|$, then the gate represented by tensor in Figure~\ref{fig:CZgate} can be written as 
\begin{align*}
    &\Big(\sum_{i,j=0}^{d-1} |i\rangle\langle ii|_{AB}\otimes |j\rangle\langle jj|_{CD} \Big)|H\rangle_{AC}\\
    =&\Big(\sum_{i,j=0}^{d-1} |i\rangle\langle ii|_{AB}\otimes |j\rangle\langle jj|_{CD}\Big)\Big( \sum_{k,l=0}^{d-1}\omega^{kl}|k\rangle_A|l\rangle_C\Big)\\
    =&\sum_{i,j=0}^{d-1} \omega^{ij}\,|i\rangle\langle i|\otimes |j\rangle\langle j|
    =\sum_{i=0}^{d-1} |i\rangle\langle i|\otimes Z^i
\end{align*}
where $|H\rangle$ is the Hadamard state. We have labelled the relevant Hilbert space factors to avoid confusion.  

As for quantum channels in the form of isometric encoding maps, one can first convert it into a quantum state and then build up its corresponding tensor network. Because the same tensor network represents the desired encoding map, the same atomic legos are sufficient for expressing any such encoding map.
\end{hproof}

\section{Details of Some Code Constructions}
\label{app:f}
\subsection{Toric code}
\label{subapp:422toric}
\begin{figure}
    \centering
    \includegraphics[width=0.45\textwidth]{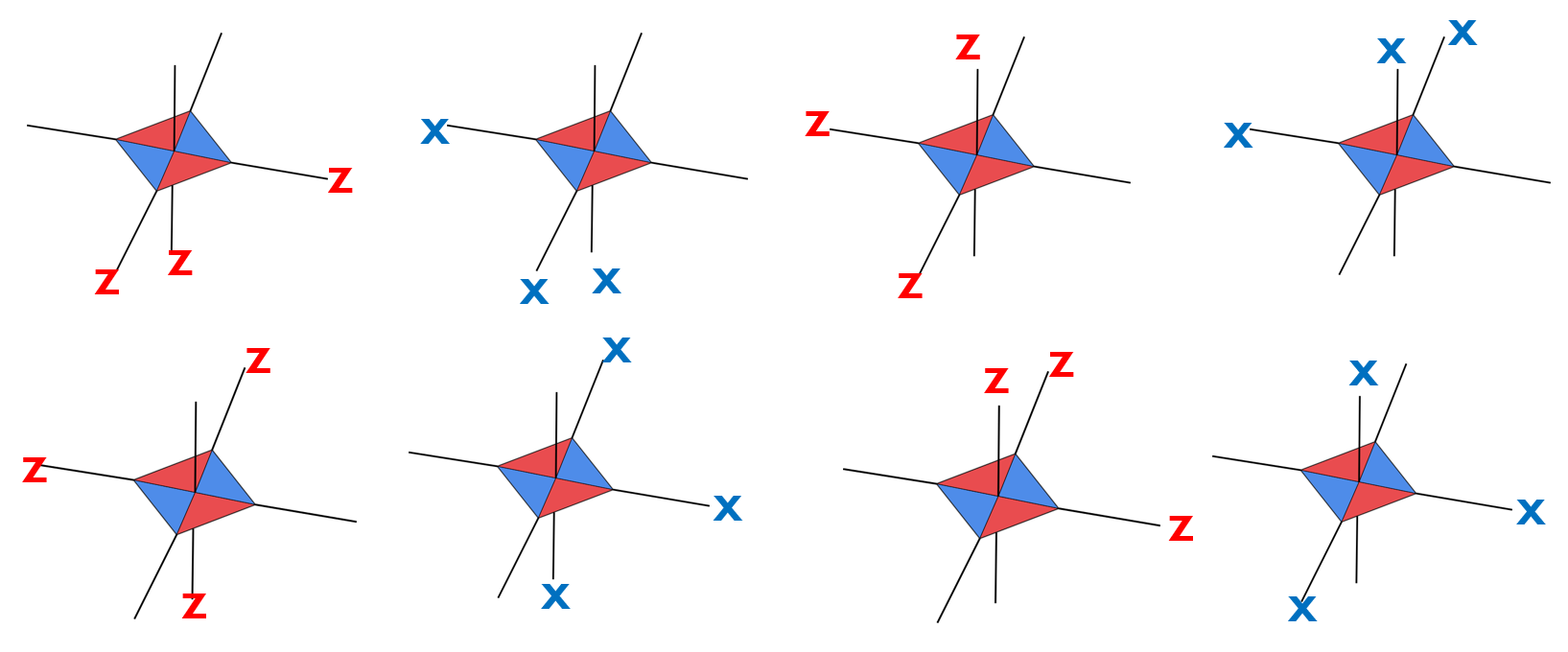}
    \caption{Some unitary product stabilizers of the rank-6 tensor derived from the $[[4,2,2]]$ encoding map. Identifying the two legs pointing up and down as the logical legs, these diagrams also produce the different representations of the logical operators $\bar{Z}\bar{I}, \bar{X}\bar{I}, \bar{I}\bar{Z},\bar{I}\bar{X}$ for the $[[4,2,2]]$ code.}
    \label{fig:422stabilizer}
\end{figure}

Starting with the $[[4,2,2]]$ code tensors, we can understand them as 6 qubit stabilizer states whose generators are given in Figure~\ref{fig:422stabilizer}. If we promote one of the legs (upward pointing) to be a logical degree of freedom, then the same tensor also represents a $[[5,1,2]]$ code where stabilizers that act non-trivially on the same leg now can be converted to logical operators. Similarly, it serves as an encoding map for the $[[4,2,2]]$ code where one promotes both the upward and downward pointing legs to logical degrees of freedom. The colours of the tensors are chosen such that a logical $Z$ ($X$) operator acting on the downward pointing logical leg is represented by weight-2 $ZZ$ ($XX$) operator acting on physical in-plane edges that are two sides of an angle coloured in red (blue). 

Now, it is tempting to simply contract these tensors into a grid of square lattices. However, to create a code with localized stabilizer generators, we need to choose the orientation of the tensors such that stabilizers pushed into a local tensor can flow through the tensor work in a small enough closed loop. Additionally, they need to act non-trivially on the physical qubits which are dangling legs. The simplest configuration was given in Figure~\ref{fig:toric_code_network} where there are small closed loops of $X$ or $Z$ types --- loops that have red triangles facing inward correspond to $Z$ type loops where pushing of the local Z stabilizer results in a weight 4 $ZZZZ$ stabilizer of the contracted tensor network. Similarly, if the inward facing triangles are blue, it results in the $X$ type stabilizers. Because by contracting CSS codes together, we end up with another CSS code, it is sufficient for us to check stabilizers that are purely $Z$ type or $X$ type. For any given type, the only remaining local stabilizer of a tensor that does not result in a closed loop when pushing operators is the weight for $XXXX,ZZZZ$ stabilizer that act on the 4 in-plane legs of the tensor. This operator leaves the physical leg untouched and continues to push any operators in one loop to another that is diagonally adjacent. However, because it is simply the product of the two stabilizers shown in Figure~\ref{fig:422stabilizer}, pushing using this stabilizer is equivalent to multiplying two of the adjacent star or plaquette operators. 
Therefore, for a tensor network with toroidal boundary condition, it does not produce any new stabilizers that have not been generated by the star or plaquette operators.

One can confirm that there are two actual logical degrees of freedom after imposing the independent constraints from pushing logical operators that result in stabilizers. We first note that the toric code has $2L^2$ physical qubits and $2L^2-2$ stabilizer generators from the plaquette and star operators \cite{Kitaev2003}. 

We now wish to know how many independent constraints one can produce by pushing only logical operators, i.e., the pushing of these logical operators end up being a stabilizer element. The set of constraints are said to be independent if the logical operators pushed in one set of constraint cannot be obtained by multiplying the logical operators pushed in the other constraints. 

In the case of the toric code tensor network, this is actually identical to counting the stabilizer generators of the toric code. We note that the tensor network has a symmetry such that the network remains invariant by exchanging $Z\leftrightarrow X$ and all the downward pointing physical and the upward pointing logical legs. For example, pushing $\bar{Z}\bar{Z}\bar{Z}\bar{Z}$ into the 4 logical legs around a closed loop with blue inward facing triangles result in the identity operator will now map to operator pushing that produces the $XXXX$ stabilizer under this transformation. Therefore, by identifying each such weight 4 logical operator with a stabilizer generator, the number of independent logical constraints must also be $2L^2-2$. Thus we arrive at the two remaining true logical degrees of freedom despite having $2L^2$ apparent logical legs in the tensor network.

From the above duality, we can also conclude that if we were to treat this tensor network as a $4L^2$ stabilizer state, i.e. all dangling legs are physical legs, then it is the unique ground state of two copies of the toric code Hamiltonian with four additional global string-like coupling terms. More explicitly, the stabilizer Hamiltonian is
\begin{equation}
    \hat{H}_{\rm state} = \hat{H}^{\rm toric}_A\otimes I_B +I_A\otimes \hat{H}_B^{\rm toric} + \hat{H}_{AB}^{\rm strings},
\end{equation}
where $A$ and $B$ label the two copies of the local toric code Hamiltonians $\hat{H}^{\rm toric}$. 
The local terms from the first two contributions in $\hat{H}_{\rm state}$ provide $4L^2-4$ local stabilizer generators. The remaining four non-local generators are given by $\hat{H}^{\rm strings}_{AB}$. From the toric code tensor network, they are precisely generated by pushing what was logical operators into strings that wrap around the torus (Figure~\ref{fig:toric_code_network}b). Therefore, in the above stabilizer Hamiltonian, the last coupling term is the sum of two X-type and two Z-type non-contractible Pauli string operators, 
\begin{equation}
    \hat{H}_{AB}^{\rm strings} = \sum_{i=1,2} L_{A}^{X}(i)\otimes L_{B}^{X}(i)+L_{A}^{Z}(i)\otimes L_{B}^{Z}(i).
\end{equation}
Each $L(i)$ is the conventional non-contractible string-like logical operator in the usual toric code construction where it wraps around the $i$-th direction of the torus. The superscript denotes the type of the logical operator while the subscript denotes on which copy of the toric code it acts.

\subsection{Boundary conditions and defects}
\label{subapp:defects}
We have seen in the previous example that by imposing the periodic boundary condition on the tensor network, we construct the toric code. One can similarly impose other boundary conditions. One possibility is to simply leave the dangling legs on the boundary open for patches shown in Figure~\ref{fig:toric_code_network}. We call this the bare boundary condition. This builds a subsystem code that is similar to the surface code in the bulk, in that it has similar stabilizer elements acting on stars and plaquettes, but is somewhat similar to the Bacon-Shor code on the boundary, where one can identify weight 2 $X$- or $Z$-type gauge operators. The logical (string) operators that connect different parts of the boundary can also be deformed to an operator that acts purely on the boundary degrees of freedom. An example of such a code is shown in Figure~\ref{fig:1dcode}. 

Alternatively, one can contract the dangling legs in the bare tensor network with ``stopper tensors'' which are simply eigenstates of Pauli operators. An example is shown in Figure~\ref{fig:surface_code}a, which reproduces the surface code. 

However, because of the flexiblity in a tensor network, we can choose other types of ``boundary tensors'' to contract with the bare network. This allows us to tune the boundary condition by using a combination of such different tensors according to our needs, which hopefully creates new codes with interesting properties. 

\begin{figure}
    \centering
    \includegraphics[width=0.45\textwidth]{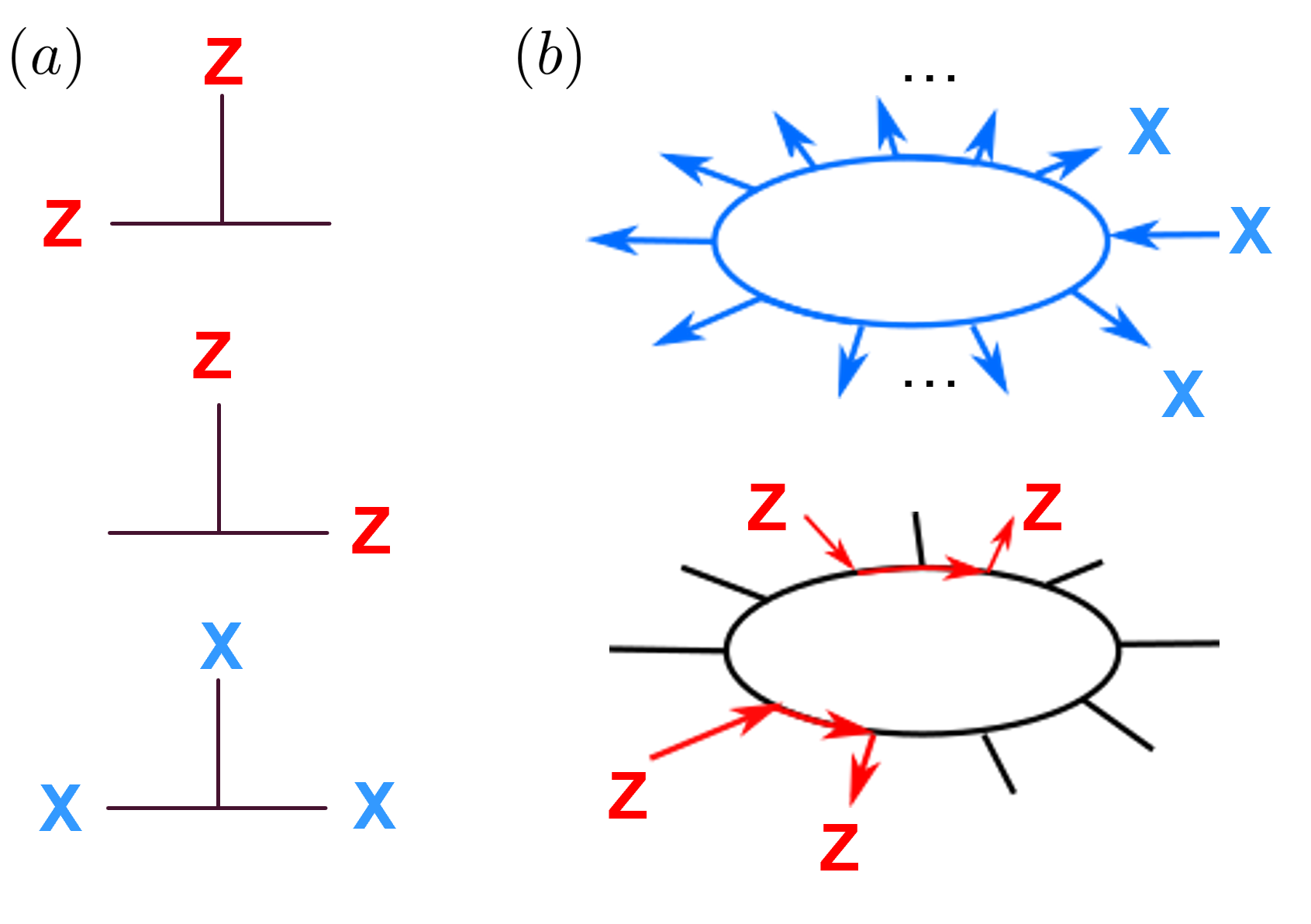}
    \caption{(a) Tensors that correspond to a 2-qubit repetition code or a 3 qubit stabilizer state stabilized by the generators shown. (b) By gluing the horizontal legs of these tensors, one can create larger tensors with more legs whose stabilizers can be determined through operator pushing.}
    \label{fig:rep_tensor}
\end{figure}

For a straightforward example, another type of tensor one can contract with the boundary is given in Figure~\ref{fig:surface_code}b. These tensors can be easily created by gluing together repetition codes (Figure~\ref{fig:rep_tensor}). In the example shown, the tensor is created by gluing 2 qubit repetition codes that are stabilized by the $ZZ$ operator. The resulting tensor is represented as the blue loop tensor in Figure~\ref{fig:surface_code}b. By exchanging $X\leftrightarrow Z$, one can also generate the tensor with a red loop. Their colours denote the type of the global stabilizer (X or Z). One can check that they simply correspond to GHZ states in different local bases. 

A subset of their legs can then be contracted with the dangling boundary legs of the bare tensor network while leaving the rest dangling. This helps create different boundary conditions while maintaining the ``bulk properties'' of the surface code. One can also understand this as a form of code concatenation, where we have encoded a subset of the boundary qubits with larger repetition codes. 

\subsubsection{XZZX code}
\label{subsubapp:xzzx}
\begin{figure}
    \centering
    \includegraphics[width=0.45\textwidth]{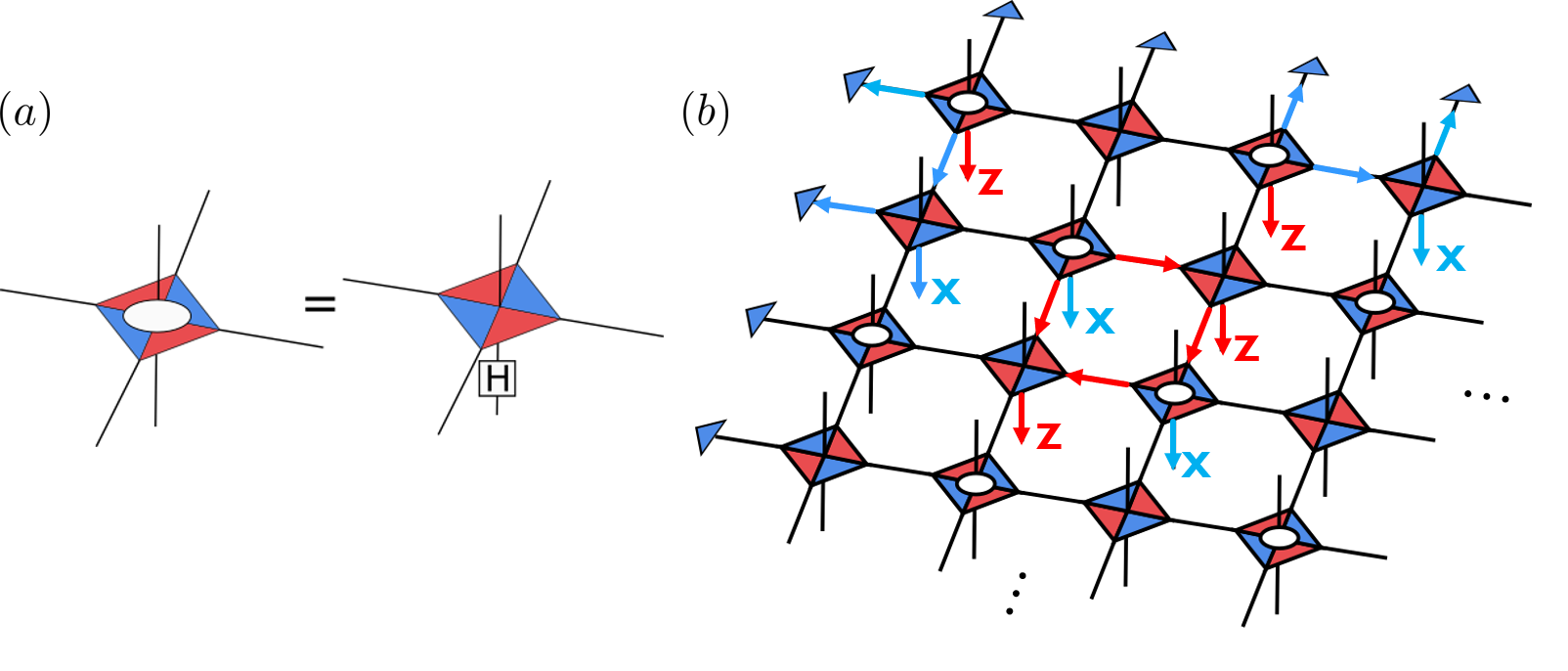}
    \caption{(a) The locally modified tensor by a Hadamard. (b) A sample XZZX code tensor network where the boundary is contracted with X stopper tensors. Some stabilizer generators are marked using operator pushing.}
    \label{fig:xzzx_app}
\end{figure}

In addition to the boundary, one can also make modifications to tensors in the ``bulk''.
It is known that the XZZX code \cite{xzzx_qm} is related to the surface code via local transformations. A tensor network construction reflects these changes. To do so, we simply modify every other tensor and pass their physical leg through a Hadamard gate (Figure~\ref{fig:xzzx_app}a). These tensors can be contracted into a network (Figure~\ref{fig:xzzx_app}b). In this example, we have contracted the boundary legs with ``$X$ stopper'' tensors which are simply $|+\rangle$. However, like our earlier example with the surface code, this is one of many different boundary tensors one can contract. The stopper tensors produce $ZX$ generators for alternating segments on the boundary. Then, pushing $X$-type stabilizers through the original tensors on one of these segments will result in an $XX$ stabilizer. However, because the $X$ on modified tensors is pushed through an extra Hadamard, a $Z$ operator comes out of the modified tensors.  
Similarly, for stabilizer operators pushed around a closed loop, the Hadamards attached to the modified tensors now flips two of the $X$s to $Z$s and vice versa. This creates the same $XZZX$ operator for every square in interior of the network. 

\subsubsection{Toric Code with a Twist}
\begin{figure*}
    \centering
    \includegraphics[width=0.9\textwidth]{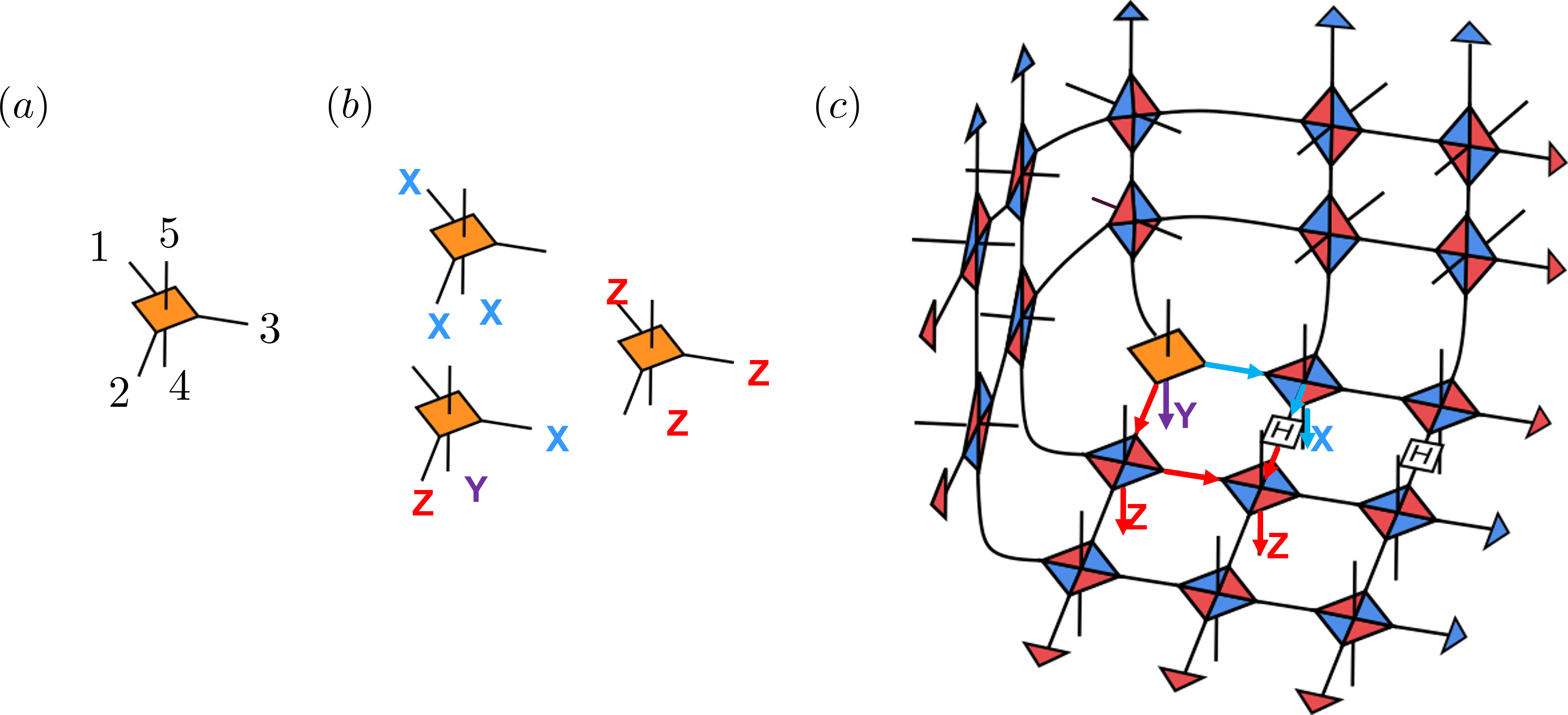}
    \caption{(a) The 4-qubit code tensor. (b) Stabilizer generators of the 4-qubit code. (c) The tensor network that corresponds to a ditance 5 triangle code. White squares marked with ``H'' are the $2\times 2$ matrix representation of the Hadamard gate. The orientation of the orange tensor is the same in all subfigures.}
    \label{fig:twistcode}
\end{figure*}

One can also use twist defects to perform the full local Clifford operations. Such defects also break the CSS nature of the surface code. One planar construction is known as the triangle code with a central twist defect \cite{surfacecode_twist}. We here present a tensor network that reproduces such triangle codes by using the $[[4,2,2]]$ code tensor, the Hadamard gate and a 4-qubit code tensor (Figure~\ref{fig:twistcode}). The same tensor network can be easily extended to larger codes, but here we have chosen to reproduce the distance 5 code defined in \cite{surfacecode_twist} for the sake of simplicity. 

As in the surface code, we take the tensor legs pointing inwards as logical legs and the ones pointing out as physical legs. This is indeed simply a variant of the surface code with a defect introduced by the orange tensor (Figure~\ref{fig:twistcode}a). Then, stabilizers can be generated through operator pushing. One such example is shown in Figure~\ref{fig:twistcode}c, where the generator acting on the square containing the orange tensor now support mixed type of Paulis. Similarly, for the plaquettes containing the Hadamard tensor, their corresponding stabilizer generator is also mixed, of the form $XXZZ$. 

Although the tensor network appears 3-dimensional, the geometry is fully planar in the form of a triangle \cite{surfacecode_twist}.

\subsection{1d code}
\label{subapp:1dcode}
\begin{figure*}
    \centering
    \includegraphics[width=\textwidth]{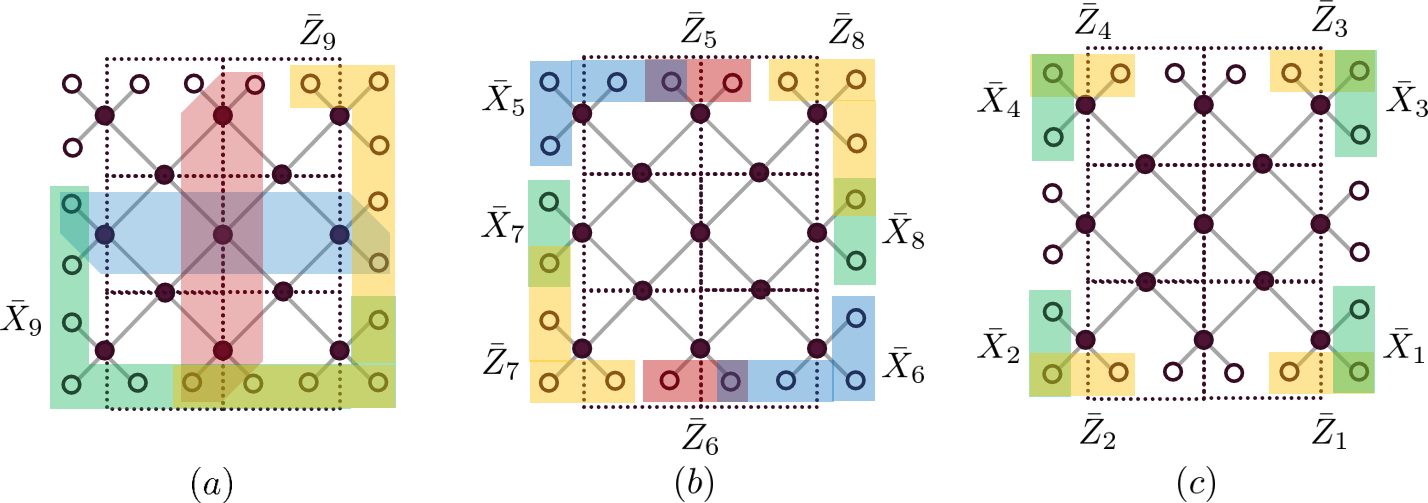}
    \caption{Starting from a 2d code with bare boundary condition, the boundary degrees of freedom we wish to preserve is denoted by circles whereas the bulk degrees of freedom are solid disks. If we promote all bulk degrees of freedom to logical legs, we then produce the encoding map of a 1d code. (a) One set of the logical (string) operators for the 2d code is given by all $Z$ (red) or all $X$ (blue) acting on the shaded qubits. Their equivalent boundary counterparts are given in yellow and green respectively. (b,c) Other logical operators for this example. Blue and green strings denotes the tensor product of Pauli $X$ acting on the shaded qubits while yellow and red indicate the product of $Z$ on the shaded qubits. For the 2d code, these strings can also be deformed to act on physical qubits in the bulk. For the dual 1d codes, they only act on the boundary in the way shown. }
    \label{fig:1dcode}
\end{figure*}

Once we have the 2d code with bare boundary condition, it is also easy to determine the relevant operators of a dual 1d code. We first identify the logical operators and stabilizers that act on the boundary (Figure~\ref{fig:1dcode}a). One can find that stabilizers are generated by the all $X$ and $Z$ Pauli strings that act solely on the boundary degrees of freedom (circles). The logical operators of the 2d code are given by different string operators that connect different segments of the boundary. One can identify several other such string operators from the 2d subsystem code then push them to the boundary through stabilizer multiplication. Their representations for such a 2d code is shown in (Figure~\ref{fig:1dcode}). 

To obtain a ``dual 1d code'', we  promote all bulk physical degrees of freedom to logical legs. We are then left with a set of qubits on the 1 dimensional boundary. For the example given, it is a $[[20,18,2]]$ code with two global stabilizer generators of all $X$ and all $Z$. One set of logical operators of an encoded qubit is shown as green and yellow strings in Figure~\ref{fig:1dcode}a. The other eight are shown as coloured strings in Figure~\ref{fig:1dcode}bc. By exchanging $X$ and $Z$ for the first nine sets of logical operators, we obtain the other 9 independent sets. Together they characterize the 18 logical qubits of this code.

Note that it is somewhat inaccurate to call this a 1d code as the stabilizer Hamiltonian constructed from its generators is not one that contains spatially local interaction terms. However, by fixing a suitable gauge, one can reduce the code to a $[[20,2,2]]$ stabilizer code that contains only 2-spatially-local stabilizer generators plus the two global all $X$ and all $Z$ generators.

This code is somewhat reminiscent of the holographic code in that it maps the bulk logical degrees of freedom to the boundary physical qubits. However, because of constraints and degrees of freedom counting, the ``bulk'' qubits are delocalized compared to the independent encoded qubits that are localized to the bulk AdS-sized regions (Figure~\ref{fig:holoRM}).
Note that its logical operators have a nested, multi-scale structure such that the support of certain logical operators is contained in the support of other logical operators. This is also somewhat similar to the hyperbolic holographic code.

Indeed, by switching all the bulk legs to logical degrees of freedom in the tensor network of the 2d code, we have converted into a 1d code which has a pseudo-holographic mapping in the form of the tensor network. 
Note that there are more apparent logical legs than there are actual logical degrees of freedom because of the constraints similar to what we have seen in the toric code example. 
In any case, the tensor network here serves as a natural connection between two distinct codes (1d vs 2d). We see that they are ``dual'' to each other by simply flipping some kets to bras (and vice versa) in the encoding map using a tensor network prescription.

In a similar fashion, we can consider other 2d codes with defects and non-trivial boundaries. By taking all the bulk legs to be logical legs, we can create their dual 1d codes with different properties which are manifestly related to the 2d codes through channel-state duality. For instance, one can repeat this exercise for the boundary that is contracted with repetition codes~(Figure~\ref{fig:surface_code}b).

\subsection{2d Bacon-Shor Code}
\label{subsec:2dbsc}
\begin{figure*}
    \centering
    \includegraphics[width=0.95\textwidth]{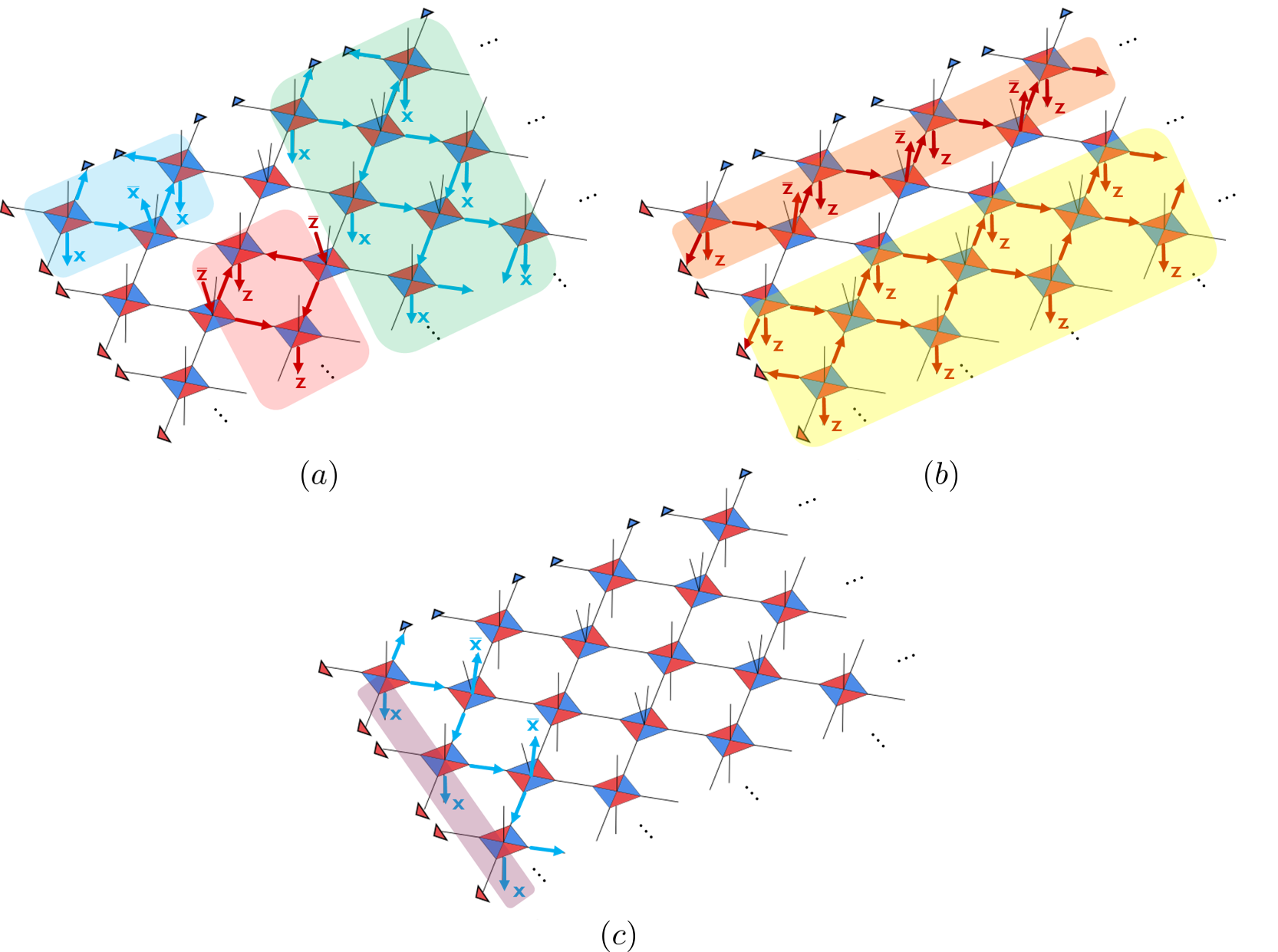}
    \caption{The tensor network of a 2d Bacon-Shor code. (a) shows the operator pushing that lead to the X (blue) and Z (red) type gauge operators by pushing logical and stabilizer operators. If we only push stabilizers of the local tensors, this results in a stabilizer generator (X type: green) and (Z type: yellow) in (b). (b) and (c) shows the operator pushing that correspond to the logical operator that has high weight. These X and Z type logical operators, shaded in purple and orange respectively, act on a column or a row.}
    \label{fig:2dbsc_tn}
\end{figure*}

Starting from the tensor network of the surface code, we can also ``re-interpret'' the dangling legs differently. One such option which preserves some degree of translation symmetry is to re-assign the physical dangling legs to be logical legs every other row. This is explicitly shown in Figure~\ref{fig:2dbsc_tn}, where we again keep the logical dangling legs pointing upwards while the physical dangling legs pointing down. We can also interpret this tensor network as constructed from alternating $[[5,1,2]]$ and $[[4,2,2]]$ codes without changing the interpretations of the dangling legs. 

Again, the physical qubits lie on the sites of an $M\times N$ lattice. By checking the stabilizer and the logical (gauge) operators through operator pushing, we can verify that it is nothing but the 2d Bacon-Shor code. Although it encodes large a number of qubits as a stabilizer code, we generally only keep the one with the largest distance while demoting the rest of the encoded qubits to gauge qubits. For $M=N$, this yields a $[[M^2, 1, M]]$ gauge code. An example for $M=3, N=4$ is constructed in Figure~\ref{fig:2dbsc_planar}. The tensors where we have both legs pointing up are coloured in yellow. Such tensors do not have a dangling physical leg, and therefore there is no physical qubit residing on those sites. Repeating the operator pushing exercises outlined in Figure~\ref{fig:2dbsc_tn}, we recover the generators of the gauge group, consisting of weight-2 $ZZ$ (acting on the two boundary sites of the vertical red edges) and weight-2 $XX$  operators (acting on the two sites adjacent to each horizontal blue edge). The $X$ logical operator (purple) is a weight $M$ Pauli X string acting on the shaded qubits while the logical $Z$ (orange) acts on a row of $N$ qubits with weight-$N$ Z strings.

This is an ``in-between'' example where some of the dangling legs in the bulk have been reassigned (c.f. Figure~\ref{fig:bulk_mod}b). Note that the apparent logical degrees of freedom (i.e. the upward pointing dangling legs) are again inter-dependent, not unlike what we encountered for the surface code tensor network. However, it is somewhat surprising that by simply ``dualizing'' the encoding map of the surface code one obtains a degenerate subsystem code related to the quantum compass model. While it is known that one can interpolate between the surface code and the Bacon-Shor code in the 2d compass code via gauge fixing\cite{compasscode}, it appears distinct from this tensor network description. 

\subsection{3d code}
\label{subapp:3dcode}
\begin{figure}
    \centering
    \includegraphics[width=0.4\textwidth]{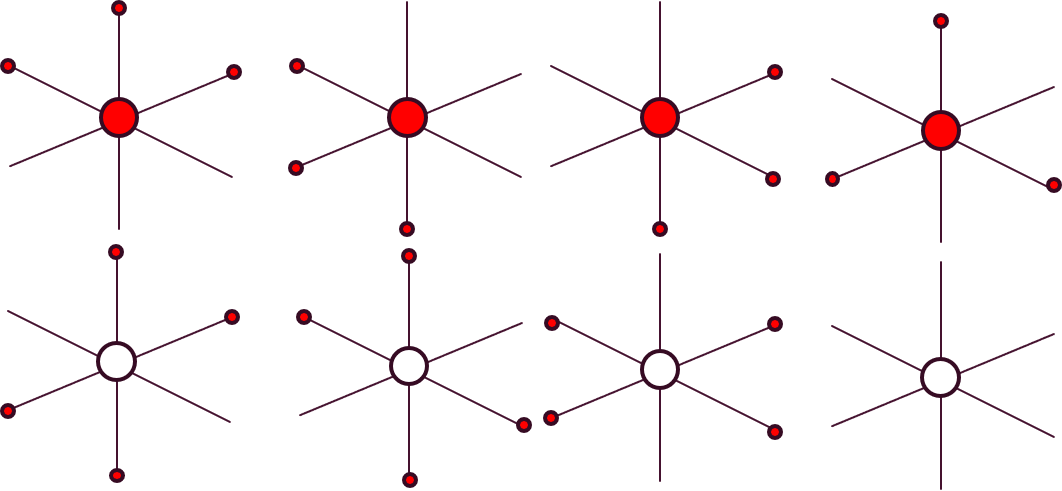}
    \caption{Graphical representation of the stabilizer elements that only contain $Z$ or $I$ operators. An identical set applies when we switch $Z$ to $X$. Disks coloured red denote Pauli $Z$s acting on the corresponding qubits.}
    \label{fig:steane_stab}
\end{figure}

We start with the $[[7,1,3]]$ Steane code as basic components in the tensor network. Each code can be represented as an 8-legged tensor (Figure~\ref{fig:steane_RM}b) among which 7 of them represent physical qubits and 1 represents the encoded qubit. For clarity, we represent such a tensor by Figure~\ref{fig:steane_stab}, where we orient 6 of the legs such that they are mutually perpendicular when embeded in 3-dimensional Euclidean space. The remaining 7th physical qubit and the logical leg are localized at the origin, where we drop them in the figure to avoid clutter. This is a CSS code, and all stabilizers that contain only Z (or X) are given by Figure~\ref{fig:steane_stab}, where a red dot indicates a Pauli Z operator acting on the corresponding physical qubit. If the interior of the circle located at the center is coloured, then we mean that a Z (or X) operator is inserted at the  physical leg located at the origin. 

\begin{figure}
    \centering
    \includegraphics[width=0.45\textwidth]{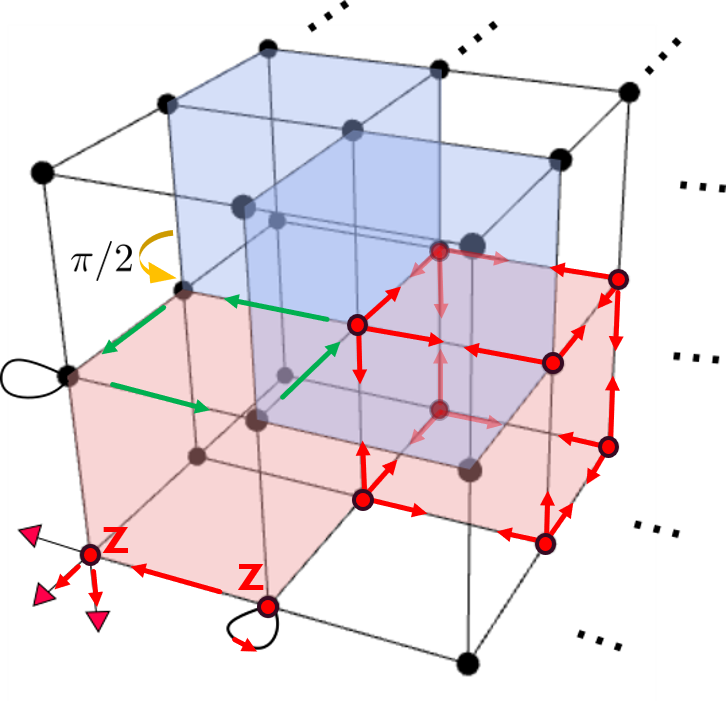}
    \caption{A tensor network that mimics a 3d cubic lattice. It represents a code where a physical qubit rests on each lattice site in the bulk. There is one logical and one physical dangling leg on each vertex. They are not shown to avoid clutter. The tensor with a yellow curly arrow hovering above is rotated relative to the central tensor by 90 degrees counterclockwise. The specific boundary conditions has to be treated separately by contracting the dangling legs for tensors on the boundary. We show an example for such contraction for the three tensors in the lower left corner. This results in $ZZ$ stabilizers that act on an edge. Localized stabilizers can be generated by pushing/matching the stabilizers of each Steane code tensor that act on the appropriate corners (red arrows). The pushing is consistent because Pauli $Z$s (or Xs) on the contracted edges match and we are left with a stabilizer that acts with the same Pauli operator on all corners of a coloured cube.}
    \label{fig:3d_stab}
\end{figure}

We then contract these tensors in a cubic lattice. They are oriented appropriately such that we end up with localized stabilizer elements (Figure~\ref{fig:3d_stab}). The tensor in the center of the network has the same orientation as the ones in Fig~\ref{fig:steane_stab} such that the stabilizers in the top row of Fig~\ref{fig:steane_stab} subtend the 4 corners of the adjacent coloured cubes. Starting with the central vertex, we move along the adjacent squares on the same horizontal plane in the counterclockwise direction. One such paths is marked by green arrows in Figure~\ref{fig:3d_stab}. When we reach the next vertex in the square, we rotate the tensor by 90 degrees counterclockwise relative to the previous tensor along the path. The tensor immediately above (or below) the central tensor is oriented such that it is a reflected version of the central tensor about the horizontal plane between them. The rest of the lattice can be generated similarly by translation and rotation. 
This creates a tensor network that has localized stabilizer elements which are $ZZZZZZZZ$ (or $XXXXXXXX$) that act on all 8 physical qubits that are located on the corners of the coloured cubes through operator pushing. One such example is shown by following the red arrows in (Figure~\ref{fig:3d_stab}). Again, the interior of a vertex is coloured red if a Pauli Z operator is acting on the dangling physical leg on the vertex, which we do not draw in the figure.

These cubes are positioned in the alternating manner as shown in Figure~\ref{fig:3d_stab}. Whether these stabilizers end up generating the entire stabilizer group of the final code will also largely depend on how we treat the boundary conditions, not unlike the surface code example we have discussed. They can be contracted differently according to our needs. As it is becoming somewhat cumbersome to track by hand, we leave the detailed analysis of this code to future work. 

In a similar way, one can examine the logical operators graphically. Because logical X and Z operators are simply the tensor product of 7 Xs or Zs, such a logical $Z$ (or $X$) operator is supported on the uncoloured qubits in Figure~\ref{fig:steane_stab}. Therefore, pushing of logical operators propagate along straight lines and acts on the physical qubits the line passes through. Alternatively, it can reach a corner and propagate in the other 2 directions orthogonal to the incoming direction. An easy way to keep track of these paths is that they traverse through the edges of the uncoloured boxes. For straight lines this is relatively trivial because every edge is adjacent to an uncolored box. However, if we use the corner terms, then they move along the corner of the transparent boxes.

\begin{figure}
    \centering
    \includegraphics[width=0.45\textwidth]{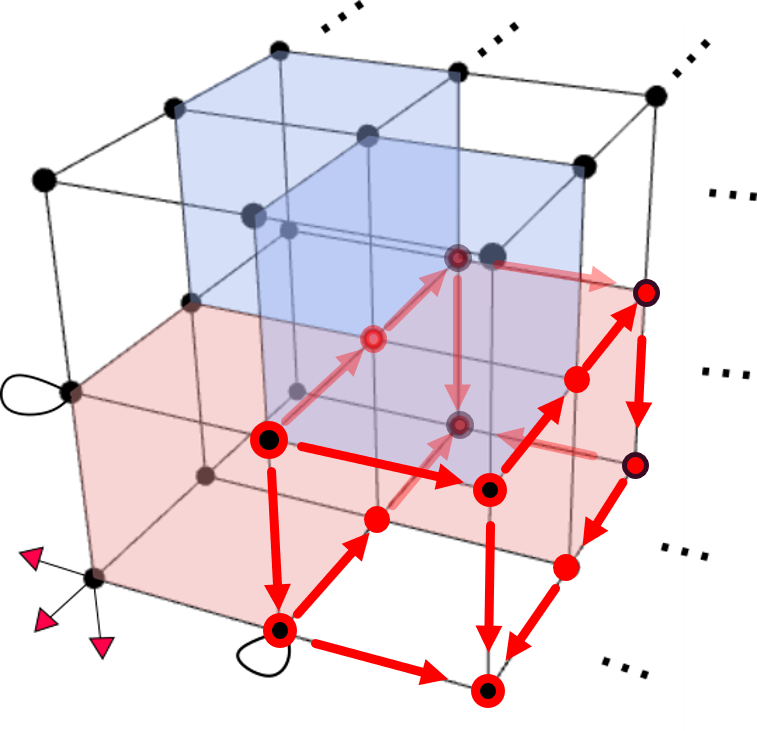}
    \caption{Pushing logical $\bar{Z}$ operators and stabilizers of the Steane tensor produces a box-like operator flow, which is nothing but an all $Z$ stabilizer acting on the dangling physical legs located at the eight corners of a red-coloured cube.}
    \label{fig:3d_latt_box}
\end{figure}

If we generate an operator flow using a combination of the Steane code logical and stabilizer operators, we can produce logical operators and possibly stabilizers with different geometric shapes. For example, one such operator is shown in Figure~\ref{fig:3d_log}, which takes on the form of a tree. Various other patterns can also be generated on a larger lattice. 
Similar to the toric code tensor network, there is a large degree of interdependence among the apparent logical degrees of freedom in the tensor network because some combination of the logical and stabilizer operator pushing produce a stabilizer. This introduces a constraint in the apparent ``logical subspace'' (Figure~\ref{fig:3d_latt_box}).

\end{document}